\documentclass[english,12pt]{article}
\pdfoutput=1
\usepackage[a4paper,bottom=4.2cm,top=1cm,head=3cm,width=18cm,dvipdfm]{geometry}
\evensidemargin=\oddsidemargin

\usepackage[T1]{fontenc}
\usepackage[latin9]{inputenc}
\usepackage{amstext}
\usepackage{babel}
\usepackage{verbatim}
\usepackage{amsfonts}
\usepackage{mathrsfs}
\usepackage{amsmath}
\usepackage{amssymb}
\usepackage{bm}
\usepackage{extarrows}
\usepackage{slashed}
\usepackage{isodateo}
\usepackage{xcolor}
\usepackage{graphicx}
\usepackage[low-sup]{subdepth}
\usepackage{isodateo}
\usepackage{caption}
\usepackage{subcaption}
\usepackage[linktocpage=true]{hyperref}
\hypersetup{
	colorlinks=true,%
	linkcolor=[rgb]{0,0,1}, 
	citecolor=[rgb]{0,0,1}
}

\numberwithin{equation}{section}

\makeatletter
\@ifundefined{textcolor}{}
{%
	\definecolor{BLACK}{gray}{0}
	\definecolor{WHITE}{gray}{1}
	\definecolor{RED}{rgb}{1,0,0}
	\definecolor{GREEN}{rgb}{0,1,0}
	\definecolor{BLUE}{rgb}{0,0,1}
	\definecolor{CYAN}{cmyk}{1,0,0,0}
	\definecolor{MAGENTA}{cmyk}{0,1,0,0}
	\definecolor{YELLOW}{cmyk}{0,0,1,0}
}

\newcommand{\red}{\color{red}}

\newcommand{\blu}{\color{blue}}

\newcommand{\fr}[2]{\mbox{$\frac{\,{#1}\,}{#2}$}}
\renewcommand{\rm}{\mathrm}
\def\bge{\begin{equation}}
	\def\ede{\end{equation}}
\def\bga{\begin{aligned}}
	\def\eda{\end{aligned}}
\newcommand{\beq}{\begin{equation}}
	\newcommand{\eeq}{\end{equation}}
\newcommand{\bq}{\begin{equation}}
	\newcommand{\eq}{\end{equation}}
\newcommand{\ba}{\begin{array}}
	\newcommand{\ea}{\end{array}}
\newcommand{\beqa}{\begin{eqnarray}}
	\newcommand{\eeqa}{\end{eqnarray}}
\newcommand{\beqs}{\begin{subequations}}
	\newcommand{\eeqs}{\end{subequations}}

\def\nn{\nonumber}

\def\dis{\displaystyle}

\def\({\left(}
\def\){\right)}

\def\deg{\circ}
\def\leqq{\leqslant}
\def\geqq{\geqslant}

\def\End{\end{document}}

\def\d{\text{d}}
\def\ii{{\tt i}}
\def\over{\overline}

\def\al{\alpha}
\def\be{\beta}
\def\ga{\gamma}

\def\ep{\epsilon}

\def\ito{\!\rightarrow\!}

\def\over{\overline}

\renewcommand{\rm}{\mathrm}
\def\bge{\begin{equation}}
\def\ede{\end{equation}}
\def\bga{\begin{aligned}}
\def\eda{\end{aligned}}

\def\nn{\nonumber}

\def\dis{\displaystyle}

\def\({\left(}
\def\){\right)}
\def\[{\left[}
\def\]{\right]}

\def\deg{\circ}
\def\End{\end{document}}

\def\over{\overline}

\def\leqq{\leqslant}
\def\geqq{\geqslant}

\def\al{\alpha}
\def\be{\beta}

\def\lam{\lambda}

\def\ga{\gamma}

\def\ep{\epsilon}
\def\lam{\lambda}

\def\si{\sigma}

\def\di{\mathrm{d}}

\def\to{\rightarrow}

\def\ii{\mathrm{i}}

\def\gaga{\gamma\gamma}

\def\cut{\Lambda}

\newcommand{\mT}{\mathcal{T}}

\newcommand{\mL}{\mathcal{L}}
\newcommand{\mO}{\mathcal{O}}

\def\gaga{\gamma\gamma}

\def\SZZ{\mathcal{Z}}

\def\gaga{\gamma\gamma}

\def\shat{\hat{s}}
\def\sbar{\bar{s}}
\def\OBW{\mathcal{O}_{\widetilde{B}W}^{}}
\def\OGP{\mathcal{O}_{G+}^{}}
\def\OGM{\mathcal{O}_{G-}^{}}
\def\OCP{\mathcal{O}_{C+}^{}}
\def\OCM{\mathcal{O}_{C-}^{}}
\def\TT{\mathcal{T}}
\def\hs{\hspace*{0.3mm}}
\def\hsx{\hspace*{0.5mm}}
\def\hsm{\hspace*{-0.3mm}}
\def\hsmx{\hspace*{-0.5mm}}

\def\to{\rightarrow}
\def\ito{\!\rightarrow\!}

\def\over{\overline}

\def\wtil#1{\widetilde{#1}}

\def\xt{\tilde{x}}
\def\yt{\tilde{y}}
\def\xb{\bar{x}}
\def\yb{\bar{y}}

\def\XX{\mathbb{X}}
\def\sigmaX{\hat{\sigma}_{\!x}^{}}
\def\sigmaY{\hat{\sigma}_{\!y}^{}}

\newlength{\halfpagewidth}
\divide\halfpagewidth by 2

\def\End{\end{document}}
\renewcommand{\thefootnote}{\fnsymbol{footnote}}
\makeatother

\begin{document}

\thispagestyle{empty}

\begin{flushright}
\vspace*{-10mm}
KCL-PH-TH/2022-35, CERN-TH-2022-089
\end{flushright}
\vspace*{2mm}

\begin{center}

{\Large\bf Probing Neutral Triple Gauge Couplings
\\[2mm] 
at the LHC and Future Hadron Colliders}

\vspace{8mm}

{{\bf John Ellis}\,$^{a,b,c,d}$,
~~{\bf Hong-Jian He}\,$^{d,e}$,
~~{\bf Rui-Qing Xiao}\,$^{a,d}$}

\vspace*{3mm}
{\small 
$^{a}$\,Department of Physics, King's College London, Strand, London WC2R 2LS, UK
\\[1.5mm]
$^{b}$\,Theoretical Physics Department, CERN, CH-1211 Geneva 23, Switzerland 
\\[1.5mm]
$^{c}$\,NICPB, R{\"a}vala 10, 10143 Tallinn, Estonia
\\[1.5mm]
$^d$\,Tsung-Dao Lee Institute and School of Physics \& Astronomy,\\ 
Key Laboratory for Particle Astrophysics and Cosmology (MOE),\\
Shanghai Key Laboratory for Particle Physics and Cosmology,\\
Shanghai Jiao Tong University, Shanghai, China
\\[1.5mm]
$^e$\,Physics Department \& Institute of Modern Physics
Tsinghua University, Beijing, China;\\
Center for High Energy Physics, Peking University, Beijing, China}
\\[1.5mm]
(\,john.ellis@cern.ch, hjhe@sjtu.edu.cn, xiaoruiqing@sjtu.edu.cn\,)
\end{center}

\vspace{2mm}
\begin{abstract}
\baselineskip 17pt
\noindent
We study probes of neutral triple gauge couplings (nTGCs)
at the LHC and the proposed 100\,TeV $pp$ colliders, 
and compare their sensitivity reaches 
with those of the proposed $e^+ e^-$ colliders.\ 
The nTGCs provide a unique window to the new physics beyond
the Standard Model (SM) because they can arise from
SM effective field theory (SMEFT) operators 
that respect the full electroweak gauge group 
$\rm{SU(2)}_{\rm L}^{}\!\otimes\hsm\rm{U(1)}_{\rm Y}^{}$ 
of the SM only at the level of dimension-8 or higher.\
We derive the neutral triple gauge vertices (nTGVs) 
generated by these dimension-8 operators in the broken phase 
and map them onto a newly generalized form factor formulation,
which takes into account only the residual U(1)$_{\rm{em}}^{}$ 
gauge symmetry.\ Using this mapping, we derive new relations 
between the form factors that guarantee a truly consistent 
form factor formulation of the nTGVs and remove large unphysical  
energy-dependent terms.\ We then analyze the sensitivity reaches 
of the LHC and future 100\,TeV hadron colliders for probing  
the nTGCs via both the dimension-8 nTGC operators and
the corresponding nTGC form factors in the reaction 
$\hs p{\hs}p{\hs}(q{\hs}\bar{q})\ito Z\gamma\hs$
with $Z\hsm\ito\ell^+\ell^-\! ,\hs\nu{\hs}\bar{\nu}{\hs}\hs$.\ 
We compare their sensitivities with the existing LHC measurements
of nTGCs and with those of the high-energy $e^+e^-$ colliders.\ 
In general, we find that the
prospective LHC sensitivities are comparable to those of
an $e^+ e^-$ collider with center-of-mass energy $\leqq\!1$\,TeV,
whereas an $e^+ e^-$ collider with center-of-mass energy 
$(3 - 5)$\,TeV would have greater sensitivities,
and a 100\,TeV $pp$ collider could provide the most sensitive probes of the nTGCs. 
\\
\begin{center}
(\,Phys.\ Rev.\ D in press, {\red Editors' Suggestion}\,)
\end{center}
\end{abstract}

\vspace{1cm}

\newpage
\baselineskip 18pt
\tableofcontents

\vspace*{5mm}

\setcounter{footnote}{0}
\renewcommand{\thefootnote}{\arabic{footnote}}

\section{\hspace*{-2.5mm}Introduction}
\label{sec:Intro}
\label{sec:1}

Neutral triple-gauge couplings (nTGCs) provide a unique window
for probing the new physics beyond the Standard Model (SM). 
It is well known that 
they do not appear among the dimension-4 terms of the
SM Lagrangian, nor are they generated by dimension-6
terms in its extension to the Standard Model Effective
Field Theory (SMEFT)\,\cite{SMEFT-Rev}.\ 
Instead, the nTGCs first appear through the gauge-invariant 
dimension-8 operators\,\cite{Gounaris:1999kf}-\cite{Ellis:2019zex}
in the SMEFT. Hence any indication of a non-vanishing
nTGC would be direct {\it prima facie} evidence for new physics 
beyond the SM, which is different in nature from anything that
might be first revealed by dimension-6 operators of the 
SMEFT\,\cite{SMEFT}-\cite{dim6B}.\ 
Moreover, searching for the effects of interference between the other
dimension-8 interactions and the SM contributions to amplitudes
must contend with possible contributions that are quadratic
in dimension-6 interactions, which is not an issue for
the nTGCs.

\vspace*{1mm}

Relatively few experimental probes of dimension-8 SMEFT
interactions have been proposed in the literature.\ 
One of them is the nTGCs mentioned 
above\,\cite{Gounaris:1999kf}-\cite{Ellis:2019zex}, 
which first arise from the dimension-8 operators of the SMEFT
and have no counterpart in the SM Lagrangian of dimension-4 
or in the dimension-6 SMEFT interactions.\   
Recent works have studied how the nTGCs can be probed 
by measuring $Z\gamma$ production at high-energy $e^+ e^-$ colliders\,\cite{Ellis:2020ljj}\cite{Ellis:2019zex}\cite{nTGC-other}
and $pp$ colliders\,\cite{pp-nTGC} under planning.\
Other examples include light-by-light scattering\,\cite{EMY},\ 
which has been measured at the LHC and could also be interesting 
for high-energy $e^+ e^-$ colliders\,\cite{EMRY}, and the processes
gluon$\hs +\hs$gluon $\ito \gamma\hsm+\hsm\gamma$\,\cite{EG} and gluon$\hs +\hs$gluon $\ito Z\hsm+\hsm\gamma$\,\cite{EGM}, 
which have been probed at the LHC.\ 
There are also recent studies on the dimension-8 operators induced
by top-like heavy vector quarks and the their probes via
$t\bar{t}h$ production at hadron colliders\,\cite{Dawson1-d8}, 
and on the dimension-8 operators induced by the heavy Higgs doublet 
of the two-Higgs-doublet model\,\cite{Dawson2-d8}.

\vspace*{1mm}

In this work, we present a systematic study of the  
sensitivity reaches of probing the 
dimension-8 nTGC interactions by measuring    
$Z\gamma$ production at the LHC{\hs}(13{\hs}TeV) and the  
$pp\hs$(100{\hs}TeV) colliders.\
The nTGCs are coupling coefficients of the
the neutral triple gauge vertices (nTGVs), 
which are often parametrized 
in terms of effective form factors 
that respect only the residual U(1)$_{\rm{em}}^{}$ gauge symmetry 
of the electromagnetism.\ 
This is in contrast with the dimension-8 nTGC operators of the
SMEFT, which respect the full electroweak gauge group
$\text{SU(2)}_{\rm L}^{}\!\otimes\hsm\text{U(1)}_{\rm Y}^{}$ 
of the SM.\  We derive the nTGVs from
these dimension-8 operators in the broken phase and map
them onto a newly generalized form factor formulation of the nTGVs.
Using this mapping, we derive new nontrivial relations among the
form factor parameters that ensure a truly consistent form factor
formulation of the nTGVs and remove unphysically  
large energy-dependent terms.\ 
Using these, we analyze systematically the sensitivity reaches 
of the LHC and future hadron colliders for   
nTGC couplings via both the dimension-8 nTGC 
operators and the corresponding nTGC form factors.\
We also make a direct comparison of our LHC analysis
with the existing LHC measurements of nTGCs in the reaction 
$\hs p{\hs}p{\hs}(q{\hs}\bar{q})\hsm\ito\hsm Z\gamma\hs$
with $Z\!\ito\nu{\hs}\bar{\nu}{\hs}\hs$ 
by the CMS\,\cite{CMS2016nTGC-FF} and ATLAS\,\cite{Atlas2018nTGC-FF} collaborations
based on the conventional nTGC form factor formulation 
that takes into account only the unbroken 
U(1)$_{\rm{em}}$ gauge symmetry\,\cite{nTGC1}\cite{Degrande:2013kka}.\ 
From this comparison, we demonstrate the importance
of using our proposed SMEFT form factor approach 
to analyze nTGC constraints at the LHC and  future 
high-energy colliders.

\vspace*{1mm}

The outline of this paper is as follows.\  
In Section\,\ref{sec:AandXS} we review the 
parametrization of nTGCs and derive the cross sections
for the reaction $\,q\hs\bar{q}\ito Z \gamma\,$ 
(followed by $Z\hsm\ito\hsm f\bar{f}\hs$ decays) 
as induced by the nTGCs.\ 
We also analyze the perturbative unitarity bounds 
on the nTGCs, showing that they are much weaker than the 
collider limits we present in Sections\,\ref{sec:4}-\ref{sec:5}.\
Then, in Section\,\ref{sec:3}{\hs} we present a newly generalized 
form factor formulation of the nTGCs and demonstrate that 
the full spontaneously-broken electroweak gauge symmetry
$\text{SU(2)}_{\rm L}^{}\!\otimes\!\text{U(1)}_{\rm Y}^{}$ 
of the SM leads to 
important restrictions on the nTGC form factors.\
As noted above, the full electroweak gauge symmetry 
is respected by the construction of the SMEFT, 
where the nTGCs appear first through 
dimension-8 operators.\ Using this formulation, we study in
Section\,\ref{sec:4}{\hs} the sensitivities of the LHC and
future $pp\hs$(100\,TeV) colliders for probes of the nTGCs
in the reaction $pp{\hs}(q\bar{q}\hsm\ito\hsm Z\ga)$
with $Z\ito \ell\hs\bar{\ell},\nu\hs\bar{\nu}\hs$.\  
We make a direct comparison of the sensitivity bounds using
our SMEFT formulation of nTGCs with the existing LHC 
measurements on the nTGCs.\ 
In Section\,\ref{sec:5}, we further present
a systematic comparison with the sensitivity reaches 
of the prospective high-energy $e^+ e^-$ colliders.\ 
Finally, we summarize our findings and 
conclusions in Section\,\ref{sec:6}.

\section{\hspace*{-2.5mm}Scattering Amplitudes and Cross Sections
for nTGCs}
\label{sec:AandXS}
\label{sec:2}

In this section, we first set up the notations and 
present the dimension-8 operators for the 
neutral triple gauge couplings (nTGCs) and the corresponding 
neutral triple gauge vertices (nTGVs).\
Then, we derive the nTGC contributions to the $Z\ga$ amplitudes
and cross sections.\ Finally, we derive the perturbative unitarity
constraints on the nTGC couplings.


\vspace*{1mm}
\subsection{\hspace*{-2.5mm}nTGCs from the Dimension-8 Operators}
\label{sec:2.1}
\vspace*{1mm}

In previous works\,\cite{Ellis:2020ljj}\cite{Ellis:2019zex} 
we studied the dimension-8 operators that generate
nTGCs and for their contributions to helicity amplitudes 
and cross sections at $e^+e^-$ colliders.
In particular, we identified a new set of 
CP-conserving pure gauge operators of dimension-8
for the nTGCs, one of which ($\mO_{G+}^{}$) can give
leading contributions to the neutral triple gauge boson vertices 
$Z\ga Z^*$ and $Z\gaga^*$ with enhanced energy-dependences
$\propto\! E^5$.
In this subsection, we recast them for
our applications to the LHC and 
future high-energy $pp$ colliders.

\vspace*{1mm}

The general dimension-8 SMEFT Lagrangian takes the following form:
%
\beqa
\Delta\mathcal{L}(\text{dim-8})
\,=\, \sum_{j}^{}
\frac{\tilde{c}_j^{}}{\,\tilde{\cut}^4\,}\mathcal{O}_j^{}
\,=\, \sum_{j}^{}
\frac{\,\text{sign}(\tilde{c}_j^{})\,}{\,\cut_j^4\,}\mathcal{O}_j^{}
\,=\, \sum_{j}^{}
\frac{1}{\,[\cut_j^4]\,}\mathcal{O}_j^{}
\,,
\label{cj}
\eeqa
where the dimensionless coefficients $\,\tilde{c}_j^{}$ 
are expected to be around ${O}(1)$
and may take either sign, 
$\,\text{sign}(\tilde{c}_j^{})\!=\!\pm$\,.\
For each dimension-8 operator $\mathcal{O}_j^{}$\,,
we have defined in Eq.\eqref{cj} the corresponding effective 
cutoff scale for new physics,
$\,\cut_j^{} \hsmx\equiv\hsm \tilde{\cut}/|\tilde{c}_j^{}|^{1/4}\,$.\
We also introduced a notation
$\,[\cut_j^4]\hsm\equiv\hsm\rm{sign}(\tilde{c}_j^{})\cut^4_j\hs$.

\vspace*{1mm}

We have analyzed the following set of dimension-8  operators\,\cite{Ellis:2020ljj} that are
relevant for our nTGC analysis:
\beqs
\vspace*{-3mm}
\label{eq:nTGC-d8}
\begin{eqnarray}
\label{eq:OG+}
g \mathcal{O}_{G+}^{} \!\!\!&=&\!\!	
\widetilde{B}_{\!\mu\nu}^{}	 W^{a\mu\rho}
( D_\rho^{} D_\lambda^{} W^{a\nu\lambda} \!+\! D^\nu D^\lambda W^{a}_{\lambda\rho}) ,	
\\[1.5mm]
\label{eq:OG-}
g \mathcal{O}_{G-}^{} \!\!\!&=&\!\! 		
\widetilde{B}_{\!\mu\nu}^{} W^{a\mu\rho}
( D_\rho^{} D_\lambda^{} W^{a\nu\lambda} \!-\! D^\nu D^\lambda W^{a}_{\lambda\rho}) ,
\\
\label{eq:OBW}
\mathcal{O}_{\widetilde{B}W}^{} \!\!\!&=&\!\!
\ii\, H^\dagger  \widetilde{B}_{\mu\nu}W^{\mu\rho}
\!\left\{D_\rho,D^\nu\right\}\! H+\text{h.c.},
\eeqa 
\beqa 
\mO_{\!C+}^{} \!\!\!&=&\!\!
\widetilde{B}_{\!\mu\nu}^{}W^{a\mu\rho}\!
\left[D_{\!\rho}^{}(\overline{\psi_{\!L}^{}}T^a\!\gamma^\nu\!\psi_{\!L}^{})
+D^\nu(\overline{\psi_{\!L}^{}}T^a\!\gamma_\rho^{}\psi_{\!L}^{})\right]\!,
\label{eq:OC+}
\\[1.5mm]
\mO_{\!C-}^{} \!\!\!&=&\!\!
\widetilde{B}_{\!\mu\nu}^{}W^{a\mu\rho}\!
\left[D_{\!\rho}^{}(\overline{\psi_{\!L}^{}}T^a\!\gamma^\nu\!\psi_{\!L}^{})
-D^\nu(\overline{\psi_{\!L}^{}}T^a\!\gamma_\rho^{}\psi_{\!L}^{})\right]\!.
\label{eq:OC-}
\end{eqnarray}
\eeqs
The fermionic operators $\hs\OCP$ and $\hs\OCM\!$ do not contribute directly to the nTGC couplings, but are connected to the three bosonic nTGC operators  $(\OGP,\hs\OGM,\hs\OBW)$ 
by the equation of motion\,\cite{Ellis:2020ljj}: 
\\[-8mm]
\beqs
\label{eq:OG-EOM}
\begin{eqnarray}
\label{eq:OCP=OGM-OBW}
\OCP \!\!\!&=&\!\! \OGM \!- \OBW \,,
\\[1.5mm]
\label{eq:OCM=OGP-HBW}
\OCM \!\!\!&=&\!\! \OGP -
\{\,\ii H^\dagger\widetilde B_{\mu\nu}{W}^{\mu\rho}\!\left[D_\rho,D^\nu\right]\! H
\!+\ii\,2(D_\rho H)^{\!\dagger}\widetilde B_{\mu\nu}{W}^{\mu\rho}\! D^\nu H+\text{h.c.}\} \hs.
\hspace*{10mm}
\end{eqnarray}
\eeqs
They both contribute to the quartic 
$\hs f\bar{f}Z\ga\hs$ vertex and thus to the on-shell amplitude
$\,\TT[f\bar{f}\!\ito\!Z\ga\hs]\,$.\ Hence they can be probed by
the reaction $f\bar{f}\hsm\ito\hsm Z\ga$\hs.\
However, we note that 
the operators $\OGP$ and $\OCM$ give exactly
the same contribution to the on-shell amplitude
$\,\TT[f\bar{f}\!\to\!Z\ga]\,$ at tree level\,\cite{Ellis:2020ljj},
because Eq.\eqref{eq:OCM=OGP-HBW} shows that the difference
$(\OCM\!\!-\OGP)$ is given by the Higgs-doublet-related term
on the right-hand side (RHS) which contains at least 4 gauge fields
and is thus irrelevant for the amplitude
$\,\TT[f\bar{f}\!\to\!Z\ga\hs ]\,$ at the  tree level.\

\vspace*{1mm}

We consider first the dimension-8
nTGC operators $\mO_{G+}, {\cal O}_{\widetilde{B}W}$ and $\mO_{G-}$.
These operators contribute to
the $Z\gamma Z^*$ and $Z\gamma \gamma^*$ vertices as follows:
\beqs
\label{eq:Vertex-G+}
\begin{eqnarray}
\label{eq:VertexZAZ-G+} 
\Gamma_{Z\gamma Z^*(G+)}^{\alpha\beta\mu}(q_1^{},q_2^{},q_3^{})
\!\!&=&\!\!
- 
\frac{\,v(q_3^2\!-\!M_Z^2)\,}{\,M_Z^{}\hs [\Lambda_{G+}^4]\,}\!
\(q_3^2\,q_{2\nu}^{}\epsilon^{\alpha\beta\mu\nu}\!
+2q_2^{\alpha} q_{3\nu}^{}q_{2\sigma}^{}\epsilon^{\beta\mu\nu\sigma}\)\!,
\hspace*{16mm}
\\[1mm]
\label{eq:VertexZAA-G+}
\Gamma_{Z\gamma \gamma^*(G+)}^{\alpha\beta\mu}
(q_1^{},q_2^{},q_3^{})
\!\!&=&\!\!
- 
\frac{\,s_W^{}v\, q_3^2\,}{\,c_W^{}M_Z^{}\hs 
[\Lambda_{G+}^4]\,}\!
\(q_3^2\,q_{2\nu}^{}\epsilon^{\alpha\beta\mu\nu}\!
+ 2q_2^{\alpha}q_{3\nu}^{} q_{2\sigma}^{}\epsilon^{\beta\mu\nu\sigma}\)\!,
\hspace*{16mm}
\\
\label{eq:VertexZAZ-BW}
\hspace*{16mm}
\Gamma^{\al\beta\mu}_{Z\gamma Z^*\!(\widetilde{B}W)}
({q}_1^{}, {q}_2^{}, {q}_3^{})
\!\!&=&\!\!
\frac{~v\hs M_Z^{}\hs ({q}_3^2\!-\!M_Z^2)\,}
{\,[\Lambda_{\widetilde{B}W}^4]\,}
\epsilon^{\alpha\beta\mu\nu} q_{2\nu}^{}  \,,
\\
\label{eq:VertexZAZ-G-}
\Gamma^{\al\beta\mu}_{Z\gamma \gamma^*(G-)}
({q}_1^{}, {q}_2^{}, {q}_3^{})
\!\!&=&\!\!
- 
\frac{\, s_W^{}\hs v\hs M_Z^{}\,}
{\,c_W^{}[\Lambda_{G-}^4]\,}
\epsilon^{\alpha\beta\mu\nu} q_{2\nu}^{}{q}_3^2 \,.
\end{eqnarray}
\eeqs
In the above and afterwards, 
the three gauge bosons are defined as outgoing.\

\vspace*{1mm}

We consider next the fermion-bilinear operator 
$\mathcal{O}_{C_+}^{}$,
which contributes to the effective contact vertex 
$q\bar{q}Z\gamma$ as follows:
\\[-7mm]
\begin{eqnarray} 
\label{eq:vertex-OC+}
\Gamma^{\al\beta}_{q\bar{q}Z\gamma(C+)}
({q}_1^{}, {q}_2^{})
\,=\,
-\text{sign}(\tilde{c}_{C+}^{})
\frac{\,2M_Z^{2}\hs T_3^{}\,}{\,\Lambda^4\,}
\epsilon^{\al\be\mu\nu} q_{2\nu^{}}^{}\gamma_\mu^{}P_L^{} \,,
\end{eqnarray}
where the four external fields are on-shell.\ 
In the above formula, we have introduced the third component of the 
weak isospin $\,T_3^{}\!=\pm\fr{1}{2}$\, and
the chirality projections
$\hs P_{L(R)}^{}\!=\!\fr{1}{2}(1\!\mp\!\gamma_5^{})\hs$.

\vspace*{1mm}
\subsection{\hspace*{-2.5mm}nTGC Contributions to \boldmath{$Z\ga$} 
Amplitude and Cross Section}
\label{sec:2.2}
\vspace*{1mm}


Next, we study the helicity amplitude for the
quark and antiquark annihilation process
$\,q\hs\bar{q}\!\to\!Z\ga\hs$,$\hs$ where the quark has 
weak isospin $\hs T_3\hs$ and electric charge $\hs Q\hs$.\ 
We can compute the SM contributions to the helicity amplitude
of $\,q\hs\bar{q}\!\to\!Z(\lam)\ga(\lam')\,$ as follows:
\beqs
\label{eq:Tsm-T+L}
\begin{eqnarray}
\mathcal{T}_{\text{sm}}^{ss'\!,\text{T}}\!\!\left\lgroup\!\!\!
\begin{array}{cc}
-- \!&\! -+ \\
+- \!&\! ++\\
\end{array}\!\!\right\rgroup
\hspace*{-2.5mm}
\!\!&=&\!\!\!
\frac{-2\hs e^2 Q}{\,s_W^{}c_W^{}(s\!-\!M_Z^2)\,}\!\!\!
\left\lgroup\!\!\!\!
\begin{array}{ll}
\left(c_L'\!\cot\!\frac{\theta}{2}\!-\!
c_R'\!\tan\!\frac{\theta}{2}\right)\!M_Z^2~
\!&\!\!
\left(-c_L'\!\cot\!\frac{\theta}{2}\!+\!
c_R'\!\tan\!\frac{\theta}{2}\right)\!s
\\[2mm]
\left(c_L'\!\tan\!\frac{\theta}{2}\!-\!
c_R'\!\cot\!\frac{\theta}{2}\right)\!s
\!&\!\!
\left(-c_L^{}\!\tan\!\frac{\theta}{2}\!+\!c_R^{}\!\cot\!\frac{\theta}{2}\right)\!M_Z^2
\end{array}
\!\!\!\right\rgroup\!\!,\hspace*{14mm}
\label{msmT}
\label{eq:Tsm-T}
\\[2mm]
\mathcal{T}_{\text{sm}}^{ss'\!,\text{L}}(0-,0+)
\hsm\!\!\!&=&\!\!\! \frac{\,-2\sqrt{2}\hs 
e^2Q\hs (c_L'\!\!+\!c_R')\hs M_Z^{}\sqrt{s\,}~}
{\,s_W^{}c_W^{}(s\!-\!M_Z^2)\,}\left(1,\,-1\right),
\label{msm}
\label{eq:Tsm-L}
\end{eqnarray}
\eeqs
for the helicity combinations
$\lam\lam'\!=\!(--,-+,+-,++)$ and $\lam\lam'\!=\!(0-,0+)$.
In the above, we have defined the coupling coefficients
$\,(c_L',\,c_R') = 
((T_3^{}\hsm -\hsm Qs_W^2)\delta_{\!s,-\frac{1}{2}}^{},\hs
-Qs_W^2\delta_{\!s,\frac{1}{2}}^{})$\, 
with the notations 
$\,(s_W^{},\,c_W^{})=(\sin\hsm\theta_W^{},\hs 
\cos\hsm\theta_W^{})\,$ 
and the subscript index
$\,s =\mp\frac{1}{2}$\, 
denoting the initial-state fermion helicities.
If the initial-state quark and antiquark masses are  
negligible, the relation $\hs s=\hsm -s'\hs$ holds.

\vspace*{1mm}

We find the following contributions to the corresponding helicity amplitudes from the dimension-8 operator
$\mO_{\!G+}^{}\,(\mO_{\!C-}^{})$:
\beqs
\label{eq:T8} 
\begin{eqnarray}
\label{eq:T8-T}
\mathcal{T}_{(8),{G+}}^{ss'\!,\text{T}}
\!\!\left\lgroup\!\!\!
\begin{array}{cc}
-- \!&\! -+ \\
+- \!&\! ++\\
\end{array}\!\!\!\right\rgroup\hspace*{-2mm}
\!\!&=&\!\!\!
\frac{\,(c'_L\!\!+\!c'_R) 
(s\!-\!M_Z^2)\hs s\sin\!\theta~}
{[\cut_{G+}^4]}\!\!
\left\lgroup\!\!
\begin{array}{cc}
1 & 0
\\[1mm]
0 & -1
\end{array}
\!\!\right\rgroup \!\!,
\label{m8T}
\\[2mm]
\mathcal{T}_{(8),{G+}}^{ss'\!,\text{L}}(0-,0+)
\!\!\!&=&\!\!\!
\frac{\,\sqrt{2}M_Z^{}(s\!-\!M_Z^2)\sqrt{s\,}\,}
{[\cut_{G+}^4]}\!
\left(\!c'_L\!\sin^2\!\frac{\theta}{2}\!-c'_R\!
\cos^2\!\frac{\theta}{2},~
c'_R\!\sin^2\!\frac{\theta}{2}\!-c'_L\!\cos^2\!\frac{\theta}{2}
\right)\!,
\hspace*{18mm}
\label{eq:T8-L}
\label{m8}
\end{eqnarray}
\eeqs
where the coupling coefficients are given by  
$\,(c'_L,\,c'_R) \!=\! -T_3(\delta_{\!s,-\frac{1}{2}}^{},\,0)$,
and we have used the notations 
$\,[\cut_{G+}^4]\hsm\equiv\hsm
\rm{sign}(\tilde{c}_{G+}^{})\cut^4_{G+}$
for $\OGP\hs$.\
We note that in Eq.\eqref{eq:T8-T} the off-diagonal amplitudes 
vanish exactly.\ 
This is because the final state
$Z(\lam)\ga(\lam')$ with helicities 
$\lam\lam'\!=\!+-,-+\hs$ should have
their spin angular momenta pointing to the same direction
in their central-of-mass frame 
and thus the sum of their spin momenta would have magnitude
equal $2\hs$.\ But this is disallowed by the $s$-channel 
spin-1 gauge boson $Z^*$ or $\ga^*$.\
\,For the same reason, the off-diagonal amplitudes 
contributed by the other dimension-8 operators in the following 
Eq.\eqref{eq:T8-TT-OG-OBW-OC+} have to vanish as well.

\vspace*{1mm}

As for the other three dimension-8 operators
$(\mO_{G-}^{},\hs \mO_{\tilde{B}W}^{},\hs \mO_{C+}^{})$, 
we derive their contributions to the helicity amplitudes of
the reaction $\,q\hs\bar{q}\ito Z\ga\,$ as follows:
\beqs
\label{eq:T8-OG-OBW-OC+} 
\begin{eqnarray}
\label{eq:T8-TT-OG-OBW-OC+}
\mathcal{T}_{(8),j}^{ss'\!,\text{T}}
\!\!\left\lgroup\!\!
\begin{array}{cc}
-- \!&\! -+ \\
+- \!&\! ++\\
\end{array}\!\!\right\rgroup
\!\!\!\!&=&\!\!\!
\frac{\,(c'_L\!+\hsm c'_R)\sin\!\theta\hs  M_Z^2\hs (s\!-\!M_Z^2)\,}{[\cut_j^4]}\!\!
\left\lgroup\!\!
\begin{array}{cr}
1 \!& 0
\\[1mm]
0 \!& -1
\end{array}
\!\!\right\rgroup \!\!,
\\[2mm]
\label{eq:T8-LT-OG-OBW-OC+}
\mathcal{T}_{(8),j}^{ss'\!,\text{L}} (0-,0+)
\!\!&=&\!\!\! 
\frac{\,\sqrt{2\,}M_Z^{}(s\!-\!M_Z^2)\sqrt{s}\,}
{[\cut_j^4]}\!
\(\!c'_L\!\sin^2\!\frac{\theta}{2}\!-c'_R\!\cos^2\!\frac{\theta}{2},~
c'_R\!\sin^2\!\frac{\theta}{2}\!-c'_L\!\cos^2\!\frac{\theta}{2}
\)\!, \hspace*{14mm}
\end{eqnarray}
\eeqs	
where $\,[\cut_j^4]\!=\rm{sign}(\tilde{c}_j^{})\cut_j^4\,$ and 
$\,j^{}\!\in\!({G_-}^{},\hs{\widetilde BW}^{},\hs {C_+}^{})\hs$.\
\,In Eq.\eqref{eq:T8-OG-OBW-OC+}, 
the coupling factors $\,(c'_L,\,c'_R)$ are given by
%
\beqs 
\vspace*{-5mm}
\label{eq:Coup-OG-OBW-OC+}
\begin{align}
(c'_L,\,c'_R) &\,=\, -Q s_W^2(\delta_{\!s,-\frac{1}{2}}^{},\,
\delta_{\!s,\frac{1}{2}}^{})\hs ,
\hspace*{-15mm}
& (\hs\text{for}~ &\mO_{G-}^{}\hs)\hs ,
\\
(c'_L,\,c'_R) &\,=\, \big( q_L^{}\delta_{\!s,-\frac{1}{2}}^{},\,
q_R^{}\hs\delta_{\!s,\frac{1}{2}}^{}\big)\hs ,
\hspace*{-15mm}
& (\hs\text{for}~ &\mO_{\!\widetilde BW}^{}\hs)\hs ,
\\
(c'_L,\,c'_R) &\,=\, -T_3(\delta_{\!s,-\frac{1}{2}}^{},\,0)\hs ,
\hspace*{-15mm}
& (\hs\text{for}~ &\mO_{C+}^{}\hs)\hs ,
\end{align}
\eeqs 
and the coefficients
$\,(q_L^{},\,q_R^{}) \!=\! (T_3^{}\!-\!Qs_W^2,\, -Qs_W^2)$
arise from $Z$ gauge boson couplings with the 
(left, right)-handed the quarks.

\vspace*{1mm}

The kinematics for the complete annihilation process
$\hs q\hs\bar q\to Z\ga\ito f\bar f\ga\hs$
are defined by the three angles  
$(\theta,\,\theta_*^{},\,\phi_*^{})$,
where $\,\theta\,$ is the polar scattering angle between 
the direction of the outgoing $Z$ and 
the initial state quark $\hs q\hs$,
$\,\theta_*^{}$ denotes the angle between the
direction opposite to the final-state $\gamma$
and the final-state fermion $f$ direction in the $Z$ rest frame,
and $\,\phi_*^{}\,$ is the angle between the scattering plane
and the decay plane of $Z$ in the $\hs q\hs\bar{q}\hs$ 
center-of-mass frame [cf.\ Eq.\eqref{eq:phi*}]. 
We note that, at a $pp$ collider,
we cannot determine which is the initial state quark (antiquark)
in each collision, so we could only determine the scattering angle 
up to an ambiguity $\theta \leftrightarrow \pi\hsm -\theta\hs$.
It follows that the determination of the angle 
between the scattering plane and $Z$-decay plane 
also has an ambiguity $\hs\phi_*^{} \leftrightarrow \hs\pi\hsm -\phi_*^{}\hs$.

\vspace*{1mm}

Taking these remarks into account, we can express the full amplitude of the 
reaction process
$\hs q\hs\bar q\to Z\ga\ito f\bar f\ga\hs$ 
in the following form:
{\small
\begin{eqnarray}
\hspace*{-6mm}
\mathcal{T}^{ss'}_{\si\si'\!\lambda}\!(f\bar f\ga) 
&\!\!=\!\!&
\frac{\,eM_Z^{}\mathcal{D}_Z^{}\,}{s_W^{}c_W^{}}
\left[\sqrt{2\,}e^{\ii\phi_*^{}}\!
\left(\!f_R^{\si}\cos^2\!\frac{\,\theta_*^{}}{2}\!-f_L^{\si}\!\sin^2\!\frac{\,\theta_*^{}}{2}\!\right)\!
\mathcal{T}_{ss'}^T(+\lam )
\right.
\nn\\
&&\hspace*{10mm}
\left.
+\sqrt{2\,}e^{-i\phi_*^{}}\!\hsm
\left(\!f_R^{\si}\sin^2\!\frac{\,\theta_*^{}}{2}\!
-f_L^{\si}\!\cos^2\!\frac{\,\theta_*^{}}{2}\!\right)\!
\mathcal{T}_{ss'}^T(-\lam )
+(f_R^{\si}\!+\!f_L^{\si})\sin\!\theta_* \mathcal{T}_{ss'}^L(0\lam )\right]\hsm\!,
\hspace*{3mm}
\label{eq:T-llgamma}
\end{eqnarray}
}
\hspace*{-2mm}
where $\,\mathcal{D}_Z^{}\hsm =\hsm 1/
(q_1^2\!-\!M_Z^2+\ii\hs M_Z^{}\Gamma_Z^{})\,$
comes from the $Z$ propagator. 
In Eq.\eqref{eq:T-llgamma}, the final-state fermions have
the electroweak gauge couplings given by
$\hs (f_L^{\sigma}, f_R^{\sigma}) \!=\! 
((T_3^{}\hsm - Qs_W^2)\hs 
\delta_{\!\sigma,-\frac{1}{2}}^{},\hs
-Qs_W^2\delta_{\!\sigma,\frac{1}{2}}^{})\hs$,
and the scattering amplitudes  
$\mathcal{T}_{ss'}^T(\pm\lam )$ and $\mathcal{T}_{ss'}^L(0\lam )$
represent the on-shell helicity amplitudes for 
the reaction $\hs q\hs\bar{q}\!\to Z\ga$\,:
\beqa
\label{eq:T-llgamma-sum}
\mathcal{T}_{ss'}^T(\pm\lam ) \!\!&=&\!\!
\mathcal{T}^{ss'\!,T}_{\text{sm}}(\pm\lam )
+ \mathcal{T}^{ss'\!,T}_{(8)}(\pm\lam )\hs ,
\nn\\[-3mm]
\\[-2mm]
\mathcal{T}_{ss'}^L(0\lam ) \!\!&=&\!\!
\mathcal{T}^{ss'\!,L}_{\text{sm}}(0\lam )
+ \mathcal{T}^{ss'\!,L}_{(8)}(0\lam )\hs ,
\nn
\eeqa
which receive contributions from both the SM and the dimension-8 operator.\

\vspace*{1mm}

Applying a lower angular cut $\,\sin\theta\!>\!\sin\delta\,$ 
for some $\delta\!\ll\! 1$,
we derive the following total cross section for the 
partonic process
$\,q\hs\bar q \ito Z\gamma\hs$,\,
including both the linear and quadratic contributions of $\mathcal{O}_{G+}^{}$ and
summing over the final-state $Z$ and $\ga$ polarizations: 
%
\begin{eqnarray}
\label{eq:CS-qq-Zgamma}
\sigma_{}^{}(Z\gamma)
\!\!&=&\!\!
\frac{\,e^4(q_L^2\!+\!q_R^2)\hs Q^2\!
\left[-(s\!-\!M^2_Z)^2\!-\!2(s^2\!+\!M_Z^4)\hsm 
\ln\sin\!\frac{\delta }{2}\,\right]\,}
{\,8\pi s_W^2c_W^2(s\!-\!M^2_Z)s^2\,}
\hspace*{15mm}
\nn\\[1mm]
\!\!&&\!\!
+\,\frac{\,e^2q_L^{} Q\hs{T_3^{}} M_Z^2\!
\left(s\!-\!M_Z^2\right)\,}{4\pi s_Wc_W  s}
\frac{1}{\,[\cut_{G+}^4]\,}
\label{eq:ZA-G+}
\\[1mm]
\!\!&&\!\!
+\,\frac{~{T_3^2}\hs (s\hsm +\!M_Z^2)\!
\left(s\!-\!M^2_Z\right)^{\!3}\,}
{\,48\pi\,s\,}\frac{1}{\,\cut_{G+}^8\,} + O(\delta)\,,
\nn
\end{eqnarray}
%
where 
the weak isospin $\, T_3^{}\hsm =\hsm \pm\fr{1}{2}\,$ 
is associated with the $\,W_3^{}\hs$ gauge coupling,  
and the coefficients
$\,(q_L^{},\,q_R^{})\!=\! (T_3^{}\hsm -\hsm Qs_W^2,\, -Qs_W^2)$
are the (left, right)-handed gauge couplings 
of the quarks to the $Z$ boson.\ 
In Eq.\eqref{eq:CS-qq-Zgamma}, $\sqrt{s\,}\hs$
denotes the center-of-mass energy of the partonic process
$\,q\hs\bar q \!\to Z\gamma\hs$, 
but for the $pp$ collider analyses 
in Section\,\ref{sec:4} we will rename the above partonic 
center-of-mass energy as 
$\sqrt{\hat{s}\,}$ for clarity. 


We define the normalized angular distribution functions as follows:
\begin{eqnarray}
f_\xi^j \, = \, \frac{1}{\sigma_{\!j}^{}}\frac{\di\sigma_{\!j}^{}}{\,\di\xi\,}\,,
\end{eqnarray}
where the angles $\,\xi \in (\theta,\,\theta_*^{},\,\phi_*^{})$,\,
and the cross sections
$\,\sigma_j^{}$ ($j=0,1,2$) represent the SM contribution ($\sigma_0^{}$),
the ${O}(\cut^{-4})$ contribution ($\sigma_1^{}$), and the
${O}(\cut^{-8})$ contribution ($\sigma_2^{}$),\, respectively.
In the following, we derive the explicit formulas for the 
normalized azimuthal angular distribution functions
$\,f_{\phi_*}^j\,$:
{\small
\beqs
\label{eq:f-phi*-OGP} 
\begin{eqnarray}
\label{eq:f0-phi*}
\hspace*{-13mm}
f_{\phi_*^{}}^{0} \!\!\!\!&=& \!\!\!\!
\frac{1}{\,2\pi\,}\hsm  +\hsm\frac{\,3\pi^2(q_L^2\!-\!q_R^2)(f_L^2\!-\!f_R^2)M_Z^{}\sqrt{s}\,
	(s\!+\!M_Z^2)\cos\hsm\phi_*^{}\!
	-8(q_L^2\!+\!q_R^2)(f_L^2\!+\!f_R^2)M_Z^2\,s \cos\hsm 2\phi_*^{}\,}
{\,16\pi(q_L^2\!+\!q_R^2)(f_L^2+f_R^2)\!
	\left[(s\!-\!M_Z^2)^2\!+2(s^2\!+\!M_Z^4)\ln\sin\!\frac{\delta}{2}
	\hs\right]\,}+O(\delta) ,\hspace*{2mm}
\nn\\[-2mm]
&&\\ 
\label{eq:f1-phi*}
\hspace*{-7mm}
f_{\phi_*^{}}^{1} \!\!\!\!&=&\!\!\!\!
\frac{1}{\,2\pi\,} \hsm -\hsm 
\frac{\,3\pi (f_L^2\!-\!f_R^2)(M_Z^2+5\hs s)
	\cos\phi_*\,}{256(f_L^2\!+\!f_R^2)M_Z\sqrt s}+
\frac{\,s\cos2\phi_*\,}{8\pi M_Z^2}\, ,
\label{f1}
\\[2mm]
\label{eq:f2-phi*}
\hspace*{-7mm}
f_{\phi_*^{}}^{2} \!\!\!\!&=&\!\!\!\!
\frac{1}{\,2\pi\,} - 
\frac{\,9\pi (f_L^2\!-\!f_R^2)M_Z^{}\sqrt{s\,}
	\cos\hsm\phi_*^{}\,}
{\,128(f_L^2\!+\!f_R^2)(s\!+\!M_Z^2)\,} \, ,
\label{eq:phi-}
\end{eqnarray}
\eeqs
}
\hspace*{-2mm}
where we denote the $Z$ couplings with the initial state quarks
as $\,(q_L^{},\,q_R^{}) \!=\! (T_3^{}\hsm -\hsm Qs_W^2,\hs -Qs_W^2)$,
and the the $Z$ couplings with the final-state fermions as
$\hs (f_L^{}, f_R^{}) \!=\! 
((T_3^{}\hsm - Qs_W^2),\hs -Qs_W^2)\hs$.


In the cases of the other nTGC operators $\mathcal{O}_j^{}$, 
we further derive their contributions to the total cross sections 
of the reaction $\,q\hs\bar q \!\to Z\gamma\,$
as follows:
%
\begin{eqnarray}
\label{eq:CS-qqZA-d8other}
\sigma(Z\gamma)
&\!\!=\!\!&
\frac{~e^4(q_L^2\!+\!q_R^2)Q^2\!\!\left[-(s\!-\!M^2_Z)^2\!-\!2(s^2\!+\!M_Z^4)
\ln\sin\!\frac{\delta }{2}\,\right]\,}
{\,8\pi s_W^2c_W^2(s\!-\!M^2_Z)\hs s^2\,}
\hspace*{15mm}
\nn\\[1mm]
&&
-\,\frac{~e^2 Q(q_L^{}x_{L}^{}\!\!-\!q_R^{}x_{R}^{})M_Z^2\!
\(s\!-\!M_Z^2\)\!\(s\!+\!M^2_Z\)\,}
{\,8\pi s_W^{}c_W^{}\hs s^2\,}\frac{1}{\,[\cut_j^4]\,}
\label{eq:CS-qq-Zgamma-Ojx}
\\[1mm]
&& +\frac{~(x_L^2\!+\!x_R^2)M_Z^2\!\(s\!+\!M^2_Z\)\!\(s\!-\!M^2_Z\)^{\!3}\,}
{\,48\pi\hs s^2\,}\frac{1}{\,\cut_j^8\,} + O(\delta)\,,
\nn
\end{eqnarray}
where we define the relevant coupling coefficients
$\hs (x_L^{},\hs x_R^{})\hs$ as
\beqs
\label{eq:xLR-OBWC+G-}
\begin{align}
(x_L^{},\,x_R^{}) &= -Qs_W^2(1,\,1),
& (\text{for}~\,&\mathcal{O}_j^{}\hsm =\mathcal{O}_{G-}^{}) ,
\label{eq:xLR-OG-}
\\
(x_L^{},\,x_R^{}) &= \(T_3^{}\!-Qs_W^2,\,-Qs_W^2\)\!,
& (\text{for}~\,&\mathcal{O}_j^{}
\hsm =\mathcal{O}_{\widetilde{B}W}^{}),
\label{eq:xLR-OBW}
\\
(x_L^{},\,x_R^{}) &= -(T_3,\,0) ,
& (\text{for}~\,&\mathcal{O}_j^{}\hsm =\mathcal{O}_{C+}^{}) .
\label{eq:xLR-OC+}
\end{align}
\eeqs
We see that in the high energy limit, 
the contributions of the SM, the interference term, and the 
squared term behave as $(s^{-1}\hsm ,\hs s^0\hsm ,\hs s^2)$ 
respectively. We can compare the above cross section with
that of Eq.\eqref{eq:CS-qq-Zgamma} for the nTGC operator
$\hs\OGP$ where the SM term, the interference term, and 
the squared term scale as 
$(s^{-1}\hsm ,\hs s^0\hsm ,\hs s^3)$ respectively.
This shows that the contribution of $\hs\OGP$ to the squared term
has higher energy power enhancement of $\hs s^3\hs$ than the factor
$\hs s^2\hs$
of the other operators.

Then, for the full  process
$\hs q\hs\bar q\ito\! Z\ga\!\ito\! f\bar f\ga\hs$, 
we further derive the following normalized 
angular distribution functions $\,f_{\phi_*}^j\,$ 
for the operators $(\OGM,\hs\OBW,\hs\OCP)$:
\beqs
\label{eq:f-phi*-Ojx}  
\begin{eqnarray}
\hspace*{-8mm}
f_{\phi_*}^{0} \!\!\!&=&\!\!
\frac{1}{2\pi} +
\frac{\,3\pi^2f_{-}^2q_-^2M_Z^{}\sqrt{s}\,(s\!+\!M_Z^2)\cos\!\phi_*^{}\!
-8f_{+}^2q_+^2M_Z^2\,s \cos\!2\phi_*^{}\,}
{\,16 \pi f_{+}^2q_+^2\!\left[(s\!-\!M_Z^2)^2\!+2(s^2\!+\!M_Z^4)
\ln\sin\frac{\delta}{2}
\right]\,}+O(\delta),
\hspace*{4mm}
\\[2mm]
\hspace*{-8mm}
f_{\phi_*}^{1} \!\!\!&=&\!\! \frac{1}{2\pi} -
\frac{\,9\pi (q_L^{}x_L^{}\!+\!q_R^{}x_R^{})
(f_L^2\!-\!f_R^2) \sqrt{s}\,\cos\!\phi_*^{}\,}
{\,128(q_L^{}x_L^{}\!-\!q_R^{}x_R^{})(f_L^2\!+\!f_R^2) M_Z^{}\,}+
\frac{s \cos\!2\phi_*^{}\,}
{\,4\pi(s\!+\!M_Z^2)\,} ,
\label{f1q}
\\[2mm]
\hspace*{-8mm}
f_{\phi_*}^{2} \!\!\!&=&\!\!
\frac{1}{2\pi} - \frac{\,9\pi (x_L^2\!-\!x_R^2)(f_L^2\!-\!f_R^2)M_Z^{}\sqrt{s}\,\cos\!\phi_*^{}\,}
{\,128(x_L^2\!+\!x_R^2)(f_L^2\!+\!f_R^2)(s\!+\!M_Z^2)\,}\,,
\end{eqnarray}
\eeqs
\hspace*{-2mm} 
where we have defined the coefficients
\,$(f_{\pm}^2,\,q_\pm^2)\!\equiv\!(f_L^2\pm f_R^2,\,q_L^2\pm q_R^2)$,
and the electroweak gauge couplings of the final state fermions
are given by
$\hs (f_L^{}, f_R^{}) \!=\! 
((T_3^{}\hsm - Qs_W^2),\hs -Qs_W^2)\hs$.

\vspace*{1mm}
\subsection{\hspace*{-2.5mm}Analysis of Unitarity Constraints on nTGCs}
\label{sec:2.3}
\vspace*{1mm}

In this subsection, we analyze the perturbative unitarity constraints
on the nTGCs, showing that these constraints are much weaker than
the sensitivity reaches of the collider probes presented in
the following Sections\,\ref{sec:3}-\ref{sec:5}.\

\vspace*{1mm}

We first make the following partial-wave expansion~\cite{unitarity} 
of the nTGC contributions to the scattering amplitude for the reaction 
$f\bar{f}\ito Z\ga\hs$: 
%
\begin{eqnarray}
a_J^{}\,=\,\frac{1}{\,32\pi\,}e^{i(\nu'-\nu)\phi}\!\!
\int_{-1}^{1}\!\!\d (\cos\theta)\,
d^{\hs J}_{\nu'\nu}(\cos\theta)
\mT_{\text{nTGC}}^{s_f^{}s_{\bar f}^{},\lambda_Z^{}\lambda_\gamma^{}},
\end{eqnarray}
where the differences of initial/final state helicities are given by
$\,\nu \!=s_f^{}\!-\!s_{\bar f}^{}\!=\pm 1$ and 
$\,\nu'\!=\lambda_Z^{}\!-\!\lambda_\gamma^{}\!=0,\pm 1\hs$,
respectively.\ 
We note that for the present collider analysis it is sufficient 
to treat the initial-state fermions $(f,\bar{f})$
(light quarks or leptons) as massless. Thus we have
$\,s_f^{}\!=\!-s_{\bar f}^{}\,$, which leads to 
$\,\nu\hsm =\!\pm 1\hs$.\ Hence the 
$\hs J\!=\!1\hs$ partial wave makes the leading contribution.\
The relevant Wigner $d$ functions are given by
\beqa 
d^1_{1,0}=-\fr{1}{\,\sqrt{2\,}\,}\sin\theta\hs,
\hspace*{8mm} 
d^1_{1,\pm1}=\fr{1}{\,2\,}(1\pm\cos\theta)\hs ,
\eeqa
and we have a general relation
$\,d^{\hs J}_{m,m'}\!=\hsm d^{\hs J}_{-m,-m'}\hs$.\

\vspace*{1mm}

In the case of the dimension-8 operator $\OGP$ (or $\OCM$), 
its leading contribution
to the amplitude $\mT_{\text{nTGC}}^{s_f^{}s_{\bar f}^{},\lambda_Z^{}\lambda_\gamma^{}}$ is given by 
Eq.\eqref{eq:T8-T}, as follows:
\begin{eqnarray}
\mathcal{T}_{(8)G+}^{s_fs_{\bar f},\text{T}}(\mp\mp) =
\pm\frac{~(c'_L\!+\!c'_R)\hs s^2\sin\hsm\theta~}
	{[\cut_{G+}^4]}\,,
\end{eqnarray}
where $\sqrt{s\,}\hsm =\!E_{\rm{CM}}^{}$ stands for the c.m.\ energy of
$\,f\bar{f}$.\
As for the other three dimension-8 operators
$\,\mO_j^{}\!\in\!\hsm(\mO_{G-}^{},\hs 
 \mO_{\tilde{B}W}^{},\hs \mO_{C+}^{})$,
their leading contributions to the amplitude $\mT_{\text{nTGC}}^{s_f^{}s_{\bar f}^{},\lambda_Z^{}\lambda_\gamma^{}}$ 
are given by Eq.\eqref{eq:T8-LT-OG-OBW-OC+}, as follows:
\begin{eqnarray}
\mathcal{T}_{(8)\hs j}^{s_fs_{\bar f},\text{L}} (0-,0+)
\!\!&=&\!\!\!
\frac{\,\sqrt{2\,}M_Z^{}\hs s^{3/2}\,}
{[\cut_j^4]}\!
\(\!c'_L\!\sin^2\!\frac{\theta}{\,2\,}\!
-\hsm c'_R\!\cos^2\!\frac{\theta}{\,2\,},~
 c'_R\!\sin^2\!\frac{\theta}{\,2\,}\!
-c'_L\!\cos^2\!\frac{\theta}{\,2\,}\!
	\)\!, \hspace*{14mm}
\end{eqnarray}
where the coupling factors $\,(c'_L,\,c'_R)$ are defined in
Eq.\eqref{eq:Coup-OG-OBW-OC+}.\ 

\vspace*{1mm}

Then, we derive the leading $p$-wave amplitude $\hs a_1^{}\hs$
for the nTGC operator $\OGP$:
\beqa
|\Re\mathfrak{e}(a_1^{G+})|\hs =\hs 
\frac{s^2}{~48\sqrt 2 \pi\cut_{G+}^4\,} \hs.
\eeqa 
For the other nTGC operators 
$\,\mO_j^{}\!\in\!\hsm(\mO_{G-}^{},\hs 
 \mO_{\tilde{B}W}^{},\hs \mO_{C+}^{})$,
we derive their leading $p$-wave amplitudes as follows:
\beq
|\Re\mathfrak{e}(a_1^{j})|\hs =\hs 
\frac{~c_{L,R}'M_Z^{}\hsx s^{3/2}~}{24\sqrt{2\,}\pi\hs\cut_{j}^4~} \hs.
\eeq 

\begin{table}[t]
\begin{center}
\begin{tabular}{c||c|c|c|c|c|c|c}
\hline\hline	
& & & & &
\\[-4mm]
$E_{\rm{CM}}^{}\hs$(TeV) & 0.25&0.5 & 1&3 &5&25&40
\\[0.3mm]
\hline\hline
& & & & & &
\\[-4.3mm]
$\Lambda_{G+}\hs$(TeV)
& 0.078 &0.16 & 0.31 &0.93&1.6&7.8&12 \\
\hline
& & & & & &
\\[-4.3mm]
$\Lambda_{\!\widetilde{B}W}\hs$(TeV)
&  0.058 & 0.098 & 0.16 & 0.37 & 0.55& 1.8&2.6\\
\hline
& & & & &&
\\[-4.3mm]
$\Lambda_{G-}\hs$(TeV)
&  0.050 & 0.084 & 0.14 & 0.32 & 0.47 & 1.6&2.2\\
\hline
& & & & &&
\\[-4.3mm]
$\Lambda_{C+}\hs$(TeV)
&  0.060 & 0.10 & 0.17 & 0.39 & 0.57& 1.9 &2.7\\
\hline\hline
& & & & &&
\\[-4.3mm]
$|h_4|^{}$ & $33$ &$2.0$ & $0.13$ & $0.0016$ &
$2.0\!\times\!10^{-4}$ & $3.3\!\times\!10^{-7}$ & $5.0\!\times\!10^{-8}$ \\
\hline
& & & & &&
\\[-4.3mm]
$|h_3^Z|$ &  $53$ & $6.6$ & $0.83$ & $0.031$
& $6.6\!\times\!10^{-3}$ & $5.3\!\times\!10^{-5}$ & $1.3\!\times\!10^{-5}$\\
\hline
& & & & &&
\\[-4.3mm]
$|h_3^\gamma|$
&  $53$ & $6.6$ & $0.83$ & $0.031$ & $6.6\!\times\!10^{-3}$
& $5.3\!\times\!10^{-5}$& $1.3\!\times\!10^{-5}$\\[0.5mm]
\hline\hline
\end{tabular}
\end{center}
\vspace*{-4mm}
\caption{\small%
{\it Unitarity bounds on the new physics scale $\cut_j^{}$ 
of the dimension-8 nTGC operators and on the nTGC form factors 
$h_j^V$, as derived for various sample values 
of the center-of-mass energy $\,E_{\rm{CM}}^{}$ of the reaction
$\,q\bar{q}\ito Z\ga\,$ or
$\,e^-e^+\ito Z\ga\,$ that are relevant to the present collider study.}
}
\label{tab:0}
\end{table}
Next, we impose the partial-wave unitarity condition 
$|\Re\mathfrak{e}(a_J^{})|\!<\!\frac{1}{\,2\,}$ for $J\!=\!1$,
and derive the following unitarity bounds 
on the new physics cutoff scales $(\cut_{G+}^{},\hs \cut_j^{})$
of the nTGC operators $\OGP$ and 
$\,\mO_j^{}\!\in\!\hsm(\mO_{G-}^{},\hs 
 \mO_{\tilde{B}W}^{},\hs \mO_{C+}^{})\hs$, respectively:
\beqs 
\label{eq:UB-Lambda8}
\beqa 
\cut_{G+}^{} \!&>&\!
\frac{\sqrt{s\,}}{~(24\sqrt{2\,}\pi)^{1/4}~}
\simeq\, 0.311\sqrt{s\,}\,,
\\
\cut_{j}^{} \!&>&\!
\(\!\frac{\,C_{L,R}'M_Z^{}~}
 {~12\sqrt{2\,}\hs\pi\,}\!\)^{\hsm\!\!\frac{1}{4}}\!\!
\(\!\sqrt{s\,}\hs \)^{\!\frac{3}{4}}
\simeq\, 0.203\(\hsm C_{L,R}'\hsm\)^{\!\frac{1}{4}} 
\(\!\rm{TeV}\sqrt{s^3\,}\)^{\hsm\!\frac{1}{4}} 
,
\eeqa
\eeqs 
where $\sqrt{s\,}\hsm =\!E_{\rm{CM}}^{}$ denotes the 
center-of-mass energy of
$\,f\bar{f}\hs$.\

\begin{figure}[t]
\includegraphics[width=8.3cm,height=6cm]{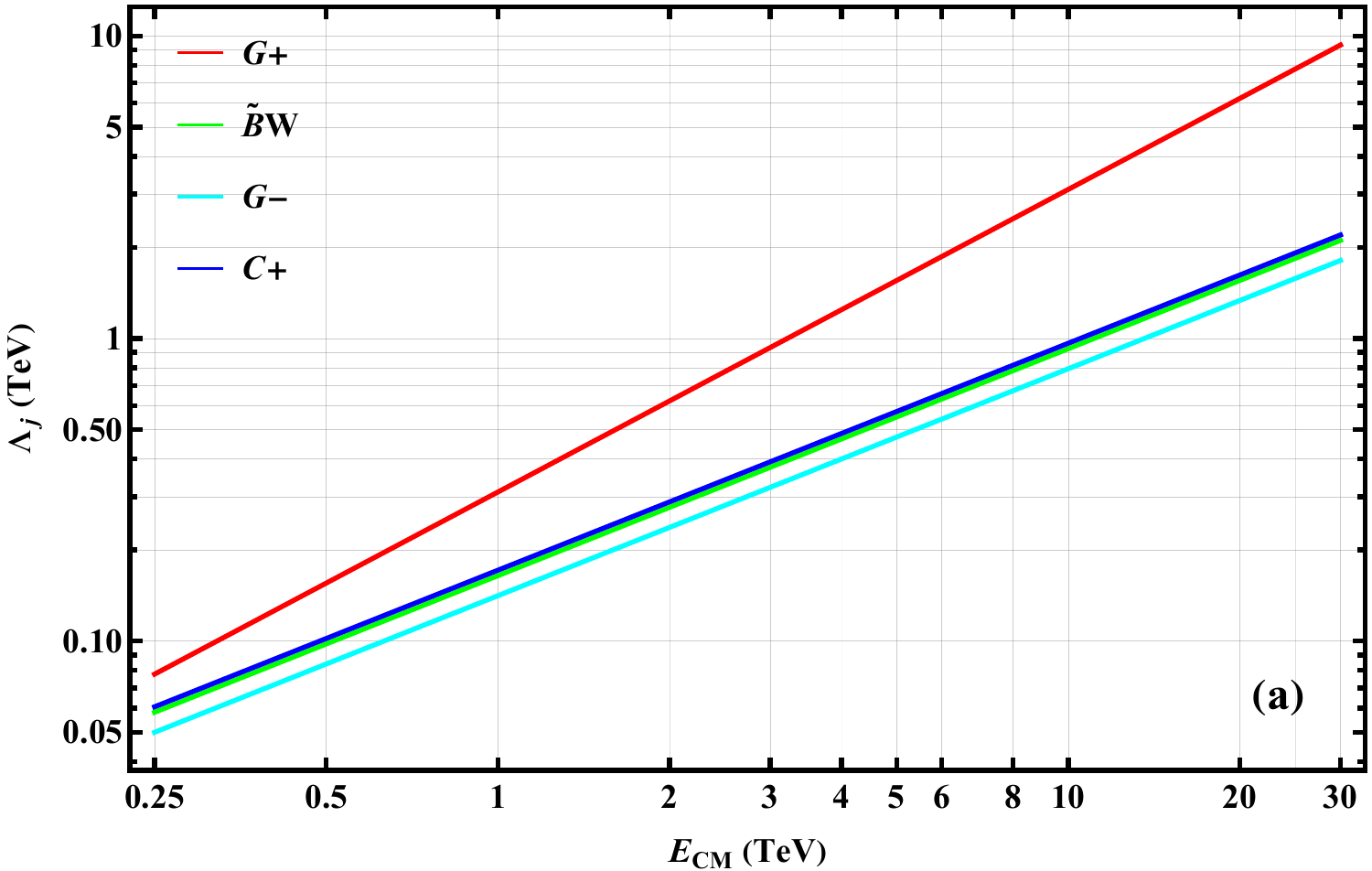}
\includegraphics[width=8.3cm,height=6cm]{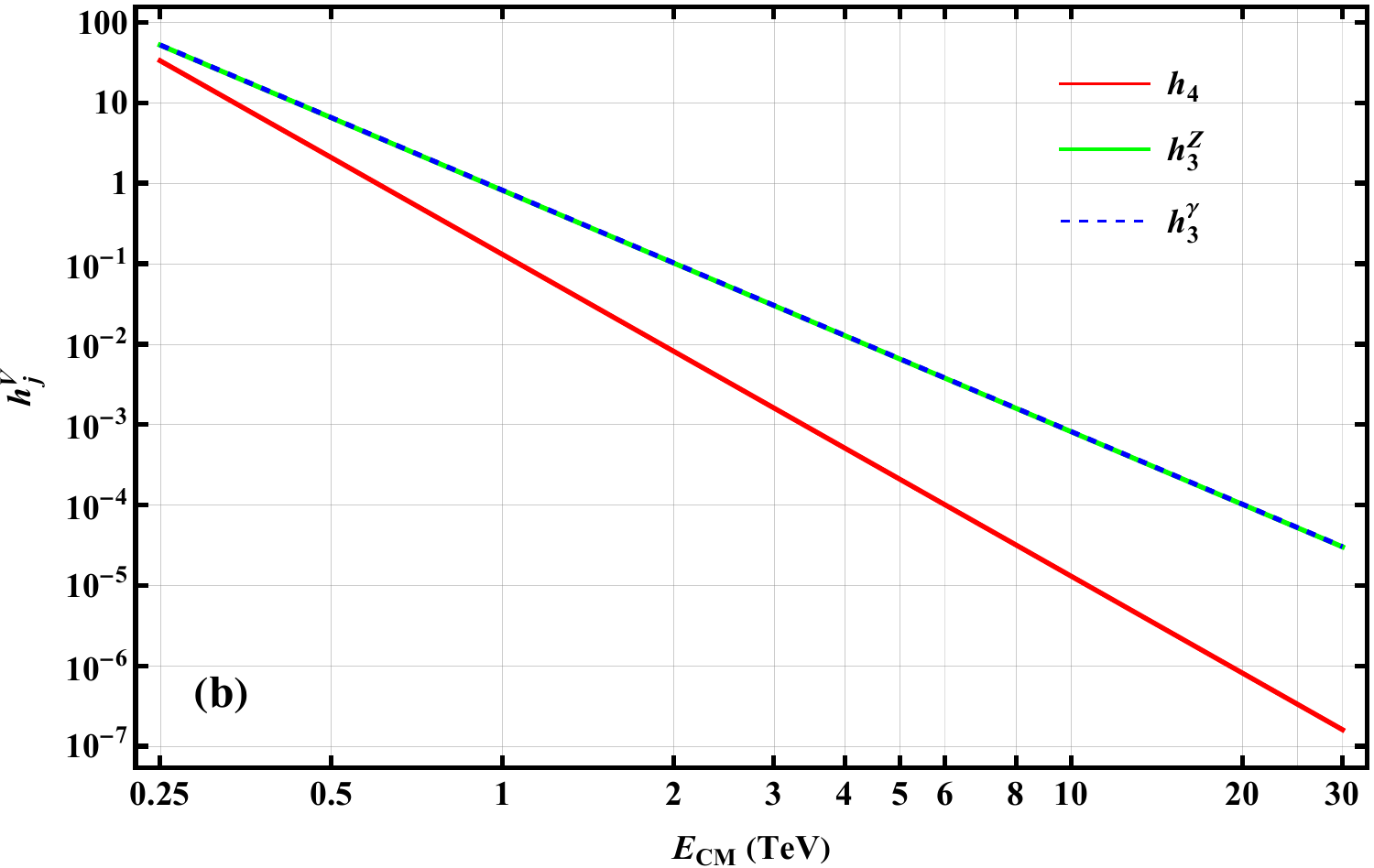}
\caption{\small%
{\it Unitarity bounds on new physics cutoff scales for the nTGC operators 
$(\OGP,\OBW,\OGM,$ $\OCP)$ in plot\,(a) and for the nTGC 
form factors $(|h_4^{}|,|h_3^Z|,|h_3^{\ga}|)$ in plot\,(b).\
These bounds are derived from the $p$-wave amplitudes of the 
reaction $\hs f\bar{f}\ito Z\ga\hs$, where
$\hs f\bar{f}\!=\!q\bar{q},e^+e^-$ with $q$ being the light quarks.\ 
}}
\label{fig:0}
\end{figure}

\vspace*{1mm}

In the cases of the nTGC form factors $(h_4^{},\hs h_3^Z,\hs h_3^{\ga})$ defined in
Eq.\eqref{eq:FF2-nTGC} 
of Section\,\ref{sec:3},
they are connected to the cutoff scales of 
$(\OGP,\hs \OBW,\hs \OGM)$ via
$\,(|h_4^{}|,|h_3^{Z},|h_3^{\ga}|)\!=
 (r_4^{}/\hsm\cut_{G+}^4,r_3^Z/\hsm\cut_{\widetilde{B}W}^4,
  r_3^{\ga}/\hsm\cut_{G-}^4)\hs$,
as given by Eq.\eqref{eq:h-dim8}.\
Thus, using Eq.\eqref{eq:UB-Lambda8} 
we further derive the following unitarity bounds 
on the nTGC form factors:
\beqs 
\label{eq:UB-h34}
\beqa 
|h_4^{}| \!&<&\!  \frac{~24\sqrt{2\,}\pi v^2M_Z^2~}{s_W^{}c_W^{}s^2}
\simeq \(\!\frac{\,0.597\,\rm{TeV}\,}{\sqrt{s\,}}\!\)^{\!\!4},
\\
|h_3^V\hsm | \!&<&\! 
\frac{\,6\sqrt{2\hs}\pi\hs\bar{r}_3^V\,}{\,s_W^{}c_W^{}C_{L,R}'\,}
\frac{\,v^2M_Z^{}\,}{\sqrt{s^{3}\,}}
\simeq \frac{\,0.350\hs\bar{r}_3^V\,}{\,C_{L,R}'\hs}\!
\(\!\hsm\frac{\,\rm{TeV}\,}{\!\sqrt{s\,}\,}\!\)^{\!\!3}
\!,
\end{eqnarray}
\eeqs 
where we have used the expressions in Eq.\eqref{eq:r4-r3} for 
the coefficients $\,(r_4^{},\hs r_3^V)$ and have defined
$\,\bar{r}_3^V\!\in\!(\bar{r}_3^Z,\hs\bar{r}_3^{\ga})
 \!=\!(1,\hs s_W^{}/c_W^{})$.\ 

Using formulae \eqref{eq:UB-Lambda8} and
\eqref{eq:UB-h34} for the unitarity bounds, we present their values in  
Table\,\ref{tab:0} for various sample values of the c.m.\ energies
$\,E_{\rm{CM}}^{}\!=\!(0.25,0.5,1,3,5,25)$TeV,
of the reactions $\,q\bar{q}\ito Z\ga\,$ and
$\,e^-e^+\!\ito Z\ga\,$
that are relevant to the present collider study.\ 
Then, in Fig.\,\ref{fig:0} we present the unitarity bounds 
on the nTGC operators and nTGC form factors as functions of the
center-of-mass energy 
$\,E_{\rm{CM}}^{}\hsm\!=\!(0.25\hsm -\hsm 30)\hs$TeV
for the reaction $\hs f\bar{f}\ito Z\ga\hs$, where 
$\hs f\bar{f}\!=\!q\bar{q},e^+e^-$ and $q$ denotes the light quarks.\
We plot the unitarity bounds on  
the new physics cutoff scales of the nTGC operators
$(\OGP,\OBW,\OGM,\OCP)$ in plot\,(a), whereas in plot\,(b)
we impose the unitarity bounds on the nTGC 
form factors $(|h_4^{}|,|h_3^Z|,|h_3^{\ga}|)$,\
as derived from the $p$-wave amplitudes.\
Finally, by comparing the unitarity bounds of Table\,\ref{tab:0}
and Fig.\,\ref{fig:0} with our collider bounds summarized
in Tables\,\ref{tab:7}-\ref{tab:8} and in 
Figs.\,\ref{fig:8}-\ref{fig:9} of Section\,\ref{sec:5},
we find that these perturbative unitarity bounds are much weaker
than our collider bounds.\ Hence, they do not affect our collider
analyses in the following Sections\,\ref{sec:4}-\ref{sec:5}.\

\section{\hspace*{-2.5mm}Form Factor Formulation for nTGCs}
\label{sec:FF}
\label{sec:3}

We study in this Section the form factor formulation
of the neutral triple gauge vertices (nTGVs)  
$Z\gamma V^*$.\ 
After imposing Lorentz invariance, the residual electromagnetic $U(1)_{\rm{em}}^{}$ 
gauge symmetry and CP conservation,
they are conventionally expressed in the following 
form\,\cite{nTGC1}\cite{Degrande:2013kka}:
\begin{eqnarray}
\label{eq:FF0-nTGC}
\Gamma_{Z\gamma V^*}^{\alpha\beta\mu}
(q_1^{},q_2^{},q_3^{})
\!\!&=&\!\!
\frac{~e\hs (q_3^2-\hsm M_V^2)\,}{\,M_Z^{2}~}\!
\(\! h_3^V\hs q_{2\nu}^{}\epsilon^{\alpha\beta\mu\nu}\!
+ \frac{h_4^V}{\,M_Z^2\,}\hs q_2^{\alpha}\hs q_{3\nu}^{}\hs q_{2\sigma}^{}\epsilon^{\beta\mu\nu\sigma}\!\)\!,
\hspace*{16mm}
\end{eqnarray}
where the gauge bosons are denoted by $\,V\!\! \equiv \! Z,\gamma$ 
and the form factor parameters
$(h_3^V,\, h_4^V)$ are treated as 
constant coefficients for the purposes of
experimental tests\,\cite{CMS2016nTGC-FF}.\footnote{$q_2^{\alpha}\hs q_{3\nu}^{}\hs q_{2\sigma}^{}\epsilon^{\beta\mu\nu\sigma}$ is equivalent to $\,q_3^{\alpha}\hs q_{3\nu}^{}\hs q_{2\sigma}^{}\epsilon^{\mu\beta\nu\sigma}$ under 
the on-shell condition $\hs (q_2^\alpha\hsm +q_3^\alpha)\epsilon_{\alpha}^*
=-q_1^\alpha\epsilon_{\alpha}^*\hsmx = 0\hsx$.}

\vspace*{1mm}

We stress that the spontaneous breaking of the SM electroweak 
gauge symmetry requires the nTGCs to be 
generated only by the gauge-invariant 
effective operators of dimension-8 or higher.\ 
This implies that the consistent form factor formulation of the 
neutral triple gauge vertices 
must map precisely the expressions for these
gauge-invariant nTGC operators in the broken phase.\  
This precise mapping between the nTGVs in 
the broken phase of these dimension-8 nTGC operators \eqref{eq:nTGC-d8}
imposes nontrivial relations between the parameters 
of the nTGVs in the form factor formulation and removes possible unphysical 
energy-dependent terms in them.\footnote{%
The spontaneous breaking of the SM electroweak 
gauge symmetry has many important physical consequences that,
most notably, 
guarantee the renormalizability\,\cite{SSBrenor} of the 
SM electroweak gauge theory.\ 
Here our new observation is that {\it the spontaneous breaking of the
electroweak gauge symmetry requires nontrivial extension of
the conventional form factor parametrization and 
imposes new restrictions on these form factors} 
that go beyond the residual U(1)$_{\rm{em}}^{}$ 
gauge symmetry alone.\ 
These considerations were not incorporated in the conventional
form factor formulation of the 
nTGVs\,\cite{nTGC1}\cite{Degrande:2013kka}.} 

\vspace*{1mm}

By direct power counting, we find that 
the dimension-8 operator $\mO_{G+}^{}$ contributes 
to the nTGVs with a leading 
$\hs E^5\hs$ energy dependence. 
Based on this and the above observations, 
we find that {\it the conventional
form factor formula \eqref{eq:FF0-nTGC} is not 
compatible with the gauge-invariant SMEFT formulation,
and a new term must be added,
labelled by $h_5^V$} in the following.\ 
With these remarks in mind, we express 
the neutral triple gauge vertices $Z\gamma V^*$ as follows:
\begin{eqnarray}
\label{eq:ZAV*-FormF} 
\Gamma_{Z\gamma V^*}^{\alpha\beta\mu(8)}
(q_1^{},q_2^{},q_3^{}) \,=\,
\frac{\,e\hs (q_3^2\hsm -\!M_V^2)\,}{\,M_Z^2\,}\!
\[\!\(h_3^V\!+h_5^V\frac{q_3^2}{\,M_Z^2\,}\)\!
q_{2\nu}^{}\epsilon^{\alpha\beta\mu\nu}
\hsm +\hsm \dis\frac{h_4^V}{\,M_Z^2\,}\hs 
q_2^{\alpha}\hs q_{3\nu}^{}\hs  q_{2\sigma}^{}\hs\epsilon^{\beta\mu\nu\sigma}\]\! ,
\end{eqnarray}
where the form factors $h_i^V$ are taken as constants
in the present study.\ 
The parametrization of the nTGVs 
in Eq.\eqref{eq:ZAV*-FormF} corresponds to 
the following effective Lagrangian: 
\begin{align}
\mL \,=\,&
\frac{e}{\,M_Z^2\,}\! \Bigg[\!
-\!\hsm\[h_3^\gamma  (\partial_\sigma F^{\sigma \rho})\hsm 
+ h_3^Z  (\partial_\sigma Z^{\sigma \rho})\hsm 
+\!\frac{h_5^\gamma}{\,M_Z^2\,}
(\partial^2\partial_\sigma F^{ \rho\sigma})
+ \frac{h_5^Z}{\,M_Z^2\,}(\partial^2\partial_\sigma Z^{ \rho\sigma})
\hsm\]\!\hsm Z^\alpha
\wtil{F}_{\rho\alpha}^{} 
\nn\\
& \hspace*{10mm}+ \hsm\left\{\!
\frac{h_4^\gamma}{\,2M_Z^2\,}\!\[\square \partial^\sigma\!
F^{\rho \alpha}\] +
\frac{h_4^Z}{\,2M_Z^2\,}\!\[(\square +\!M_Z^2) \partial^\sigma\!
Z^{\rho \alpha}\] \!\right\}\!Z_\sigma \wtil{F}_{\rho \alpha }
\Bigg].
\label{eq:LnTGV}
\end{align}
which differs from the conventional  nTGV form factor 
Lagrangian~\cite{Gounaris:1999kf} by the new 
$h_5^V$ terms.

\vspace*{1mm}

We now compare our modified nTGV formula  
\eqref{eq:ZAV*-FormF} with the nTGVs in
Eqs.\eqref{eq:VertexZAZ-G+}-\eqref{eq:VertexZAZ-G-}
as predicted by the gauge-invariant dimension-8 nTGC operators 
$(\OGP,\,\OGM,\,\OBW)$ in Eqs.\eqref{eq:OG+}-\eqref{eq:OBW},
which should match exactly case by case.
In the case of the operator $\OGP$, this matching leads to the
following two restrictions on the form factors 
in Eq.\eqref{eq:ZAV*-FormF}:
\beqs
\label{eq:h4h5-h4ZA}
\begin{align}
\label{eq:h4-h5} 
h_4^V & \,=\, 2\hs h_5^V\,,
\\[1mm]
\label{eq:h4Z-h4A} 
h_4^Z\hs & =\hs \frac{\,c_W^{}\,}{s_W^{}}\hs h_4^\gamma \,,
\end{align}
\eeqs 
where henceforward we denote 
$\hs h_4^{}\equiv h_4^Z\hsx$ for convenience.\ 
These conditions demonstrates that {\it there are only
three independent form-factor parameters} 
$(h_3^Z,\hs h_3^\gamma,\hs h_4^{})$.\ 
Applying the condition \eqref{eq:h4-h5}, we can express 
the $Z\gamma V^*$ vertex \eqref{eq:ZAV*-FormF} as follows:
\begin{eqnarray}
\label{eq:FF2-nTGC}
\Gamma_{Z\gamma V^*}^{\alpha\beta\mu(8)}(q_1^{},q_2^{},q_3^{})
\,=\,
\frac{\,e\hs (q_3^2\hsm -\!M_V^2)\,}{\,M_Z^{2}\,}\!
\[\!\(\! h_3^V\!+\!\frac{\,h_4^V\,}{\,2M_Z^2\,}q_3^2\hsm\)\!
q_{2\nu}^{}\hs\epsilon^{\alpha\beta\mu\nu}\hsm 
+\hsm \frac{h_4^V}{\,M_Z^2\,}\hs q_2^{\alpha}\hs q_{3\nu}^{}\hs  q_{2\sigma}^{}\hs\epsilon^{\beta\mu\nu\sigma}\]\! .
\end{eqnarray}
Comparing the nTGVs \eqref{eq:Vertex-G+} predicted by
the dimension-8 operators \eqref{eq:OG+}-\eqref{eq:OBW}
with the form factor formulation \eqref{eq:FF2-nTGC} 
of the nTGVs, we can connect the three independent form-factor
parameters $(h_3^Z,\hs h_3^\gamma,\hs h_4^{})$ to the 
cutoff scales
$(\cut_{G+}^{},\hs\cut_{G-}^{},\hs\cut_{\widetilde{B}W}^{})$
of the corresponding dimension-8 operators
$(\mO_{G+}^{},\hs\mO_{G-}^{},\hs\mO_{\widetilde{B}W}^{})$,
as follows:
\\[-6mm]
%
\beqs	
\label{eq:h-dim8}
\begin{align}
\label{eq:h-OG+}
h_4^{} &=-
\frac{~\text{sign}(\tilde{c}_{G+}^{})\,}{\,\Lambda^4_{G+}\,}
\frac{v^2\hsm M_Z^2}{\,s_W^{}c_W^{}\,}
\equiv\!\frac{r_4^{}}{\,[\Lambda^4_{G+}]\,}
\hs , 
&& \hspace*{-10mm}
h_3^V\!=0\hs, 
\hspace*{-12mm}
&& \hs\text{for}~\OGP\hs \hs, 
\hspace*{5mm}
\\
\label{eq:h-OBW}
h_3^Z &= 
\frac{\,\text{sign}(\tilde{c}_{\!\widetilde BW}^{})\,}
{\,\Lambda_{\!\widetilde BW}^4\,}
\frac{v^2\hsm M_Z^2}{~2s_W^{}c_W^{}\,} 
\equiv\!
\frac{r_3^Z}{\,[\Lambda^4_{\widetilde{B}W}]\,} \hs ,
&& \hspace*{-10mm}
h_3^\gamma,h_4^V\!=0\hs, 
\hspace*{-12mm}
&& \hs\text{for}~\OBW\hs \hs, 
\\
\label{eq:h-OGM}
h_3^\gamma &= 
-\frac{\,\text{sign}(\tilde{c}_{G-}^{})\,}
{\,\Lambda_{G-}^4\,}
\frac{v^2\hsm M_Z^2}{~2c_W^2\,}
\equiv\!\frac{r_3^{\ga}}{\,[\Lambda^4_{G-}]\,}
\,.
&& \hspace*{-10mm}
h_3^Z,h_4^V\!=0\hs, 
\hspace*{-12mm}
&& \hs\text{for}~\OGM\hs \hs, 
\end{align}
\eeqs
where the form factor $h_4^{}$ is defined below
Eq.\eqref{eq:h4h5-h4ZA} and we have used the notations:
\beqs
\label{eq:[Lambda-4]-r34}
\begin{align}
\label{eq:[Lambda-4]}
&
[\Lambda^4_{G+}] =
\text{sign}(\tilde{c}_{G+}^{})
\Lambda^4_{G+}\hs,~~~
[\Lambda^4_{\widetilde{B}W}] = 
\text{sign}(\tilde{c}_{\!\widetilde BW}^{})
\Lambda^4_{\widetilde{B}W}\hs,~~~
[\Lambda^4_{G-}] =
\text{sign}(\tilde{c}_{G-}^{})
\Lambda^4_{G-}\hs,
\\[1mm]
\label{eq:r4-r3}
&
r_4^{} = -\frac{~v^2\hsm M_Z^2~}{\,s_W^{}c_W^{}\,}\hs, 
~~~~~
r_3^Z=\frac{v^2\hsm M_Z^2}{~2s_W^{}c_W^{}\,} \hs,
~~~~~
r_3^{\ga} = -\frac{~v^2\hsm M_Z^2~}{~2c_W^2\,}\hs.
\end{align} 
\eeqs 
From the above, we see that only the operator $\OGP$ 
can directly contribute to the form factor $h_4^V$, 
as in Eq.\eqref{eq:h-OG+}, which can be understood 
from the explicit formulae \eqref{eq:VertexZAZ-G+}.
We note that the operator $\hs\OBW\hsm$ contains Higgs-doublet fields 
and thus cannot contribute to the $\hs h_4^V$ term in Eq.\eqref{eq:FF2-nTGC},\ 
but $\hs\OBW\!$ can contribute to the $\hs h_3^Z$ term 
through the $Z\gamma Z^*$ vertex and leaves $\hs h_3^\gamma\!=0\hs$,
as shown in Eq.\eqref{eq:h-OBW}.\
The operator $\OGM$ also cannot contribute to $h_4^V$
due to the equation of motion \eqref{eq:OCP=OGM-OBW}, 
$\OGM\!\!=\hsm \OBW\! +\OCP$,
where $\OCP$ contains a bilinear fermion factor and 
cannot contribute directly to the nTGC.
The fact that $\OGM$ is irrelevant to $h_4^V$ is also shown explicitly in Eq.\eqref{eq:VertexZAZ-G-}.\  The explicit formula
\eqref{eq:VertexZAZ-G-} further shows that 
$\OGM$ makes a nonzero contribution to $h_3^\gamma\hs$, 
but leaves $\hs h_3^Z\!=0\hs$, 
as we find in Eq.\eqref{eq:h-OGM} above.

\vspace*{1mm}

Using Eq.\eqref{eq:ZAV*-FormF} or \eqref{eq:FF2-nTGC}
and by direct power counting,
we infer the following leading energy-dependences of the $h_i^V$ 
contributions to the helicity amplitudes  
$\hs \TT[f\bar f\ito Z\ga]\,$:
\beqs
\label{eq:FT8}  
\begin{eqnarray}
\label{eq:FT8-ZT}
\mathcal{T}_{(8)}^{ss'\!,\text{T}}
\hspace*{-2mm}
&=&\!\! h_3^V O(E^2)+h_5^V O(E^4) \, ,
\\[1mm]
\label{eq:FT8-ZL}
\mathcal{T}_{(8)}^{ss'\!,\text{L}}
\hspace*{-2mm}
&=&\!\! h_3^V O(E^3)+h_4^V O(E^5)+h_5^V O(E^5) \, .
\hspace*{18mm}
\end{eqnarray}
\eeqs
We note in Eq.\eqref{eq:FT8-ZT} that the form factor 
$\,h_4^V\hs$ does not contribute to the production of a 
transversely polarized $Z$ boson in the final state, because 
the $s$-channel momentum $\,q_3^\alpha\,$ has no spatial component
and the $Z$ boson's transverse polarization vector 
$\hs\epsilon_{T\alpha}^{}\hs$  has no time component,
and thus $\,q_3^\alpha\epsilon_{T\alpha}^*\!=\hsm 0\hs$.

\vspace*{1mm}  

Inspecting Eq.\eqref{eq:FT8},
it would appear that the leading energy-dependence of $\,\mathcal{T}_{(8)}^{ss'\!,\text{L}}$
should be $\mO(E^5)$.\  
However, we observe that the helicity amplitudes including a
final-state longitudinal $Z$ boson
as contributed by the gauge-invariant dimension-8 nTGC
operators must obey 
the equivalence theorem (ET)\,\cite{ET}.
At high energies $E\!\gg\! M_Z^{}\hs$, the ET 
takes the following form:
\beqa
\vspace*{-2mm}
\mathcal{T}_{(8)}^{}[Z_L^{},\ga_T^{}] \,=\,
\mathcal{T}_{(8)}^{}[-\ii\pi^0,\ga_T^{}] + B\,,
\label{eq:ET}
\eeqa
where the longitudinal gauge boson $Z_L^{}$ absorbs 
the would-be Goldstone boson $\pi^0$   
through the Higgs mechanism, and
the residual term 
$\,B\hsm =\hsm\mathcal{T}_{(8)}^{}[v^\mu Z_\mu^{},\ga_T^{}]\,$
is suppressed by the relation
$\,v^\mu\!\hs\equiv\!\epsilon_L^\mu -$
$q_Z^\mu/M_Z^{}\hsm =\hsm O(M_Z^{}/E_Z^{})$ \cite{ET}. 
However, we cannot apply the ET \eqref{eq:ET} directly to the 
form factor formulation \eqref{eq:ZAV*-FormF}, because it 
does not respect the full electroweak gauge symmetry of the SM
and contains no would-be Goldstone boson.
We stress again that the  electroweak gauge-invariant formulation
of the nTGCs can be derived only from the dimension-8 operators
as in Eq.\eqref{eq:nTGC-d8}.
Hence, we study the allowed 
leading energy-dependences of the helicity amplitudes 
\eqref{eq:FT8} by applying the ET to the contributions
of the dimension-8 nTGC operators \eqref{eq:nTGC-d8}.\ 
Then, we find that only the operator
$\mathcal{O}_{\widetilde{B}W}^{}$ could give a nonzero
contribution to the Goldstone amplitude   
$\,\mathcal{T}_{(8)}^{}[-\ii\pi^0,\ga_T^{}]\,$,
with a leading energy-dependence $\mO(E^3)$
that corresponds to the form factor $h_3^Z\hs$.\ 
The operator $\hs\mathcal{O}_{G_+}^{}\!$ does not contribute
to the Goldstone amplitude 
$\mathcal{T}_{(8)}^{}[-\ii\pi^0,\ga_T^{}]$,  
but can contribute the largest residual term $B=\mO(E^3)$. 
From these facts, we deduce that in Eq.\eqref{eq:FT8-ZL} 
{\it the $\mO(E^5)$ terms due to the form factors 
$h_4^V$ and $h_5^V$ must exactly cancel each other,}  
from which we derive the following condition,
\beq 
\label{eq:E5cancel-h4h5} 
h_4^V/\hs h_5^V \,=\, 2\,,
\eeq
which agrees with Eq.\eqref{eq:h4-h5}. 
Then, using our improved form factor
formulation \eqref{eq:FF2-nTGC} of the nTGCs, 
we can compute the corresponding helicity amplitudes
of $\hs f\bar f\ito Z\ga\hs$ 
from the nTGC contributions: 
\beqs
\label{eq:T8h}
\begin{align}
\hspace*{-3mm}
&	\mathcal{T}_{(8),\text{F}}^{ss'\!,\text{T}}
\!\!\left\lgroup\!\!\!
\begin{array}{cc}
-- \!&\! -+ \\
+- \!&\! ++\\
\end{array}\!\!\!\right\rgroup\!\!
=
\frac{\,(c_L^V\!\!+\!c_R^V)e^2(2h^V_3M_Z^2+h_4^Vs) (s\!-\!M_Z^2)\sin\!\theta\,}{4M_Z^4 c_Ws_W}\!\!
\left\lgroup\!\!
\begin{array}{cr}
1 ~& 0
\\[1mm]
0 ~& -1
\end{array}
\!\!\right\rgroup \!\!,
\\[2mm]
\hspace*{-3mm}
& \mathcal{T}_{(8),\text{F}}^{ss'\!,\text{L}}(0-,0+)
=
\frac{\,\sqrt{2\,}\hs e^2(s\!-\!M_Z^2)\sqrt{s\,}\,}
{4M_Z^3c_Ws_W}\hsm
(2h_3^V\!+\hsm h_4^V)\!
\(\!c_L^V\!\sin^2\!\frac{\theta}{2}\!
-c_R^V\hsm\cos^2\!\frac{\theta}{2},~
c_R^V\hsm\sin^2\!\frac{\theta}{2}\!
-c_L^V\hsm\cos^2\!\frac{\theta}{2}
\)\!,
\end{align}
\eeqs
where the coupling coefficients are defined as
$\,(c_L^Z,\hs c_R^Z)\!=\hsm (T_3^{}\hsm -Qs_W^2,-Qs_W^2)$
for $V\!\!=\hsmx Z\hs$ and 
$\,c_L^A\hsm =\hsm c_R^A=Qc_W^{} s_W^{}$ for
$V \!=\gamma\hs$.\ 
On the right-hand-side of the above formulas, 
the subscript ``\,F\,'' indicates contributions 
given by the form factors.\  
From the above, we see that the helicity amplitude
$\hs\mathcal{T}_{(8)}^{ss'\!,\text{T}}$
for the transverse $Z_T^{}$ final state contains 
the $O(E^2)$ contribution from the form factor $\hs h_3^V$ 
and the leading contribution of $O(E^4)$ from the form factor 
$\hs h_4^V$, while the the helicity amplitude
$\hs\mathcal{T}_{(8)}^{ss'\!,\text{L}}$
for the longitudinal $Z_L^{}$ final state has a
leading contribution of $O(E^3)$ from the form factor combination 
$\hs (2h_3^V\!+h_4^V)\hs$.

\vspace*{1mm}

We note that the operators $\mO_{C+}^{}$ and $\mO_{C-}^{}$ 
both contain only left-handed fermions,
and recall that the operators $\OGP$ and $\OCM$ give the same contributions
to the amplitude $\TT[f\bar f\ito Z\ga]\hs$, due to
the equation of motion \eqref{eq:OCM=OGP-HBW}.\ 
Thus, we find that the ratio $\hs h_4^Z/h_4^\gamma\hs$ 
must be fixed to cancel their contributions to the amplitude
$\,\mathcal{T}[f\bar f\!\!\to\! Z^*\!\!\to\!\! Z\gamma]+ 
\mathcal{T}[f\bar f\!\!\to\! \gamma^*\!\!\to\!\! Z\gamma]$ via right-handed fermions\,\cite{Ellis:2020ljj}.\ 
This imposes the following condition on the two form factors
$(h_4^Z,\hs h_4^\gamma)$: 
\beq 
\label{eqx:h4Z-h4A}
h_4^{}\,\equiv\, h_4^Z\hs =\hs \frac{\,c_W^{}\,}{s_W^{}}\hs 
h_4^\gamma \hs,
\eeq
for the $\OGP$ operator.\ 
This condition agrees with Eq.\eqref{eq:h4Z-h4A},
which we derived earlier by matching the prediction
of the operator $\OGP$ with the nTGV formulation 
\eqref{eq:ZAV*-FormF}.
Hence, using the gauge-invariant dimension-8 nTGC operators
to derive the form factor formulation \eqref{eq:ZAV*-FormF}, 
we deduce that there are only three 
independent form-factor parameters $(h_3^Z,\hs h_3^\gamma,\hs h_4^{})$,
where $\hs h_4^{}\hsm\equiv\hsm h_4^Z$ and 
$\hs h_4^\gamma\hs$ are connected by the condition 
\eqref{eqx:h4Z-h4A}.\


The fermionic dimension-8 operators $\mO_{C+}^{}$ and
$\mO_{C-}^{}$ contribute to the quartic vertex $f\bar{f}Z\ga\hs$,
but do not contribute directly to the
nTGC vertex $Z\ga V^*$ in Eq.\eqref{eq:FF2-nTGC}.\ 
We can factorize their contribution to the on-shell quartic vertex 
$f\bar{f}Z\ga\hs$ as follows:
\beq
\label{eq:Vertex-ffZA}
\Gamma^{\al\beta}_{\!f\bar fZ\gamma} (q_1^{}, q_2^{})
\,=\,
\sum_V\Gamma_\mu(f\bar{f}V^*)P_L^{}\hsm\times\hsm  
(q_3^2\hsm -\! M_V^2)^{-1}\!\times\hsm
\Gamma_{Z\gamma V^*}^{\alpha\beta\mu}(q_1^{},q_2^{},q_3^{})\,,
\eeq 
which includes effectively an nTGC vertex 
$\hs\Gamma_{Z\gamma V^*}^{\alpha\beta\mu}\hs$.\ 
This effective nTGC vertex function 
$\hs \Gamma_{Z\gamma V^*}^{\alpha\beta\mu}\hs$
contains the form factor parameters $(h_3^Z,\hs h_3^\gamma)$
for the operator $\OCP\hs$.\  Since $\OCP$ involves
purely left-handed fermions, we find that
the ratio $\hs h_3^Z/h_3^\gamma\hs$ 
must be fixed, so as to cancel its contributions to the amplitude
$\,\mathcal{T}[f\bar f\!\!\to\! Z^*\!\!\to\!\! Z\gamma]\hsm +\hsm 
\mathcal{T}[f\bar f\!\!\to\! \gamma^*\!\!\to\!\! Z\gamma]$ via right-handed fermions. This imposes the following condition between
form factors $(h_3^Z,\hs h_3^\gamma)$:
\beq 
\label{eq:h3Z-h3A}
h_3^{}\,\equiv\, h_3^Z\hs =\hs \frac{\,c_W^{}\,}{s_W^{}}\hs 
h_3^\gamma \hs, \hspace*{12mm}
\hs\text{for}~\OCP\hs \hs .
\eeq
We note that the above relation holds only for the
fermionic operator $\hs\OCP\hs$.\
For the other fermionic operator $\OCM$, its contribution
to the effective nTGC vertex function 
$\hs \Gamma_{Z\gamma V^*}^{\alpha\beta\mu}\hs$ 
in Eq.\eqref{eq:Vertex-ffZA}
contains the same form factors $(h_4^Z,\hs h_4^\gamma)$
as that of the operator $\OGP$,
because the equation of motion guarantees\,\cite{Ellis:2020ljj}
that both of the operators $\OGP$ and $\OCM$ give the same contributions
to the  on-shell quartic vertex $f\bar{f}Z\ga\hs$.\
Thus, the form factors $(h_4^Z,\hs h_4^\gamma)$ of the effective nTGC vertex function 
$\hs \Gamma_{Z\gamma V^*}^{\alpha\beta\mu}\hs$
of the left-handed fermionic operator $\OCM$ 
obey the same cancellation
condition Eq.\eqref{eq:h4Z-h4A}.

\section{\hspace*{-2.5mm}Probing nTGCs at the LHC and Future  \boldmath{$pp$} Colliders}
\label{sec:4}
\vspace*{1mm}

In this Section we will analyze the sensitivity reaches on
probing the nTGCs at the LHC and future $p{\hs}p$ colliders
via the reactions 
$\hs p{\hs}p{\hs}(q{\hs}\bar{q})\ito Z\gamma\hs$
with $Z\hsm\ito\ell^+\ell^-\! ,\hs\nu{\hs}\bar{\nu}{\hs}\hs$.\ 
In Section\,\ref{sec:4.0}, we give the setup for the analyses.\
In Sections\,\ref{sec:4.1}-\ref{sec:4.2}, we present the 
analyses of nTGCs at ${\cal O}(\Lambda^{\!-4})$ and
${\cal O}(\Lambda^{\!-8})$ respectively.\ 
In the analysis of Section\,\ref{sec:4.3}, we further include
the decay channel of $Z\ito\nu{\hs}\bar{\nu}$.\  
Then, we study the probes of the nTGV form factor in
Section\,\ref{sec:4.4}, and the correlations between the
nTGC sensitivities in Section\,\ref{sec:4.5}.\
Finally, we compare in Section\,\ref{sec:4.7}
our predicted LHC sensitivity reaches on
the nTGCs with the published LHC experimental limits
by both the ATLAS and CMS collaborations.

\vspace*{1mm}
\subsection{\hspace*{-2.5mm}Setup for the Analyses at Hadron Colliders}
\label{sec:4.0}
\vspace*{1mm}

The distributions of quark and antiquark momenta in protons 
are given by parton distribution functions (PDFs). 
At leading order, 
the total cross section of $\,p{\hs}p\ito Z\ga\hs$  
at the LHC is calculated by integrating 
the convolved product of the quark and antiquark PDFs 
and the parton-level cross section of 
the $\hs q\hs\bar{q}\ito Z\ga\hs$ subprocess:
\beqa
\sigma ~=~
\sum_{q,\bar{q}\,}\!\int\!\! \d x_1^{}\hs\d x_2^{} \[
\mathcal{F}_{\!q/p}^{}(x_1^{},\mu)\hs
\mathcal{F}_{\!\bar{q}/p}^{}(x_2^{},\mu)\hsx
{\sigma}_{\!q\bar q}^{}(\hat s)+(q\leftrightarrow\bar q)\]\!,
\eeqa
where the functions $\mathcal{F}_{\!q/p}^{}$ and 
$\mathcal{F}_{\!\bar{q}/p}^{}$ are the PDFs of the quark and 
antiquark in the proton beams, and 
$\,\hat s\!=\!x_1^{}\hs x_2^{}\hs s\,$ with the collider energy 
$\sqrt{s\,}\hsmx\!=\hsmx\! 13\,\text{TeV}\hs$.\  
The PDFs depend on the factorization scale $\mu$, 
which is set to be $\,\mu\hsm\! =\hsmx\!\sqrt{\hat{s}\,}/2\,$ 
in our leading-order analysis.\
We use the PDFs of the quarks 
$\hs q\hsm =u,d,s,c,b\hs$ and their antiquarks
determined by the CTEQ collaboration\,\cite{Lai:1999wy}. 

\vspace*{1mm}

During LHC Run-2 the ATLAS measurements of the 
$\ell^+ \ell^- \gamma$ and $\bar \nu \nu \gamma$ 
final states reached a maximum
value of $M_{\ell\ell\ga} \!\sim\! 3$\,TeV.\footnote{%
We thank our ATLAS colleague Shu Li for discussions of the ATLAS
measurements during LHC Run-2.}  
Accordingly, we set  $\hat{s}\hsmx\lesssim\hsmx 3$\,TeV 
for our LHC analysis and use an upper limit
$\hat{s}\lesssim 23$\,TeV for the 100\,TeV $pp$ collider.

\vspace*{1mm}

We compute the production cross section
of $\hs q\hs\bar q\ito Z\ga\hs$ at leading order (LO) 
in QCD and ${\cal O}(\alpha^2)$ for the SM, and 
$O(\alpha^{1.5}\tilde{c}_j^{})$ or
$O(\alpha\tilde{c}_j^2)$ for the nTGCs,
where $\alpha\!=\!\alpha_{\rm{em}}^{}$ 
or $\alpha_{\rm{w}}^{}$, 
as the possible
high-order contributions are not important
for our study.\ 
There are next-to-leading-order (NLO) QCD corrections from
the gluon-induced loop diagrams for $\hs q\hs\bar q\to Z\ga\hs$ 
and the real emission of a gluon: $\hs q\hs\bar q\ito Z\ga\hsm +\hsm g\hs$, and there are also NLO QCD contributions from 
$g\hs q \ito Z\ga\hsm +\hsm q$
$(\hs g\hs\bar{q}\ito Z\ga\hsm+\bar{q}\hs )\hs$.\
In these cases the NLO/LO ratio is ${\cal O}(\alpha_s^{})$, and
it was found numerically that the effect of adding the full 
NNLO corrections is less than 10\% \cite{Campbell:2017aul}\cite{Grazzini:2017mhc}\cite{Aad:2019gpq}.\ 
We define a QCD $K$-factor for the nTGC signal by 
$\hs K_{\rm{S}}^{} \!\hsm \equiv \! S\hs /S_{\rm{LO}}^{}\!\hsm =\!\hsm
1\!+\!\Delta K_{\rm{S}}^{}\hs$ 
and for the SM background by 
$\hs K_{\rm{B}}^{} \! \equiv \! B/B_{\rm{LO}}^{}\!=\!
1\!+\!\Delta K_B^{}\hs$.\
We have checked the $K$-factors for $\hs pp\ito Z\ga\hs$ 
by using {\tt Madgraph5@NLO}~\cite{Alwall:2014hca}, 
and find that they depend on the kinematic cuts.\ 
The corrections $\Delta K$ can be larger than one 
if only basic cuts are made, but we find that adding a cut to remove the small  $P_T^{}(\ga)$ region and vetoing extra jets in the final state
reduces $\Delta K\hs$ 
to only a few percent, which may be neglected.

\vspace*{1mm}

We note in addition that $Z\ga$ production by the gluon fusion
process is formally a next-to-next-to-leading-order (NNLO) contribution, 
and is found to be generally less than 1\% \cite{Adamson:2002rm}.\ 
The nTGC contributions via gluon fusion is also found to be
negligible\,\cite{Adamson:2002rm}.\

\vspace*{1mm}

Next, we discuss the statistical significance and its optimization
for our present analysis of sensitivity reaches on the nTGCs.\
Since the SM contribution $\sigma_0^{}$ could be small, the ratio 
$\hs S/\hsm\sqrt{B\,}\hs$ is not an optimal measure of the 
statistical significance.\ 
We use instead the following formula for the 
background-with-signal hypothesis\,\cite{Z0}:
\beqa
\label{eq:SS-Z}
\mathcal{Z} \,=\, \sqrt{2\!\(\!B\ln\!
\frac{\,B\,}{\,B\hsm+\hsm S\,} +  S\hsm\)\,}
\,=\,\sqrt{2\!\left(\!\sigma_0^{}\ln\!
\frac{\,\sigma_0^{}\,}{\,\sigma_0^{}\hsm+\hsm\Delta\sigma\,} +   \Delta\sigma\!\)\,}\times\sqrt{\mL\times\epsilon~},
\eeqa
where  
$\hs\Delta\sigma \hsm =\sigma\hsm -\hsm\sigma_0^{}\hs$
denotes the part of the cross section beyond the SM contribution, 
$\hs\mL\hs$ is the integrated luminosity, and 
$\hs\epsilon\hs$ is the detection efficiency.\
When $\hs B\!\gg\!S\hs$, we can expand (\ref{eq:Z>2}) 
in terms of $\,S/B\,$ and
find that it reduces to the form
$\,\SZZ \!\simeq\! S/\!\sqrt{B\,}$, 
whereas for $\hs S\!\gg\! B\hs$ it reduces to 
$\,\SZZ\!\simeq\!\sqrt{2S\,}\hs$.\
If the signal $S$ is dominated by the interference contribution 
of $O(\cut^{-4})$,  
we can deduce that the sensitivity reach on the new physics scale:
\beqs 
\beqa
&& \Lambda \propto (\mL\hsmx\times\hsmx\ep)^{1/8},\,
\hspace*{14mm} (\text{for}~B\!\gg\! S\hs),
\\
&& \Lambda \propto (\mL\hsmx\times\hsmx\ep)^{1/4},\,
\hspace*{14mm} (\text{for}~S\!\gg\! B\hs).
\eeqa 
\eeqs  
If the signal $S$ is dominated by the squared contribution 
of $O(\cut^{-8})$,  
we can deduce that the sensitivity reach on the new physics scale:
\beqs 
\beqa
&& \Lambda \propto (\mL\hsmx\times\hsmx\ep)^{1/16},\,
\hspace*{14mm} (\text{for}~B\!\gg\! S\hs),
\\
&& \Lambda \propto (\mL\hsmx\times\hsmx\ep)^{1/8},\,
\hspace*{15.5mm} (\text{for}~S\!\gg\! B\hs).
\eeqa 
\eeqs  
%
In either case, we see that the bound on the 
new physics scale $\cut$ is
not very sensitive to the integrated luminosity $\hs\mL\hs$ 
and the detection efficiency $\hs\ep\hs$. 
For instance, in the case of $B\!\gg\! S$,
if the integrated luminosity $\hs\mL\,$ increases 
by a factor of $10\hsx$,
we find that the sensitivity reach of $\hs\cut\hs$ is enhanced
by about 33\% when the interference contribution dominates the signal
and 15\% when the squared contribution dominates the signal.\ 
If the detection efficiency $\hs\ep\hs$ is reduced from the
ideal value of $\hs\ep \!=\!1\hs$ to $\hs\ep\!=\!0.5\hs$,
we find that the sensitivity reach of $\cut$ is weakened 
by only about 8\% when the interference contribution dominates the signal
and 4\% when the squared contribution dominates the signal.

\vspace*{1mm}

In order to achieve higher sensitivity, 
we can discriminate between the signal and background 
by using the photon $P_T^{}$ distribution, employing the following measure of significance:
\beqa
\mathcal{Z}_{\rm{total}}^{}\,=\,
\sqrt{\,\sum\!\mathcal{Z}_{\rm{bin}}^2~} \,.
\label{eq:Zbin}
\eeqa
In the above, we impose the optimal cut on the photon 
$P_T^{}$ for each bin
and compute the corresponding significance 
$\mathcal{Z}_{\rm{bin}}^{}$ of each bin.\ 
By doing so, we maximize the significance
$\hs \mathcal{Z}_{\rm{total}}^{}\hs$ given in  
Eq.\eqref{eq:Zbin}.\

\vspace*{1mm}
\subsection{\hspace*{-2.5mm}Analysis of nTGCs at \boldmath{${\cal O}(\Lambda^{\!-4})$}}
\label{sec:4.1}
\vspace*{1mm}

We compute analytically the parton-level 
cross section of the annihilation process
$\hs q\hs\bar{q}\ito Z\ga\hs$, and
then perform the convolved integration over the 
product of the quark and antiquark 
PDFs to obtain the cross section for 
$\hs p{\hs}p{\hs}(q{\hs}\bar{q})\ito Z\gamma\hs$.

\vspace*{1mm} 

\begin{figure}[t]
\vspace*{-5mm} 
\begin{center}
\includegraphics[width=10.4cm]{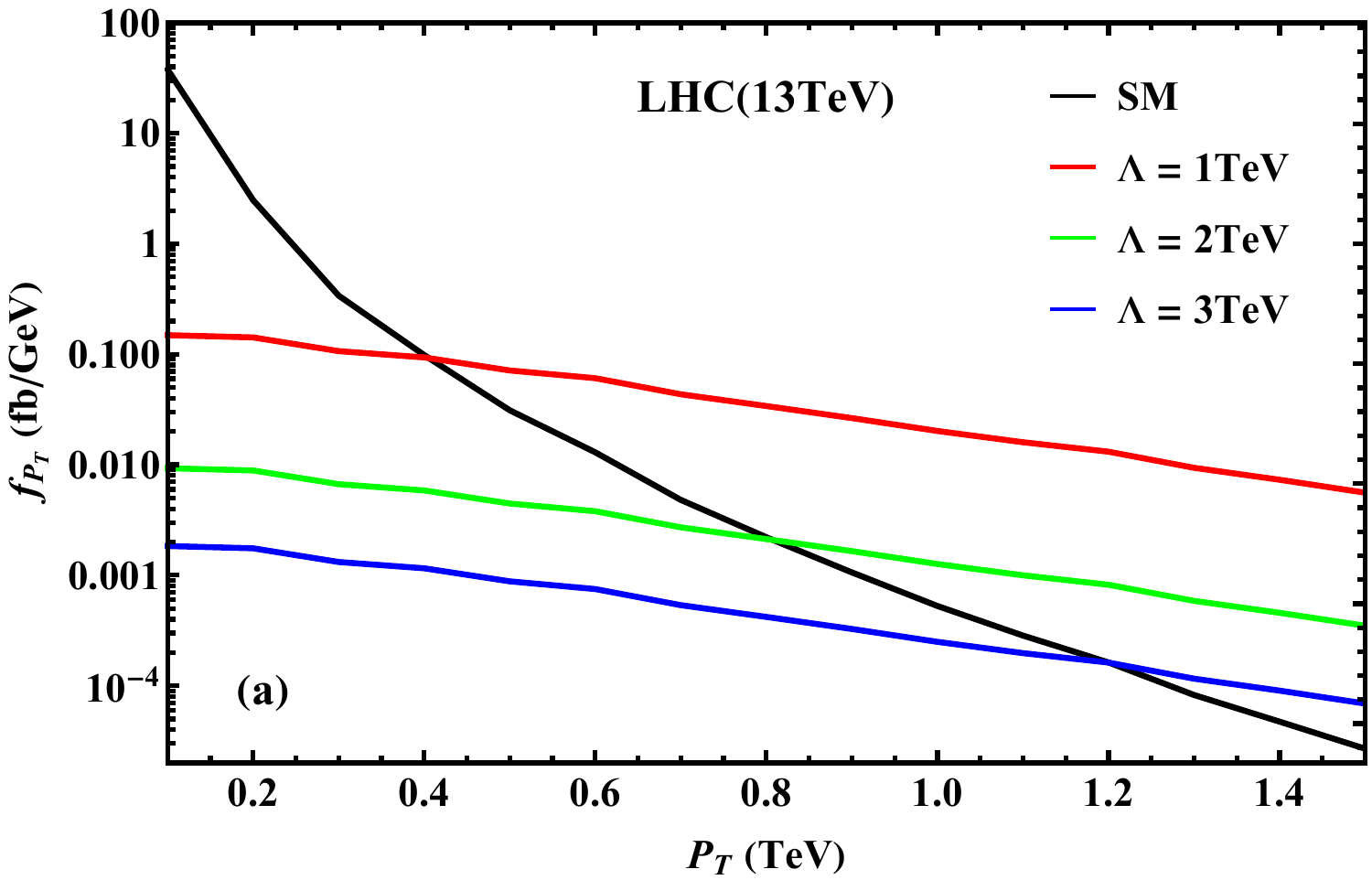}
\\[2mm]
\hspace*{2mm} 
\includegraphics[width=10.7cm]{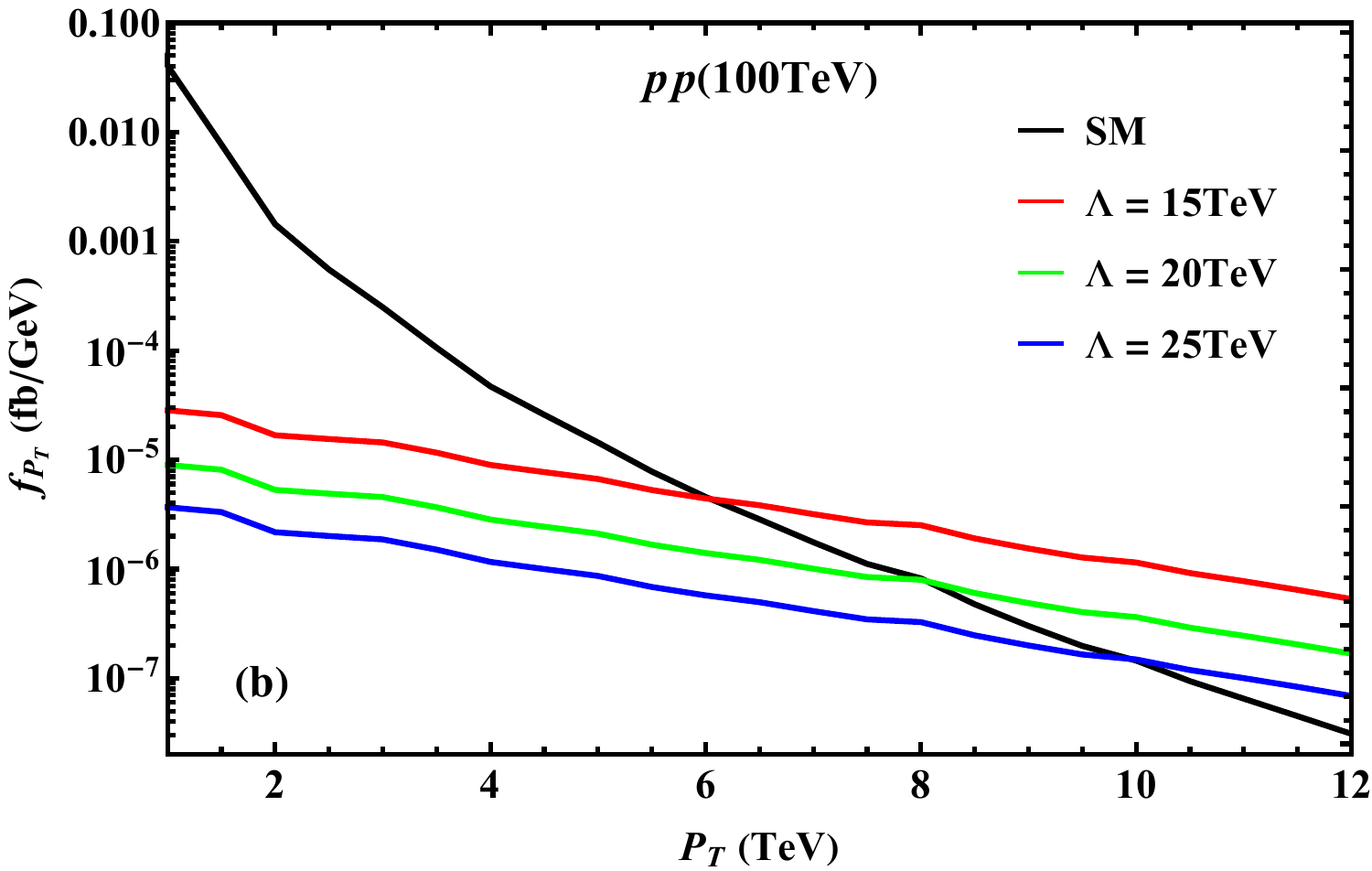}
\vspace*{-2mm}
\caption{\small{\it\hspace*{-2mm} 
		Photon transverse momentum 
		$P_T$ distributions at the azimuthal angle
		$\,\phi_*^{}\hsm\!=\hsmx 0\,$ for the reaction 
		$\hs p{\hs}p{\hs}(q{\hs}\bar{q})\ito Z\gamma\hs$
		followed by $\hs Z\!\to \ell\hs \bar\ell\hs$ decays, 
		as contributed by the SM (black curve) and
		by the nTGC operator $\OGP$ at $O(\cut^{-4})$ 
		(colored curves for the indicated values of $\Lambda\hs$) at the LHC}\,(13\,TeV) {\it in the upper panel 
		and at the} 100\,TeV $pp$ {\it collider in the lower panel.}
}
\label{figpt}
\label{fig:1}
\vspace*{-4mm}
\end{center}
\end{figure}

Inspecting the azimuthal angular distributions in 
Eq.\eqref{eq:f-phi*-OGP}, we note that the 
SM $\phi_*^{}$ distribution $\hs f_{\phi_*^{}}^{0}$
is nearly flat,
whereas the maximum of the nTGC contribution 
$\hs f_{\phi_*^{}}^{1}$ is at $\hs\phi_*\hsm =0\hsx$.\   
We consider the double differential cross section 
with respect to the photon transverse momentum $P_T^{}$ and $\phi_*^{}$  at  
$\hs\phi_*^{}\!=0\hs$,\footnote{%
In our study we define
the angles $\hs\theta\hs$ and $\hs\phi_*^{}$ 
and the momenta in the center-of-mass frame of the
$\hs\ell\bar\ell\ga\hs$ system, rather than in the laboratory frame.}
\\[-8mm]
\beqa
\label{eq:PT-photon}
f_{{P_T^{}}}^j\,=\,
\left. \frac{2\pi\hs \d^2\sigma_j}{\,\d{P_T^{}}\,\d\phi_*\,}
\right|_{\phi_*^{}=\hs 0}.
\eeqa
Eq.\eqref{eq:f0-phi*} gives
$\,\d\sigma_0^{}/\d\phi_*^{}\hsm\simeq\hsm \sigma_0^{}/(2\pi)\hs$ for the SM contribution, so we can deduce:
\beqa
f_{P_T}^0\,\simeq \frac{\,\d\sigma_0^{}\,}{\,\d P_T^{}\,} \,.
\eeqa
We present in Fig.\,\ref{figpt} the photon $P_T^{}$ distribution
\eqref{eq:PT-photon} at the LHC (upper panel) and a
100~TeV $pp$ collider (lower panel), 
where in each plot the SM contribution is shown as a
black curve and the $\OGP$ new physics contributions
for different values of $\Lambda$
are shown as the colored curves.
We find that the SM contribution to the photon $P_T^{}$ distribution $f_{{P_T^{}}}^0$ decreases more rapidly with the increase of
$P_T^{}$, whereas the nTGC contribution to $f_{{P_T^{}}}^1$
reduces much more slowly with $P_T^{}\hs$.\

\begin{figure}[t]
\centering
\includegraphics[width=10cm]{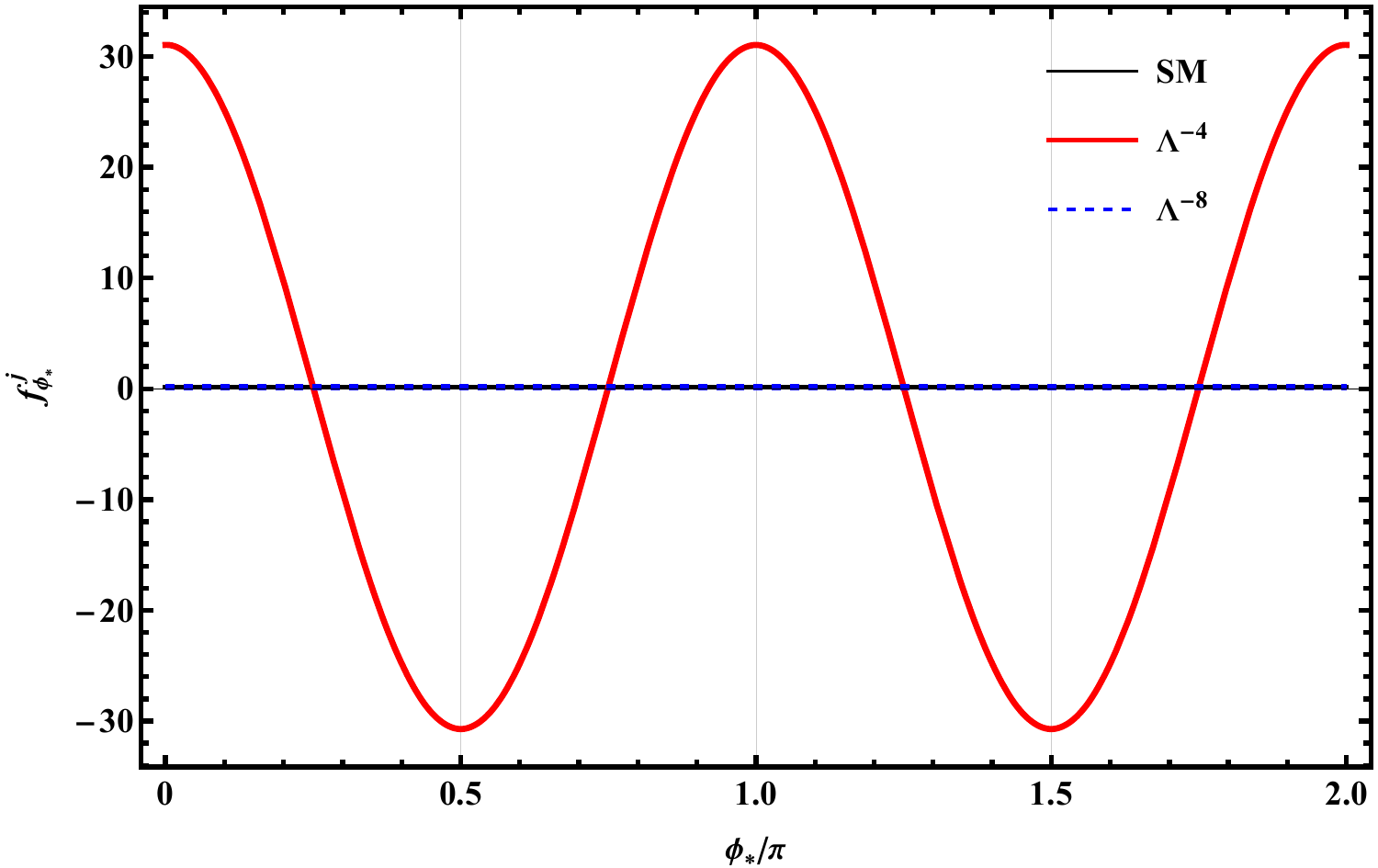}
\vspace*{-2mm}
\caption{\small{%
	\it Normalized distributions in the azimuthal angle
	$\phi_*^{}$ for the reaction 
	$\hs p{\hs}p{\hs}(q{\hs}\bar{q})\ito Z\gamma\hs$ 
	followed by $Z\!\to \ell\hs\bar{\ell}\hs$ decays,
	as generated by the dimension-8 nTGC operator $\,\mO_{G+}^{}$\! 
	at the LHC}\,(13\,TeV). {\it The angular distribution
	$\,f_{\phi_*^{}}^{1}$ of the interference contribution
	of $\hs O(\cut^{-4})\hs$ is shown as a red curve; 
	the angular distribution $\hs f_{\phi_*^{}}^{2}\hs$ of 
	the squared contribution of $\hs{\cal O}(\cut^{-8})$ is shown
	as the blue curve that is flat like the SM distribution
	$\hs f_{\phi_*^{}}^0\hs$ (black curve).}
}
\label{fig:2}
\end{figure}

According to our definition of the
azimuthal angle $\phi_*^{}$ in Section~\ref{sec:2},
we have
\beqa
\label{eq:phi*}
\cos\phi_*^{} \,=\,
\frac{~(\mathbf{p}_q^{}\!\times\!\mathbf{p}_Z^{})\cdot
(\mathbf{p}_f^{}\!\times\!\mathbf{p}_{\bar f}^{})~}
{|\mathbf{p}_q^{}\!\times\! \mathbf{p}_Z^{}|\hs 
|\mathbf{p}_f^{}\!\times\! \mathbf{p}_{\bar f}^{}|} \, .
\eeqa
We note that the quark $q$ can be emitted 
from either proton beam, so the direction of 
$\,\mathbf{p}_q^{}\,$ is subject to a $180^\deg$ ambiguity.\ 
This means that the normal direction of the scattering plane of 
$\hs q\hs\bar{q}\ito Z\ga\hs$ is also subject to a $180^\deg$ ambiguity,
so that $\cos\phi_*^{}$ can take either sign in each event 
and the $\cos\phi_*$ terms in $f_{\phi_*}^j$ 
cancel out when the statistical average is taken. 
However, the angular terms $\hs\propto\hsmx \cos(2\hs\phi_*^{})\!=\!2\cos^2\hsm\phi_*^{}\!-\!1\hs$ 
are not affected by this ambiguity and 
survive statistical average.\ 
Thus, for the nTGC operator $\OGP$ 
and also the related contact operator $\OCM$,
we derive the following {effective distributions}
of $\hs\phi_*^{}$ after averaging:
\beqs
\label{eq:f-phi*OG+eff}
\begin{eqnarray}
\label{eq:f0-phi*OG+eff}
\hspace*{-12mm}
\bar{f}_{\phi_*^{}}^{\hs 0} \!\!&=&\!\!
\frac{1}{\,2\pi\,} 
-\frac{\hat{s}\hs M_Z^2 \cos\hsm 2\phi_*^{}\,}
{\,2\pi\!\left[(\hat s\!-\!M_Z^2)^2\!+2(\hat s^2\!+\!M_Z^4)
\ln\sin\!\frac{\delta}{2}
\right]\,}+O(\delta) \hs ,
\hspace*{4mm}
\\[1mm]
\label{eq:f1-phi*OG+eff}
\hspace*{-8mm}
\bar{f}_{\phi_*^{}}^{\hs 1} \!\!&=&\!\!
\frac{1}{\,2\pi\,}+
\frac{~\hat s\cos2\phi_*^{}~}{8\pi M_Z^2} \hs ,
\label{f1}
\\[1mm]
\label{eq:f2-phi*OG+eff}
\hspace*{-8mm}
\bar{f}_{\phi_*^{}}^{\hs 2} \!\!&=&\!\!
\frac{1}{\,2\pi\,}  \hs .
\label{eq:phi-n}
\end{eqnarray}
\eeqs
We see that the interference term 
$\hs \bar{f}_{\phi_*^{}}^{\hs 1}\hs$ 
has a nontrivial angular dependence
$\,\propto \hsm\cos(2\phi_*^{})\hs$ 
that is enhanced by the energy factor
$\hs s/M_Z^2\hs$ relative to the nearly flat SM distribution
$\hs \bar{f}_{\phi_*^{}}^{\hs 0}\!\!\simeq\! 1/2\pi\hs$.
We present the angular distributions of $\hs\phi_*^{}\hs$ in Fig.\,\ref{fig:2}, where 
the angular distribution
$\hs \bar{f}_{\phi_*^{}}^{\hs 1}\hs$ (red curve)
from the interference contribution of $O(\cut^{-4})$
dominates over the nearly flat SM distribution
$\hs \bar{f}_{\phi_*^{}}^{\hs 0}\hs$ (black curve)
and the distribution 
$\hs \bar{f}_{\phi_*^{}}^{\hs 2}\hs$ (blue curve) 
of the squared contribution of $O(\cut^{-8})$,
which is flat and behaves like the SM distribution.\ 
In this figure, for illustration we have imposed a selection cut
on the parton-parton collision energy, $\sqrt{\hat s\,}\!>\!2$\,TeV.

\vspace*{1mm}

For the other operators $\hs(\OGM,\hs\OBW,\hs\OCP)\hs$,
inspecting their angular distributions in Eq.\eqref{eq:f-phi*-Ojx}
we find that $(f_{\phi_*^{}}^0,\,f_{\phi_*^{}}^1)$
have the leading energy contributions given by the
$\cos\phi_*^{}$ terms and the $\hs\cos(2\phi_*^{})\hs$ terms only have
subleading energy-dependence. In addition, their contributions
to $f_{\phi_*^{}}^2$ contain no $\hs\cos(2\phi_*^{})\hs$ term.
After statistically averaging over the two possible directions 
of the scattering plane at $pp$ colliders, we derive the following
effective distributions:
\beqs
\label{eq:f-phi*-other-AV}
\begin{eqnarray}
\hspace*{-8mm}
\bar{f}_{\phi_*}^{\hs 0} \!\!\!&=&\!\!
\frac{1}{\,2\pi\,} -
\frac{\,\hat{s}\hs M_Z^2 \cos\hsm 2\phi_*^{}\,}
{~2\pi \!\left[(\hat{s}\!-\!M_Z^2)^2\!+2(\hat{s}^2\!+\!M_Z^4)
\ln\sin\frac{\delta}{2}
\right]\,}+O(\delta),
\hspace*{4mm}
\\[1mm]
\hspace*{-8mm}
\bar{f}_{\phi_*}^{\hs 1} \!\!\!&=&\!\! \frac{1}{\,2\pi\,} +
\frac{\hat{s}\cos\hsm 2\phi_*^{}\,}{~4\pi(\hat{s}\!+\!M_Z^2)~} ,
\label{eq:f1q-AV}
\\[1mm]
\hspace*{-8mm}
\bar{f}_{\phi_*}^{\hs 2} \!\!\!&=&\!\! \frac{1}{\,2\pi\,} \,,
\label{eq:f2q-AV}
\end{eqnarray}
\eeqs
where the SM contribution $\hs \bar{f}_{\phi_*^{}}^{\hs 0}$ 
is the same as that of Eq.\eqref{eq:f0-phi*OG+eff}. 
For operators  $\hs(\OGM,\hs\OBW,\hs\OCP)\hs$,
under the statistical average, their angular distribution
$f_{\phi_*^{}}^1$ has a high-energy dependence 
of $\hs\hat{s}^0\hs$, while
the angular distribution $f_{\phi_*^{}}^2$
becomes a constant and is independent of both 
the energy and $\phi_*^{}$.\
These should be compared to the statistically averaged angular
distributions \eqref{eq:f1-phi*OG+eff}-\eqref{eq:f2-phi*OG+eff} 
for the nTGC operator $\OGP$, where its angular distribution
$\hs f_{\phi_*^{}}^1\hs$ has higher-energy dependence 
of $\hs\hat{s}^1\hs$ for the $\hs\cos(2\phi_*^{})\hs$ term, 
while the angular distribution $f_{\phi_*^{}}^2$
also becomes constant.

\vspace*{1mm}

Based on the effective angular distributions \eqref{eq:f-phi*OG+eff}
and Fig.\,\ref{fig:2}, we construct the following observable $\mathbb{O}_1^{}$\,:
\\[-9mm]
\begin{eqnarray}
\mathbb{O}_1^{} \,=\, 
\left|\hs\sigma_1^{}\!\int\!\!\di\phi_*^{}\,
f_{\phi_*^{}}^1\!\hsm\times\hsm\text{sign}(\cos\hsm 2\phi_*^{})
\right|,
\label{eq:O1-G+}
\end{eqnarray}
where $\hs\sigma_1^{}$ is the total cross section from
the interference contribution of ${\cal O}(\cut^{-4})\hs$.\ 
Then, we use the formula \eqref{eq:SS-Z} to derive the
significance:
\beqa
\label{eq:Z>2}
\mathcal{Z} \,=\, \sqrt{2\!\(\!B\ln\!
\frac{\,B\,}{\,B\hsm +\hsm S\,} + S\)}
\,=\,\sqrt{2\!\(\!\sigma_0^{}\ln\!
\frac{\,\sigma_0^{}\,}{\,\sigma_0^{}\hsm +\hsm\mathbb{O}_1^{}} + \mathbb{O}_1^{}\!\)}\times\sqrt{\mL\times\epsilon~},
\eeqa
where $\hs\mL\hs$ is the integrated luminosity and 
$\hs\epsilon\hs$ denotes the detection efficiency.\

\begin{table}[t]
\begin{center}
\begin{tabular}{c||ccc|ccc}
	\hline\hline
	& & & & &
	\\[-4mm]
	$\sqrt{s\,}$ & &
	\hspace*{-5mm}LHC{\,}(13{\,}TeV)\hspace*{-5mm}
	& & & \hspace*{-5mm}$pp${\,}(100{\,}TeV)\hspace*{-5mm} \\
	\hline
	& & & & &
	\\[-4mm]
	$\mL$~(ab$^{-1}$) & ~0.14 &    0.3  &  3~ & ~3 & 10 & 30~ \\
	\hline
	& & & & &
	\\[-4mm]
	~$\cut_{G+}^{2\sigma}$\,(TeV)~ 
	& ~2.1 & 2.4 & 3.3~ & ~14 & 17 & 19~\\
	\hline
	& & & & &
	\\[-4mm]
	~$\cut_{G+}^{5\sigma}$\,(TeV)~ 
	& ~1.6 &1.8 & 2.6~ & ~10 & 12 & 15~\\
	\hline\hline
\end{tabular}
\end{center}
\vspace*{-3mm}
\caption{\small{%
	\it Sensitivities to the new physics scale $\hs\cut\hs$
	at $O(\cut^{-4})$
	of the nTGC operator $\,\OGP$
	at the $2\hs\sigma$ and $5\hs\sigma$ levels,
	as obtained by analyzing the reaction
	$\,p{\hs}p{\hs}(q{\hs}\bar{q})\ito Z\ga\ito\ell\hs\bar{\ell}\ga$\,
	at the LHC}\,(13\,TeV) {\it and the} $pp$\,(100\,TeV) 
{\it collider respectively, with the indicated integrated luminosities.}
}
\vspace*{1mm}
\label{tab:1}
\end{table}

\vspace*{1mm}

To achieve the optimal sensitivity, we apply the formula
\eqref{eq:Zbin} to compute the total significance
$\mathcal{Z}_{\rm{total}}^{}$ from the 
contributions of the significances 
$\{\mathcal{Z}_{\rm{bin}}^{}\}$ of 
all the individual bins.\ 
In our analysis, we choose the bin size to be
$\hs\Delta P_T^{}\hsm\! =\hsm\! 100$\,GeV for the LHC\,(13{\hs}TeV) 
and $\Delta P_T^{}\!=\!500$\,GeV for 
the $pp{\hs}$(100{\hs}TeV) collider. But we find that 
$\hs\mathcal{Z}_{\rm{total}}^{}$ is not very sensitive 
to such choice.\ 
For instance, if we choose $\Delta P_T^{}\!=\!50$\,GeV or 
$\Delta P_T^{}\!=\!200$\,GeV at the LHC, 
we find that the significance
$\hs\mathcal{Z}_{\rm{total}}^{}$ 
only varies by about 1\%\hs.


We present prospective sensitivity reaches 
for probing the new physics 
scale $\cut$ of the nTGC operator $\OGP$ in Table\,\ref{tab:1}.\ 
For instance, given an integrated luminosity 
$\,\mathcal{L}\!=\!300\,$fb$^{-1}$\,($3\,$ab$^{-1}$) 
at the LHC and choosing the ideal detection efficiency 
$\epsilon\!=\!1\hs$, 
we find the $2{\hs}\sigma$ sensitivity reach 
$\,\cut_{G+}^{2\sigma}\!\simeq\! 2.6$\,TeV 
($\,\cut_{G+}^{2\sigma}\!\simeq\! 3.6$\,TeV).
At the 100{\hs}TeV $pp$ collider with $\mL\!=\!3$\,ab$^{-1}$\,(30\,ab$^{-1}$), we
derive the $2{\hs}\sigma$ sensitivity reach 
$\,\cut_{G+}^{2\sigma}\!\simeq\! 15$\,TeV
($\hs\cut_{G+}^{2\sigma}\!\simeq\hsm 21$\,TeV\hs).

\vspace*{1mm}
\subsection{\hspace*{-2.5mm}nTGC Analysis Including  \boldmath{${\cal O}(\Lambda^{\!-8})$} Contributions}
\label{sec:4.2}
\vspace*{1mm}

In this subsection, we further analyze the squared contributions of 
$\mO(\cut^{\!-8})$ and study their impact on the sensitivity reaches
at the LHC and the $pp\hs$(100{\hs}TeV) collider.\ 
Inspecting the effective angular distributions \eqref{eq:f-phi*OG+eff},
we find that requiring the differential cross section of 
the interference contribution of $\mO(\cut^{\!-4})$ to be larger than 
that of the squared contribution of $\mO(\cut^{\!-8})$ 
would impose the following condition:
\beqa
\left|\sigma_1^{}f^1_{\phi_*^{}}\right|\,>\,
\sigma_2^{}f^2_{\phi_*^{}}\!=\!\frac{\sigma_2}{~2\pi~}\, ,
\eeqa
which gives a lower bound of 
$\,\cut\hsm\!>\hsm\!1.3\sqrt{\hat s\,}\,$ for 
the reaction channel $\,u\hs\bar u\ito Z\ga\,$   
and $\,\cut\hsm\!>\hsm\!1.5\sqrt{\hat s\,}\,$ for
the $d\bar d\ito Z\ga\,$ channel.\ 
These bounds are comparable or somewhat stronger 
than the LHC sensitivity limits 
of the new physics scale $\hs\cut\hs$ given in 
Table\,\ref{tab:1}, whereas they are satisfied by the
sensitivity limits of the 100{\hs}TeV $pp$ collider.\
Thus, to improve the sensitivities for the LHC probe of the nTGCs,
we consider the full contributions of the nTGC operators 
including their squared terms of ${\cal O}(\Lambda^{\!-8})$.\
We note that including the full contributions of the nTGC operators  
also allows a consistent mapping of the current analysis to 
the form factor approach given in the following 
Section\,\ref{sec:4.4} which always includes the full contributions
of the form factors to the cross sections.\

\vspace*{1mm}

\begin{figure}[t]
\centering
\includegraphics[width=10.5cm]{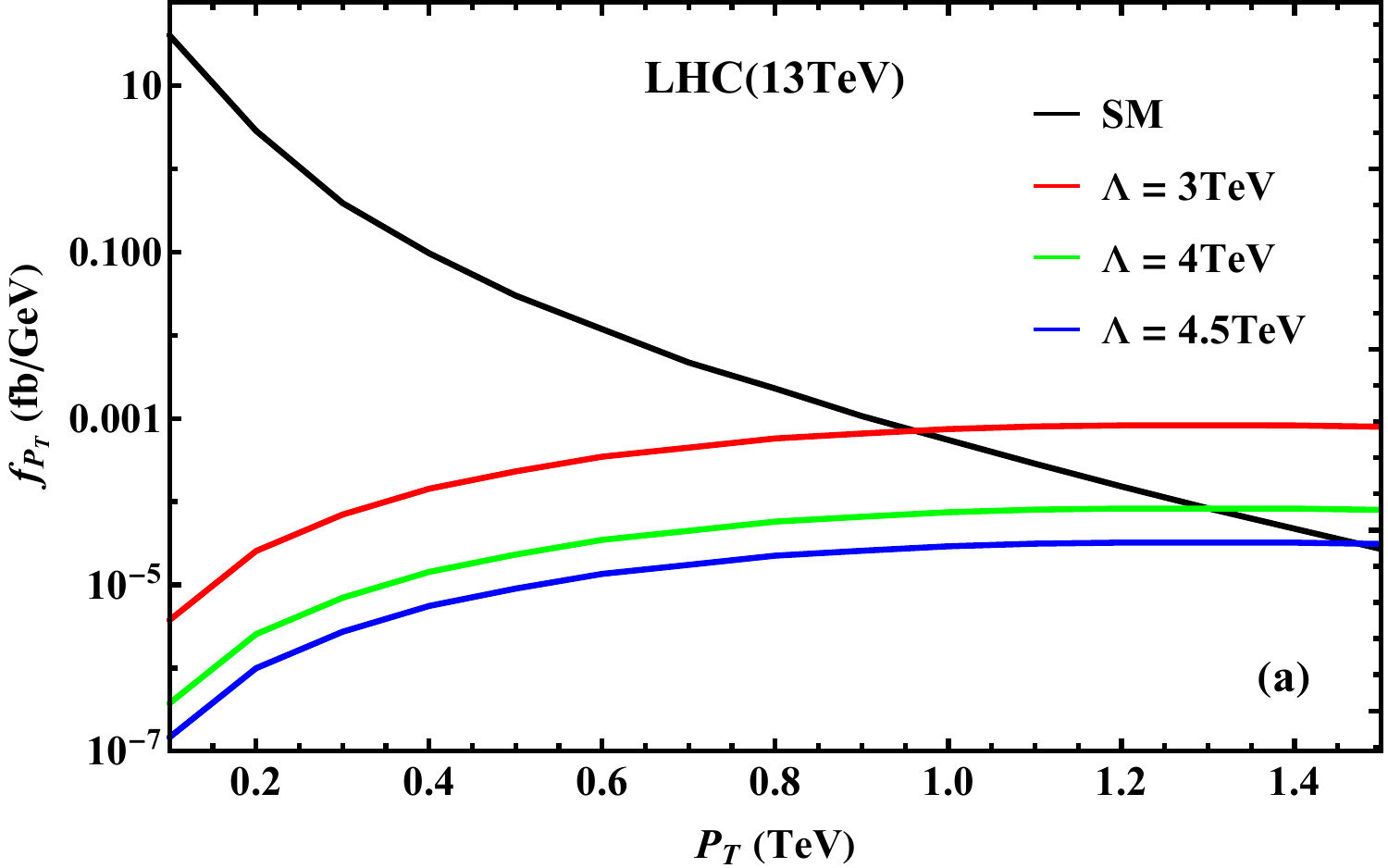}
\\[3mm]
\hspace*{3mm}
\includegraphics[width=10.9cm]{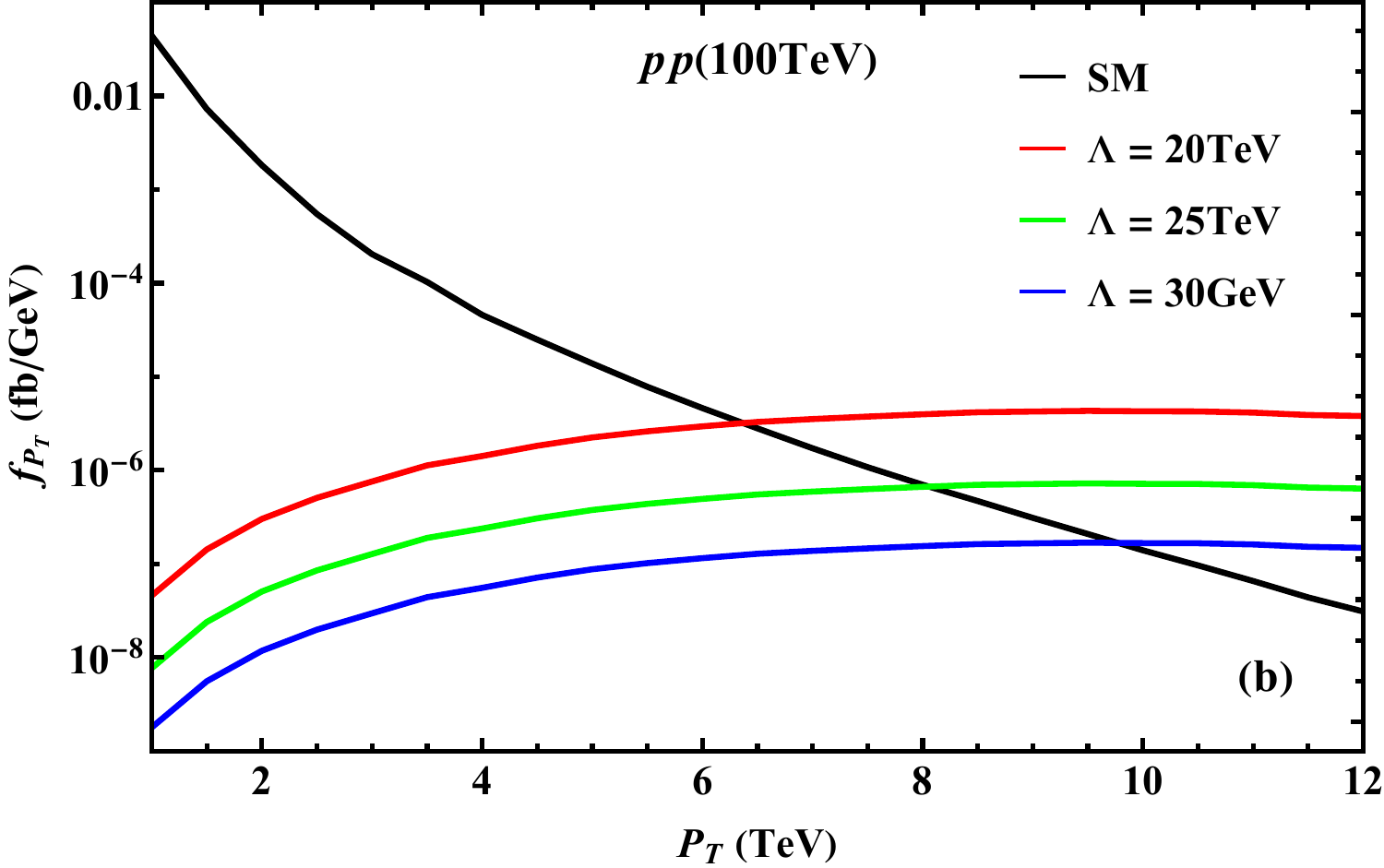}
\vspace*{-2mm}
\caption{\small{%
	\it Photon transverse momentum 
	$P_T^{}$ distributions at the azimuthal angle
	$\phi_*^{}\!=\!0$ for the reaction 
	$\hs p{\hs}p{\hs}(q{\hs}\bar{q})\ito Z\gamma\hs$
	followed by $\hs Z\!\to \ell\bar\ell\hs$ decays, 
	as contributed by the SM (black curve) and
	by the nTGC operator $\OGP$ up to $\hs{O}(\cut^{-4})$ 
	and $\hs{O}(\cut^{-8})$
	(colored curves) at the LHC}\,(13\,TeV) {\it and the} 
$pp$\,(100\,TeV) {\it collider in the lower panel.}
}
\label{figpt8}
\label{fig:3}
\vspace*{3mm}
\end{figure}

We present in Fig.\,\ref{fig:3} the photon $P_T^{}$ distribution
including the contribution of 
${\cal O}(\Lambda^{\!-8})$.\  
Since the ${\cal O}(\Lambda^{\!-8})$ contribution 
can be larger than $ {\cal O}(\Lambda^{\!-4})$ 
for large $\hat{s\,}$,  we choose here a set of larger values 
$\,\Lambda\hsm\!=\hsm\!(2,4,6)$\,TeV for the LHC distributions
and $\,\Lambda\hsm\!=\hsm\!(20,25,30)$\,TeV for the distributions
at the $pp\hs$(100{\hs}TeV) collider,
instead of the previous values of 
$\,\Lambda \!=\!(1 , 2, 3)$\,TeV for the LHC and 
$\,\Lambda\hsm\!=\hsm\!(15,20,25)$\,TeV 
for the $pp\hs$(100{\hs}TeV) collider
chosen for 
Fig.\ref{fig:1}. Also, Fig.\,\ref{fig:3} extends
to a larger range of the photon $P_T^{}$\hs.

\vspace*{1mm} 

For the high-energy hadron colliders such as the LHC
and $pp\hs$(100{\hs}TeV), we have 
$\hs |\sigma_1^{}|\hsm\ll\hsm 2\pi 
|\sigma_1^{} f^1_{\phi_*}|\!<\!\sigma_2^{}\hs$, and thus 
$\sigma_1^{}$ may be neglected. 
Following the procedure in Section\,\ref{sec:4.1}, 
we use the same method and cuts on $P_T^{}$ 
to divide events into a set of bins. 
Because the $\phi_*^{}$ distribution is rather flat for both the 
SM and $O(\cut^{-8})$ contributions, 
we do not need to impose an angular cut on $\phi_*^{}\hs$. 
We analyze the sensitivity reaches of $\hs\cut\hs$
by using Eq.(\ref{eq:Zbin}), and present the results for probing
the nTGC operator $\hs\OGP\hs$ 
up to ${\cal O}(\Lambda^{\!-8})$ in Table\,\ref{tab:2}.\
The sensitivity reaches at ${\cal O}(\Lambda^{\!-8})$
appear significantly better than those at 
${\cal O}(\Lambda^{\!-4})$ shown in Table\,\ref{tab:1}. 

\vspace*{1mm} 

For instance, given an integrated luminosity 
$\,\mathcal{L}\!=\!300\,$fb$^{-1}$\,($3\,$ab$^{-1}$) 
at the LHC and choosing the ideal detection efficiency 
$\epsilon\!=\!1\hs$, 
we find from Table\,\ref{tab:2} that 
the $2{\hs}\sigma$ sensitivity reach is given by 
$\,\cut_{G+}^{2\sigma}\!\!\simeq\! 3.4$\,TeV 
($\,\cut_{G+}^{2\sigma}\!\!\simeq\! 4.1$\,TeV).
At the 100{\hs}TeV $pp$ collider with $\mL\!\!=\hsm\!3$\,ab$^{-1}$\,(30\,ab$^{-1}$), we
obtain the $2{\hs}\sigma$ sensitivity reach 
$\,\cut_{G+}^{2\sigma}\!\simeq\! 22$\,TeV
($\hs\cut_{G+}^{2\sigma}\!\simeq\hsm 27$\,TeV\hs).

\begin{table}[t]
\begin{center}
\begin{tabular}{c||ccc|ccc}
	\hline\hline
	& & & & &
	\\[-4mm]
	$\sqrt{s\,}$ && \hspace*{-5mm}LHC{\,}(13{\hs}TeV)\hspace*{-5mm}
	& & & \hspace*{-5mm}$pp${\,}(100{\,}TeV)\hspace*{-5mm} \\
	\hline
	& & & & &
	\\[-4mm]
	$\mL$~(ab$^{-1}$) & 0.14 &  0.3  & 3 & ~3 & 10 & 30~ \\
	\hline
	& & & & &
	\\[-4mm]
	~$\cut_{G+}^{2\sigma}$\,(TeV)~ & 3.0 & 3.2 & 3.9 & ~21 & 24 & 26~\\
	\hline
	& & & & &
	\\[-4mm]
	$\cut_{G+}^{5\sigma}$\,(TeV) & 2.6 & 2.8 & 3.4 & ~17 & 20 & 22~\\
	\hline\hline
\end{tabular}
\end{center}
\vspace*{-3mm}
\caption{\small{%
	\it Sensitivities to the new physics scale $\cut$ at $O(\cut^{-8})$
	of the nTGC operator $\,\mO_{G+}^{}$
	at the $2\hs\sigma$ and $5\hs\sigma$ levels,
	as obtained by analyzing the reaction
	$\,p{\hs}p{\hs}(q{\hs}\bar{q})\ito Z\ga\!\to\! \ell\bar{\ell}\ga$\,
	at the LHC}\,(13\,TeV) {\it and the} $pp\hs$(100\,TeV) 
{\it collider respectively, with the indicated integrated luminosities.}
}
\label{tab:2}
\end{table}
\begin{table}[h]
\begin{center}
\begin{tabular}{c||ccc|ccc}
	\hline\hline
	& & & & &
	\\[-4mm]
	$\sqrt{s\,}$ && \hspace*{-5mm}LHC{\,}(13 {\,}TeV)\hspace*{-5mm}
	& & & \hspace*{-5mm}$pp${\,}(100{\,}TeV)\hspace*{-5mm} \\
	\hline
	& & & & &
	\\[-4mm]
	$\mL$~(ab$^{-1}$) & ~0.14 &    0.3   &  3~ & ~3 & 10 & 30~ \\
	\hline
	& & & & &
	\\[-4mm]
	~$\cut_{G+}^{2\sigma}$\,(TeV)~ 
	& ~3.3 & 3.6 & 4.2~ & ~23 & 26 & 28~ \\
	\hline
	& & & & &
	\\[-4mm]
	~$\cut_{G+}^{5\sigma}$\,(TeV) 
	& ~2.9 & 3.1 & 3.7~ & ~20 & 22 & 24~ \\
	\hline\hline
\end{tabular}
\end{center}
\vspace*{-4mm}
\caption{\small{%
	\it Sensitivity reaches on the new physics scale $\cut$ at $O(\cut^{-8})$
	of the nTGC operator $\,\OGP$
	at the $2{\hs}\sigma$ and $5{\hs}\sigma$ levels,
	as obtained from the reactions
	$\,p{\hs}p{\hs}(q{\hs}\bar{q})\ito Z\ga\!\to\! \ell\bar{\ell}\ga$\,
	and 
	$\,p{\hs}p{\hs}(q{\hs}\bar{q})\ito Z\ga\ito \nu\bar{\nu}\ga$
	at the LHC}\,(13\,TeV) {\it and the} $pp$\,(100\,TeV) 
{\it collider, with the indicated integrated luminosities.}
}
\vspace*{5mm}
\label{tab:3}
\end{table}
\begin{table}[b]
\vspace*{2mm}
\begin{center}
\begin{tabular}{c||ccc|ccc||ccc|ccc}
	\hline\hline	
	& & & & & & & & & & &  
	\\[-4mm]  
	$\sqrt{s\,}$ & & \hspace*{-12mm}13\,TeV\,($\ell\hs\bar{\ell}$)
	\hspace*{-12mm}
	& & &
	\hspace*{-12mm}13\,TeV\,($\ell\hs\bar\ell,\nu\bar{\nu}$)\hspace*{-10mm}
	& & & \hspace*{-12mm}100\,TeV\,($\ell\hs\bar{\ell}$)\hspace*{-12mm}
	& & &
	\hspace*{-10mm}100\,TeV\,($\ell\hs\bar\ell,\nu\bar{\nu}$)
	\hspace*{-10mm}
	\\
	\hline
	& & & & & & & & & & &  
	\\[-4.3mm]
	$\mL$(ab$^{-1}$)
	& 0.14 & 0.3 & 3 & 0.14 & 0.3 & 3 & 3 & 10 & 30 & ~3 & 10 & 30~
	\\
	\hline
	& & & & & & & & & & &  
	\\[-4.3mm] 
	$\Lambda_{\hsm\widetilde{B}W}$\,(TeV) & 1.2 & 1.3 & 1.5 & 1.3 & 1.4 & 1.7 & 5.1 & 5.6 & 6.1 & ~5.6 & 6.1 & 6.7~~\\
	\hline
	& & & & & & & & & & &  
	\\[-4.3mm]
	$\Lambda_{G-}^{}$\,(TeV) & 1.0 & 1.1 &  1.3 & 1.1 & 1.2 & 1.4 & 4.2 & 4.7 & 5.1 & ~4.6 & 5.1 & 5.5~ \\
	\hline
	& & & & & & & & & & &  
	\\[-4.3mm]
	$\Lambda_{C+}^{}$\,(TeV) & 1.3 & 1.4 & 1.6 & 1.4 & 1.5 & 1.7 & 5.4 & 5.9 & 6.5 & ~5.9 & 6.5 & 7.1~ \\
	\hline\hline
\end{tabular}
\end{center}
\vspace*{-3mm}
\caption{\small%
{\it Sensitivity reaches on the new physics scales of the nTGC
	operators $\hs(\OBW,\hs\OGM,\hs\OCP)\hs$
	at the $2\hs\sigma$ level, as obtained from analyzing the reactions
	$\,p{\hs}p{\hs}(q{\hs}\bar{q})\ito Z\ga\!\to\! \ell\bar{\ell}\ga$
	and 
	$\,p{\hs}p{\hs}(q{\hs}\bar{q})\ito Z\ga\ito \nu\bar{\nu}\ga$		
	at the LHC}\,(13\,TeV) {\it and the} $pp$\,(100\,TeV) 
{\it collider, with the indicated integrated luminosities.\ 
	The third and fifth columns indicated by ($\hs\ell\hs\bar{\ell},\hs\nu\hs\bar{\nu}$)
	present the combined limits
	including both the charged-lepton and neutrino final states. 
}
}
\vspace*{2mm}
\label{tab:BW}
\label{tab:4}
\end{table}
%

\subsection{\hspace*{-2.5mm}nTGC Analysis Including the Invisible Decays \boldmath{$Z\ito\nu\bar\nu$}}
\label{sec:4.3}
\vspace*{1mm}

In this subsection, we study the probe of nTGCs 
via the $Z\ga$ production with invisible decays 
$\hs Z\!\ito\!\nu\hs\bar\nu\hs$.\
In this case, the final-state photon is 
the only signature of $Z\ga$ production 
that can be detected, and we will use the jet-vetoing
to effectively remove all the reducible SM backgrounds having 
the final state {\hs}jet$+\ga\hs$.\ 
Then, we can use the same strategy as that for probing the  
${\cal O}(\Lambda^{-8})$
contribution via the leptonic $Z$-decay channels, 
where the kinetic cut on the photon $P_T^{}$ distribution 
will play the major role to enhance
the sensitivity to the nTGC contributions.

\vspace*{1mm}

Following this strategy, 
we perform combined analyses 
for both the $\hs Z\!\ito \ell\hs\bar{\ell}\,$ final state and 
the $Z\!\ito\nu\hs\bar\nu\,$ final state.\ 
We present in Table\,\ref{tab:3} 
a summary of the prospective sensitivity reaches on
the new physics scale $\hs\Lambda\hs$ of the nTGC operator
$\hs\OGP\hs$,
where we have combined the limits from both the charged-lepton
final state and the neutrino final state.\  
We find that the combination of both leptonic and invisible $Z$-decay 
channels can enhance the sensitivity to the new physics scale 
$\hs\Lambda\hs$ by about 10\% over that using 
the leptonic channels alone.

\vspace*{1mm}

Using the sensitivity bounds of Table\,\ref{tab:3} and 
comparing them with our study for $e^+e^-$ colliders\,\cite{Ellis:2020ljj}
(which will be summarized later in Table\,\ref{tab:7} 
of Section\,\ref{sec:5}),
we find that for probing the nTGC operator $\OGP$
the sensitivity reaches with the current LHC luminosity
($\hs\mL\hsm =\hsm 140\,$fb$^{-1}$)
are already better than those at future 250\,GeV and 500\,GeV
$e^+e^-$ colliders\,\cite{Ellis:2020ljj},
and that the HL-LHC (with $\hs\mL\! =\! 3\hs$ab$^{-1}$) 
should have comparable sensitities to a 1\,TeV $e^+e^-$ 
collider\,\cite{Ellis:2020ljj}.\ 
The future $pp$\,(100{\hs}TeV) collider can have much stronger
sensitivities than an ($3\!-\!5$)\,TeV $e^+e^-$ collider.\ 
A systematic comparison with the high-energy $e^+e^-$ colliders
will be presented in the following Section\,\ref{sec:5}. 

\vspace*{1mm}

Next, we extend the above analysis to the three other nTGC operators
$\hs(\OGM,\hs\OBW,\hs\OCP)\hs$. We present the 2$\hs\sigma$ sensitivities
to their associated new physics scales in Table\,\ref{tab:4}.\ 
The third and fifth columns of this Table, marked with  ($\ell\hs\bar{\ell},\nu\hs\bar{\nu}$),
present the combined limits including both 
the charged-lepton and neutrino final states.\  
We see that these sensitivities are significantly weaker than
those of the operators $\hs\OGP\hs$ and $\hs\OCM\hs$.\ 
At the LHC, they are generally below 2\,TeV, but the proposed 
100\,TeV $pp$ collider could improve the sensitivities substantially,
reaching new physics scales  
$\hs\cut\hs$ over the $(5\hsm -\hsmx 7)$\,TeV range.\
Finally, we compare the collider sensitivity limits presented
in Tables\,\ref{tab:2}-\ref{tab:4} with the perturbative
unitarity limits given in Table\,\ref{tab:0} and Fig.\hs\ref{fig:0}.\ 
We find that our collider limits  
are much stronger than the unitarity limits 
of Table\,\ref{tab:0} and Fig.\hs\ref{fig:0}.\
Hence, our current collider analyses of probing the nTGCs via the
SMEFT formulation hold well the perturbation expansion.

\vspace*{1mm}

\begin{table}[b]
\begin{center}
\begin{tabular}{c||cccc}
\hline\hline
&&&& \\[-4mm]
Operators
 & $\mO^{(3)}_L(q_L^{})$ & $\mO^{}_L(q_L^{})$ & $\mO^{}_R(u_R^{})$ 
 & $\mO^{}_R(d_R^{})$
 \\[1mm]
 \hline\hline
&&&& \\[-4mm]
$\Lambda\hs$[PDG]\,(TeV)  & 4.7 & 4.7 & 2.9 & 2.4
 \\[0.3mm]
\hline
&&&& \\[-4mm]
$\Lambda\hs$[CEPC]\,(TeV)  & 9.1 & 9.1 & 5.5 & 5.1
\\[0.4mm]
\hline\hline
\end{tabular}
\end{center}
\vspace*{-4mm}
\caption{\small%
{\it Precision constraints at the $\hs 2\sigma$ level 
on the indicated dimension-6 operators
that contribute to the $q$-$\bar{q}$-$Z$ coupling.\ 
The bounds $\Lambda\hs$}[PDG] {\it are derived from the 
existing electroweak data}\,\cite{PDG2022}, {\it whereas the bounds
$\Lambda\hs$}[CEPC] {\it are the projected sensitivities of the
future $e^+e^-$ collider} 
CEPC$\hs$($250\hs$GeV)\,\cite{dim6HEFT-GHX}.}
\label{tab:6EW}
\end{table}

As a final remark, we emphasize that the reaction 
$\hs q\bar{q}\!\to\! Z\gamma\,$ is a unique process
for probing the nTGCs via $s$-channel 
at the LHC and future $pp$ colliders.\ 
We note, however, that certain dimension-6 operators can contribute
to the process $q\bar{q}\hsm\ito\hsm Z\gamma$  via $t$-channel diagrams by
modifying the $q$-$\bar{q}$-$Z$ vertex.\ Such contributions
are constrained separately by existing electroweak precision data
via other reactions, and
future $e^+e^-$ colliders will place more severe constraints
on the $q$-$\bar{q}$-$Z$ coupling via $Z$-pole measurements.\
These measurements are {\it independent of} the reaction 
$\,q\bar q\hsm\ito\hsm  Z\gamma\,$, 
and may be obtained from global fits to 
$(\alpha,\hs G_F^{},\hs M_Z^{},\hs M_W^{})$ and 
other $Z$-pole observables\,\cite{LEP1}\cite{dim6HEFT-GHX}\cite{PDG2022}.\  
We take values of these observables from 
the current electroweak precision data\,\cite{PDG2022} 
and from the projected CEPC sensitivities\,\cite{dim6HEFT-GHX}.\ 
For contributions to the $q$-$\bar{q}$-$Z$ coupling, 
we consider the following dimension-6 Higgs-related operators:
\begin{align}
\mO^{(3)}_L &=\, (\text{i} H^\dagger\sigma^a\!\! \stackrel \leftrightarrow D_\mu^{}\!\! H)
(\overline \Psi_L^{} \gamma^\mu \sigma^a \Psi_L^{})\hs ,
\nn\\
\label{eq:dim6-Zqq} 
\mO_L^{} &=\, 
(\text{i} H^\dagger\! \stackrel \leftrightarrow D_\mu^{}\! H)
(\overline \Psi_L^{} \gamma^\mu \Psi_L^{})\hs ,
\\
\mO_R^{} &=\, 
(\text{i} H^\dagger\!\! \stackrel \leftrightarrow D_\mu^{}\!\! H)
(\overline \psi_R^{} \gamma^\mu \psi_R^{})\hs .
\nn
\end{align}
Then, using the method of \cite{dim6HEFT-GHX}
we make a global fit and obtain
the electroweak precision constraints 
on the cutoff scale $\cut$ of these operators, which we summarize 
in Table\,\ref{tab:6EW}, assuming
for simplicity that the dimension-6 operators are universal 
for the three families of fermions.\ Table\,\ref{tab:6EW} shows that
the dimension-6 operators \eqref{eq:dim6-Zqq} can be constrained 
independently through different processes and observables.\  
The existing bounds $\Lambda\hs$[PDG] derived in Table\,\ref{tab:6EW} 
are already strong and the projected sensitivities 
on the cutoff scale $\Lambda\hs$[CEPC] 
for the future $e^+e^-$ collider CEPC$\hs$($250\hs$GeV)
are much stronger than the corresponding bounds on the
cutoff scale of the dimension-8 nTGC operators
at the same $e^+e^-$ collider (as we show below
in Table\,\ref{tab:7} of Section\,\ref{sec:5}).

\vspace*{1mm}
\subsection{\hspace*{-2.5mm}Probing the Form Factors of nTGVs}
\label{sec:4.4}
\vspace*{1mm}

In this Section we analyze the sensitivity reaches of 
the LHC and the $pp$\,(100\,TeV) collider for probing the nTGCs
by using the form factor formulation given in Section\,\ref{sec:3}.\
We will also clarify the {nontrivial difference} between
our consistent form factor formulation \eqref{eq:FF2-nTGC}
(based upon the fully gauge-invariant SMEFT approach) 
and the conventional form factor formulation \eqref{eq:FF0-nTGC}
[retaining only the residual gauge symmetry U(1)$_{\rm{em}}^{}\hs$],
where the latter leads to erroneously strong sensitivity limits.

\vspace*{1mm}

From Eqs.\eqref{eq:CS-qq-Zgamma}, 
\eqref{eq:CS-qq-Zgamma-Ojx}-\eqref{eq:xLR-OBWC+G-} and \eqref{eq:h-dim8}, we can further derive the partonic cross section 
of the reaction $\,q\hs\bar q \ito Z\gamma\hs$  
in terms of the form factors. As before, we decompose the
partonic cross section into the sum of three parts,
$\,\sigma(Z\ga )=\sigma_0^{}+\sigma_1^{}+\sigma_2^{}\,$,
where $(\sigma_0^{},\hs\sigma_1^{},\hs\sigma_2^{})$
correspond to the SM contribution, the interference contribution,
and the squared contribution, respectively. 
The cross section terms $(\sigma_1^{},\hs\sigma_2^{})$ are
contributed by the form factors and take the following expressions: 
\\[-8mm]
%
\begin{eqnarray}
\label{eq:CS1-qq-Zgamma-H34}
\sigma_{1}^{} 
\!\!&=&\!\!	
-\frac{\,e^2Qq_L^{}\hs{T_3^{}}\!
\left(\shat\!-\!M_Z^2\right)\,}{4\pi v^2\,\shat} h_4^{}
-\frac{\,e^2 Q(q_L^{}x_L^Z\!\!-\!q_R^{}x_R^Z)
(\shat^2\!\hsm-\!M_Z^4)\,}
{\,4\pi v^2\hs \shat^2\,}h_3^Z
\nn\\
&& + 
\frac{\,e^2c_W^{}Q(q_L^{}x_L^A\!\!-\!q_R^{}x_R^A)
(\shat^2\!\hsm-\!M_Z^4)\,}
{\,4\pi s_W^{}v^2\hs\shat^2\,}h_3^\gamma \,,
\hspace*{10mm}
\eeqa
\\[-8mm]
and
\\[-9mm]
\beqs 
\label{eq:CS2-qq-Zgamma-H34}
\beqa 
\label{eq:sigma2=sum}
\sigma_{2}^{}\!\!&=&\!\!
\sigma_{2}^{44}+\sigma_{2Z}^{33} + \sigma_{2A}^{33}
+\sigma_{2Z}^{43}+\sigma_{2A}^{43}+\sigma_{2Z\hsm A}^{33}\,,
\\[1.5mm]
\label{eq:sigma2-44}
\sigma_{2}^{44}
\!\!&=&\!\!
\frac{~e^4 T_3^2 (\shat\hsm +\!M_Z^2)(\shat\hsm -\!M_Z^2){}^3~}
{~768\hs\pi s_W^2c_W^2M_Z^8\,\shat~}(h_4^{}\hsm )^2\,,
\\[1mm]
\label{eq:sigma2Z-33}
\sigma_{2Z}^{33}
\!\!&=&\!\!
\frac{~e^4[Q^2 s_W^4\!+\!(T_3^{}\!-\hsm Q s_W^2\hsm )^{2}]
(\shat\!+\!M_Z^2)(\shat\!-\!M_Z^2)^3\,}
{~192\hs\pi\hs s_W^2c_W^2M_Z^6\, \shat^2~} (h_3^Z)^2 \,,
\\[1mm]
\label{eq:sigma2A-33}
\sigma_{2A}^{33}
\!\!&=&\!\!
\frac{~e^4Q^2(\shat\!+\!M_Z^2)(\shat\!-\!M_Z^2)^3\,}
{96\hs\pi M_Z^6\,\shat^2} (h_3^\gamma)^2 \,,
\\[1mm]
\label{eq:sigma2Z-43}
\sigma_{2Z}^{43}
\!\!&=&\!\!
\frac{~e^4 T_3^{}(T_3^{}\!-\!Qs_W^2)(\shat\!-\!M_Z^2)^3\,}
{96\hs\pi\hs s_W^2c_W^2 M_Z^6\,\shat} h_4^{}h_3^Z\,, 
\eeqa
\beqa 
\label{eq:sigma2A-43}
\sigma_{2A}^{43}
\!\!&=&\!\!
\frac{~e^4Q\hs T_3^{}\hs (\shat\!-\!M_Z^2)^3\,}
{~96\hs\pi\hs s_W^{}c_W^{}M_Z^6\,\shat~}
h_4^{}h_3^\gamma \,,
\\[1mm]
\label{eq:sigma2ZA-43}
\sigma_{2ZA}^{33}
\!\!&=&\!\!
\frac{~e^4 Q(T_3^{}\!-\!2Qs_W^2)
(\shat\!+\!M_Z^2)(\shat\!-\!M_Z^2)^3\,}
{~96\hs\pi\hs s_W^{}c_W^{}M_Z^6 \,\shat^2~}
h_3^Z h_3^\gamma \,,
\end{eqnarray}
\eeqs
where the coefficients
$(q_L^{},\,q_R^{})\!=\! (T_3^{}\!-\!Qs_W^2,\hs -Qs_W^2)$
denote the (left,\,right)-handed gauge couplings between 
the quarks and $Z$ boson.\ 
The form factor $\,h_3^Z$ is contributed
by the operator $\OBW$ as in Eq.\eqref{eq:h-OBW} 
and the coupling coefficients 
$(x_L^Z,\hs x_R^Z)\!=\!(T_3^{}\!-\! Qs_W^2,\hs -Qs_W^2)$
are given by Eq.\eqref{eq:xLR-OBW},
whereas the form factor $\,h_3^\gamma$ is contributed
by the operator $\OGM$ as in Eq.\eqref{eq:h-OGM}
and the coupling coefficients 
$(x_L^A,\hs x_R^A)\!=\!-Qs_W^2(1,\hs 1)$
are given by Eq.\eqref{eq:xLR-OG-}.\ 
Inspecting 
Eqs.\eqref{eq:CS1-qq-Zgamma-H34}-\eqref{eq:CS2-qq-Zgamma-H34},
we find that the cross section terms $(\sigma_1^{},\hs\sigma_2^{})$
have the following scaling behaviors in the high energy limit:
\\[-9mm]
\beqs 
\label{eq:CS12-h4h3-size}
\begin{align}
\label{eq:CS1-h4h3-size}
\sigma_1^{} &=\,
O(\shat^0)\hs h_4^{} + O(\shat^0)\hs h_3^Z + O(\shat^0)\hs h_3^\gamma\,,
\\[1mm]
\label{eq:CS2-h4h3-size}
\sigma_2^{} &=\, 
O(\shat^3)(h_4^{}\hsm )^2 +
O(\shat^2)(h_3^V\hsm )^2 + O(\shat^2)(h_4^{}h_3^V\hsm ) + 
O(\shat^2)(h_3^Zh_3^\gamma\hsm )\,,
\end{align}
\eeqs 
where we have used the notation 
$V\!=\!Z,\gamma\hsx$.

\vspace*{1mm}

If we consider instead the conventional parametrization
\eqref{eq:FF0-nTGC} 
with the nTGC form factors $(h_3^V,\hs h_4^V)$ only, 
we would obtain their contributions to the total cross section
$\,\widetilde{\sigma}(Z\ga )\!
=\sigma_0^{}+\widetilde{\sigma}_1^{}+\widetilde{\sigma}_2^{}\,$.\ 
The form factors $h_3^V$ are not subject to the constraints
\eqref{eq:h4h5-h4ZA} imposed by the dimension-8 nTGC operators 
of the SMEFT, so they contribute to 
$(\hs\widetilde{\sigma}_1^{},\hs\widetilde{\sigma}_2^{})$
in the same way as in our 
Eqs.\eqref{eq:CS1-qq-Zgamma-H34}-\eqref{eq:CS2-qq-Zgamma-H34}.\ 
However, the $h_4^V$ contributions to
the interference and squared cross sections 
$(\hs\widetilde{\sigma}_1^{},\hs\widetilde{\sigma}_2^{})$
have vital differences from 
Eqs.\eqref{eq:CS1-qq-Zgamma-H34}-\eqref{eq:CS12-h4h3-size}.\ 
For simplicity of illustration, we set $\hs h_3^V\!=\hsm 0\hs$ 
and express the $h_4^V$ contributions to 
$(\hs\widetilde{\sigma}_1^{},\hs\widetilde{\sigma}_2^{})$
as follows:
\beqs	
\label{eq:Xsigma12-dim8}
\begin{align}
\label{eq:xh4v}
\widetilde{\sigma}_1^{}(h_4^V) & =
\frac{\,e^4 Q (\shat\!-\!M_Z^2)^2\!\,}
{~32\hs\pi\hs s_W^{2}c_W^{2}M_Z^4\,\shat~}\!
\left[(q_L^{}x_L^Z\!\!-\!q_R^{}x_R^Z)h_4^Z
-(q_L^{}x_L^A\!\!-\!q_R^{}x_R^A)\frac{c_W}{s_W}h_4^\gamma\right] \!,
\\[1mm]
\widetilde{\sigma}_2^{}(h_4^V) & =
\frac{e^4(\shat\!-\!M^2_Z)^5\,}
{~768\hs\pi\hs s_W^{2}c_W^{2}M_Z^{10}\,\shat~}\!\!
\left[
X_{LR}^{ZZ}(h_4^Z)^2+X_{LR}^{AA}\frac{c_W^2}{s_W^2}(h_4^\gamma)^2
-2X_{LR}^{ZA}\frac{c_W^{}}{s_W^{}}(h_4^Zh_4^\gamma)\right] \!,
\end{align}
\eeqs
where we have defined the notations
\beqa
X_{LR}^{ZZ}\equiv (x_L^Z)^2+(x_R^Z)^2,~~~
X_{LR}^{AA}\equiv (x_L^A)^2+(x_R^A)^2,~~~
X_{LR}^{Z\hsm A}\equiv x_L^Zx_L^A+x_R^Zx_R^A\,.
\eeqa 
Taking the high-energy limit, we find that the cross sections 
$(\hs\widetilde{\sigma}_1^{},\hs\widetilde{\sigma}_2^{})$
scale as follows:
\beqs 
\label{eq:CS12-Xh4V-size}
\begin{align}
\widetilde{\sigma}_1^{}(h_4^V) &=\,
O(\shat)\hs h_4^Z + O(\shat)\hs h_4^\gamma\,,
\\[1mm]
\widetilde{\sigma}_2^{}(h_4^V) &=\, 
O(\shat^4)(h_4^Z\hsm )^2 +
O(\shat^4)(h_4^\gamma\hsm )^2 + 
O(\shat^4)(h_4^Zh_4^\gamma\hsm )\,.
\end{align}
\eeqs 
Comparing Eq.\eqref{eq:CS12-Xh4V-size} 
with Eq.\eqref{eq:CS12-h4h3-size},
we see that the $h_4^V$ contributions to the cross sections 
$(\hs\widetilde{\sigma}_1^{},\hs\widetilde{\sigma}_2^{})$
in the conventional form factor parametrization \eqref{eq:FF0-nTGC} 
have an additional high-energy factor of $s^1$ beyond
the $h_4^{}$ contributions to $(\sigma_1^{},\hs\sigma_2^{})$
in our improved parametrization 
\eqref{eq:FF2-nTGC}.\

\begin{table}[t]
\begin{center}
\begin{tabular}{c||ccc|ccc}
	\hline\hline	
	& & & & & &
	\\[-4mm]
	$\sqrt{s\,}$ & & \hspace*{-12mm}13\,TeV\,($\ell\hs\bar{\ell}$)
	\hspace*{-12mm}
	& & &
	\hspace*{-12mm}13\,TeV\,($\ell\hs\bar\ell,\nu\bar{\nu}$)
	\hspace*{-10mm}
	\\
	\hline
	& & & & & &
	\\[-4.3mm]
	$\mL$(ab$^{-1}$) & 0.14 & 0.3 & 3 & 0.14 & 0.3 & 3 			
	\\
	\hline
	& & & & & &
	\\[-4.3mm]
	\hspace*{-2.5mm}
	$|h_4^{}(\hs\mathbb{O}_1)|\!\times\!10^{5}$
	\hspace*{-2.5mm}
	& 5.8\,({\blu 18}) & 3.7\,({\blu 11}) & 1.0\,({\blu 2.8}) & 5.8\,({\blu 18}) & 3.7\,({\blu 11}) & 1.0\,({\blu 2.8})
	\\
	\hline
	& & & & & &
	\\[-4.3mm]
	$|h_4^{}|\!\times\!10^{6}$
	& 14\,({\blu 28}) & 11\,({\blu 21}) & 5.2\,({\blu 9.1}) & 9.6\,({\blu 18}) & 7.5\,({\blu 14}) & 3.8\,({\blu 6.4})
	\\
	\hline
	& & & & & &
	\\[-4.3mm]
	$|h_3^Z|\!\times\!10^{4} $
	& 2.7\,({\blu 5.0}) & 2.1\,({\blu 3.8}) & 1.1\,({\blu 1.8}) & 1.9\,({\blu 3.4}) & 1.5\,({\blu 2.7}) & 0.80\,({\blu 1.3})
	\\
	\hline
	& & & & & &
	\\[-4.3mm]
	$|h_3^\gamma|\!\times\!10^{4}$
	& 3.1\,({\blu 5.8}) & 2.5\,({\blu 4.5}) & 1.3\,({\blu 2.1}) & 2.2\,({\blu 4.0}) & 1.8\,({\blu 3.1}) & 0.97\,({\blu 1.6})
	\\
	\hline\hline
	& & & & & &
	\\[-4.3mm]
	$\sqrt{s\,}$ & & \hspace*{-12mm}100\,TeV\,($\ell\hs\bar{\ell}$)
	\hspace*{-12mm}
	& & &
	\hspace*{-12mm}100\,TeV\,($\ell\hs\bar\ell,\nu\bar{\nu}$)\hspace*{-10mm}
	\\
	\hline
	& & & & & &
	\\[-4.3mm]
	$\mL$(ab$^{-1}$) &  3 &  10 &  30 &  ~3 &  10  &  30~
	\\
	\hline
	& & & & & &
	\\[-4.3mm]
	\hspace*{-2.5mm}
	$|h_4^{}(\hs\mathbb{O}_1)|\!\times\!{\hs  10^8}$
	\hspace*{-2.5mm}
	&  3.4\,({\blu 11}) &  1.6\,({\blu5.0}) &  0.85\,({\blu 2.6}) &  3.4\,({\blu 11}) &  1.6\,({\blu5.0}) &  0.85\,({\blu 2.6})  \\
	\hline
	& & & & & &
	\\[-4.3mm]
	$|h_4^{}|\!\times\!10^{9}$
	&  6.1\,({\blu13}) &  3.9\,({\blu7.8}) &  2.6\,({\blu 5.1}) &  4.0\,({\blu 8.1}) &  2.6\,({\blu 5.1}) &  1.9\,({\blu 3.4})\\
	\hline
	& & & & & &
	\\[-4.3mm]
	$|h_3^Z|\!\times\!10^{7} $
	&  8.9\,({\blu 17}) &  6.0\,({\blu 11}) &  4.2\,({\blu 7.5}) &  6.1\,({\blu 11}) &  4.2\,({\blu 7.5}) &  3.0\,({\blu 5.2})\\
	\hline
	& & & & & &
	\\[-4.3mm]
	$|h_3^\gamma|\!\times\!10^{7}$
	&  10\,({\blu 20}) &  6.8\,({\blu 13}) &  4.9\,({\blu 8.7}) &  7.2\,({\blu 13}) &  4.9\,({\blu 8.7}) &  3.5\,({\blu 6.1}) \\
	\hline\hline
\end{tabular}
\end{center}
\vspace*{-3mm}
\caption{\small\hspace*{-1.5mm}
{\it Sensitivity reaches on the nTGC form factor parameters 
	at the $2\hs\sigma$ (black color) 
	and $5\hs\sigma$ (blue color) levels,
	as derived by analyzing the reactions
	$\,p{\hs}p{\hs}(q{\hs}\bar{q})\ito Z\ga\!\to\! \ell\bar{\ell}\ga$\,
	and 
	$\,p{\hs}p{\hs}(q{\hs}\bar{q})\ito Z\ga\ito \nu\bar{\nu}\ga$		
	at the LHC}\,(13\,TeV) {\it and the} $pp$\,(100\,TeV) 
{\it collider, with the indicated integrated luminosities.\ 
	In the third and ninth rows,
	the sensitivity limits for $|h_4^{}(\hs\mathbb{O}_1)|$
	are derived by using the observable \eqref{eq:O1-G+} 
	from the interference contributions,
	whereas the $|h_4^{}|$ limits in the fourth and tenth rows
	are derived including the squared contributions.\
	The third and fifth columns marked 
	($\hs\ell\hs\bar{\ell},\hs\nu\hs\bar{\nu}$)
	present the combined limits including both 
	the charged-lepton and neutrino final states. 
}}
\label{tab:h}
\label{tab:5}
\end{table}
%


We present in Table\,\ref{tab:h} the sensitivities of probes of the
form factor parameters $h_i^V$ at the LHC\,(13{\hs}TeV) and 
a 100\,TeV $pp$ collider (marked in blue), 
with the indicated integrated luminosities.\
We recall that the form factors 
and dimension-8 operators are connected via Eq.\eqref{eq:h-dim8}.\ 
We find that the most sensitive probes are those of the form factor $\hs h_4^{}\hs$, 
which is generated by the nTGC operator $\mO_{G+}^{}$.\  
The sensitivities of probes of $h_3^Z$ (via the operator 
$\OBW$) and $h_3^\gamma$ (via the operator $\OGM$) 
are smaller.\ 
In the case of $h_4^{}$, we present in the third row
the sensitivities obtained from the interference contributions 
using the observable 
$\mathbb{O}_1^{}$ of Eq.\eqref{eq:O1-G+},  
and in the fourth row the sensitivities 
from the squared contributions.\ 
The sensitivity limits in the third row are not improved 
by including the invisible decays of $Z\ito\nu{\hs}\bar\nu$
because the angular distribution of $\phi_*^{}$ cannot be measured 
for the invisible channel.\
We see that the sensitivity bounds on $|h_4^{}|$
in the fourth row are significantly stronger than those in the
third row.\ This is because the squared contributions
have stronger energy dependence and thus are enhanced.
The sensitivities of probes to $|h_3^Z|$ and $|h_3^\gamma|$
are shown in the last two rows of Table\,\ref{tab:h},
and are found to be much weaker than the bounds on 
$|h_4^{}|$ (third and fourth rows).
We also see from Table\,\ref{tab:5} that the
sensitivities of probes of these nTGC form factors 
at 100\,TeV $pp$ colliders
are generally much stronger than those at the LHC by
large factors of $\hs{O}(10^2\!-\!10^3)$.\
In passing, we note that the current collider limits 
on the nTGC form factors given in Table\,\ref{tab:5}   
are much stronger than the unitarity limits 
of Table\,\ref{tab:0} and Fig.\hs\ref{fig:0}.\

\begin{table}[t]
\begin{center}
\begin{tabular}{c||ccc|ccc||c|ccc|ccc}
	\hline\hline	
	& & & & & & & & & & &
	\\[-4mm]
	$\sqrt{s\,}$ & & \hspace*{-12mm}13\,TeV\,($\ell\hs\bar{\ell}$)
	\hspace*{-12mm}
	& & &
	\hspace*{-12mm}13\,TeV\,($\ell\hs\bar\ell,\nu\bar{\nu}$)\hspace*{-10mm}
	& & $\sqrt{s\,}$ & & 
	\hspace*{-12mm}100\,TeV\,($\ell\hs\bar{\ell}$)\hspace*{-12mm}
	& & &
	\hspace*{-10mm}100\,TeV\,($\ell\hs\bar\ell,\nu\bar{\nu}$)
	\hspace*{-10mm}
	\\
	\hline
	& & & & & & & & & & &
	\\[-4.3mm]
	$\mL\hs$(ab$^{-1}$)
	& 0.14 & 0.3 & 3 & 0.14 & 0.3 & 3 & $\mL\hs$(ab$^{-1}$)
	& 3 & 10 & 30 & ~3 & 10 & 30~
	\\
	\hline
	& & & & & & & & & & &
	\\[-4.3mm]
	\red$|h_4^{}|\!\times\!10^{6}$ 
	&\red 14 &\red 11 &\red 5.2 & \red9.6 &\red 7.5 &\red 3.8&	\red$|h_4|\!\times\!10^{9}$
	&\red 6.1 &\red 3.9 &\red 2.6 &\red 4.0 &\red 2.6 &\red 1.9\\
	\hline
	& & & & & & & & & & &
	\\[-4.3mm]
	\blu$|h_4^Z|\!\times\!10^{7}$ 
	& \blu 7.5 &\blu 5.7 &\blu 2.8 &\blu 5.2 &\blu 4.0 &\blu 2.0 &	\blu$|h_4^Z|\!\times\!10^{11}$
	&\blu 4.3 &\blu 2.7 &\blu 1.9 &\blu 2.8 &\blu 1.9 &\blu 1.3\\
	\hline
	& & & & & & & & & & &
	\\[-4.3mm]
	\blu$|h_4^\gamma|\!\times\!10^{7} $ 
	& \blu 8.7 &\blu 6.7 &\blu 3.2 &\blu 5.9 &\blu 4.7 &\blu 2.4 
	&\blu	$|h_4^\gamma|\!\times\!10^{11}$&\blu 4.9 &\blu 3.2 &\blu 2.1 &\blu 3.3 &\blu 2.1 &\blu 1.5 \\
	\hline\hline
\end{tabular}
\end{center}
\vspace*{-3mm}
\caption{\small%
{\it Comparisons of the $2\hs\sigma$ sensitivities
	to the form factor $\hs h^{}_4\hs$ formulated in 
	the SMEFT (marked in red color) and the conventional form 
	factors $h_4^V\!$ respecting only U(1)$_{\rm{em}}^{}$ 
	(marked in blue color),
	derived from analyses of the reactions
	$\,p{\hs}p{\hs}(q{\hs}\bar{q})\ito Z\ga\!\to\! \ell\bar{\ell}\ga$\,
	and 
	$\,p{\hs}p{\hs}(q{\hs}\bar{q})\ito Z\ga\ito \nu\bar{\nu}\ga$		
	at the LHC}\,(13\,TeV) {\it and the} 100\,TeV $pp$ 
{\it collider, with the indicated integrated luminosities.\ 
	As discussed in the text, the form-factor limits (in blue color) 
	are included for illustration only,
	as they do not respect the full SM gauge symmetry, 
	and hence are invalid.
}}
\label{tab:6}
\end{table}

\vspace*{1mm}

Next, we present in Table\,\ref{tab:6} a 
comparison of the $2{\hs}\sigma$ sensitivities
to the form factor $h_4^{}$ defined in Eq.\eqref{eq:FF2-nTGC}
(based on the SMEFT formulation and marked in red color,
taken from Table\,\ref{tab:5}) and the conventional form 
factors $h_4^V\!$ in Eq.\eqref{eq:FF0-nTGC}
(respecting only U(1)$_{\rm{em}}^{}$ and marked in blue color).\
These limits were derived by analyzing the reactions
$\,q\hs\bar{q}\ito Z\ga \ito \ell\hs\bar{\ell}\ga$\,
and $\,q\hs\bar{q}\ito Z\ga\ito\nu\hs\bar{\nu}\ga$\,		
at the LHC (13~TeV) and a 100\,TeV
$pp$ collider, 
with the indicated integrated luminosities.\ 
We see that the sensitivities to the conventional
form factor $h_4^V$ (marked in blue color) 
are generally stronger than those of the SMEFT form factor
$h_4^{}$ (marked in red color) by large factors, ranging
from ${O}(20)$ at the LHC to ${O}(10^2)$ at a 100\,TeV $\!pp$ collider.
However, they are incorrect for the reasons discussed earlier.\ 
By comparing the energy-dependences of the $h_4^V$-induced 
cross sections between Eqs.\eqref{eq:CS12-h4h3-size} 
and \eqref{eq:CS12-Xh4V-size}, we have explicitly clarified   
why the sensitivity limits based on the conventional form factor 
parametrization \eqref{eq:FF0-nTGC} are
spuriously much stronger than those given by our improved 
form factor approach \eqref{eq:FF2-nTGC}.\ 
The comparison of Table\,\ref{tab:6} demonstrates 
the importance of using our consistent
form factor approach \eqref{eq:FF2-nTGC}
based on the fully gauge-invariant SMEFT formulation.

\vspace*{1mm}

\subsection{\hspace*{-2.5mm}Correlations between the
nTGC Sensitivities at Hadron Colliders}
\label{sec:4.5}
\vspace*{1mm}

In this Section, we analyze the correlations 
between the sensitivities of
probes of the nTGCs at hadron colliders
using both the dimension-8 SMEFT operator approach
and the improved formulation of the 
form factors presented earlier.

\vspace*{1mm}

We first analyze the correlations of sensitivity reaches 
between each pair of the nTGC form factors $(h_4^{},\,h_3^Z)$,
$(h_4^{},\,h_3^\gamma)$, and $(h_3^Z,\,h_3^\gamma)$
at the  LHC{\hs}(13{\hs}TeV) and the 100{\hs}TeV $pp$ collider.\
We compute the contributions of a given pair of form factors
to the following global $\chi^2$ function:
\begin{eqnarray}
\label{eq:chi2}
\chi^2 \,=\, \,\sum_{\text{bin}}\!\frac{S^2_{\rm{bin}}}{~B_{\rm{bin}}^{}~}
=\,\sum_{\text{bin}}\!
\frac{(\sigma^{\rm{bin}}\!-\!\sigma_0^{\rm{bin}})^2}
{\sigma_0^{\rm{bin}}}\!\times\!(\mL\!\times\!\epsilon\hs )
=\,\sum_{\text{bin}}\!
\frac{(\sigma_1^{\rm{bin}}\!+\!\sigma_2^{\rm{bin}})^2}
{\sigma_0^{\rm{bin}}}\!\times\!(\mL\!\times\!\epsilon\hs ),
\hspace*{6mm}
\end{eqnarray}
where $\sigma_0^{\rm{bin}}$ is the SM contribution,
and $(\sigma_1^{\rm{bin}},\,\sigma_2^{\rm{bin}})$ are
the (interference,~squared) terms of the
form factor contributions.\ 
These cross sections are computed for each bin 
and then summed up.\
We minimize the $\chi^2$ function \eqref{eq:chi2} 
for each pair of form factors 
at each hadron collider with a given integrated luminosity
$\mL$, assuming an ideal detection efficiency  
$\hs\epsilon \hsm =\!1\hs$.\

\begin{figure}[t]
\centering
\includegraphics[height=7cm,width=7.5cm]{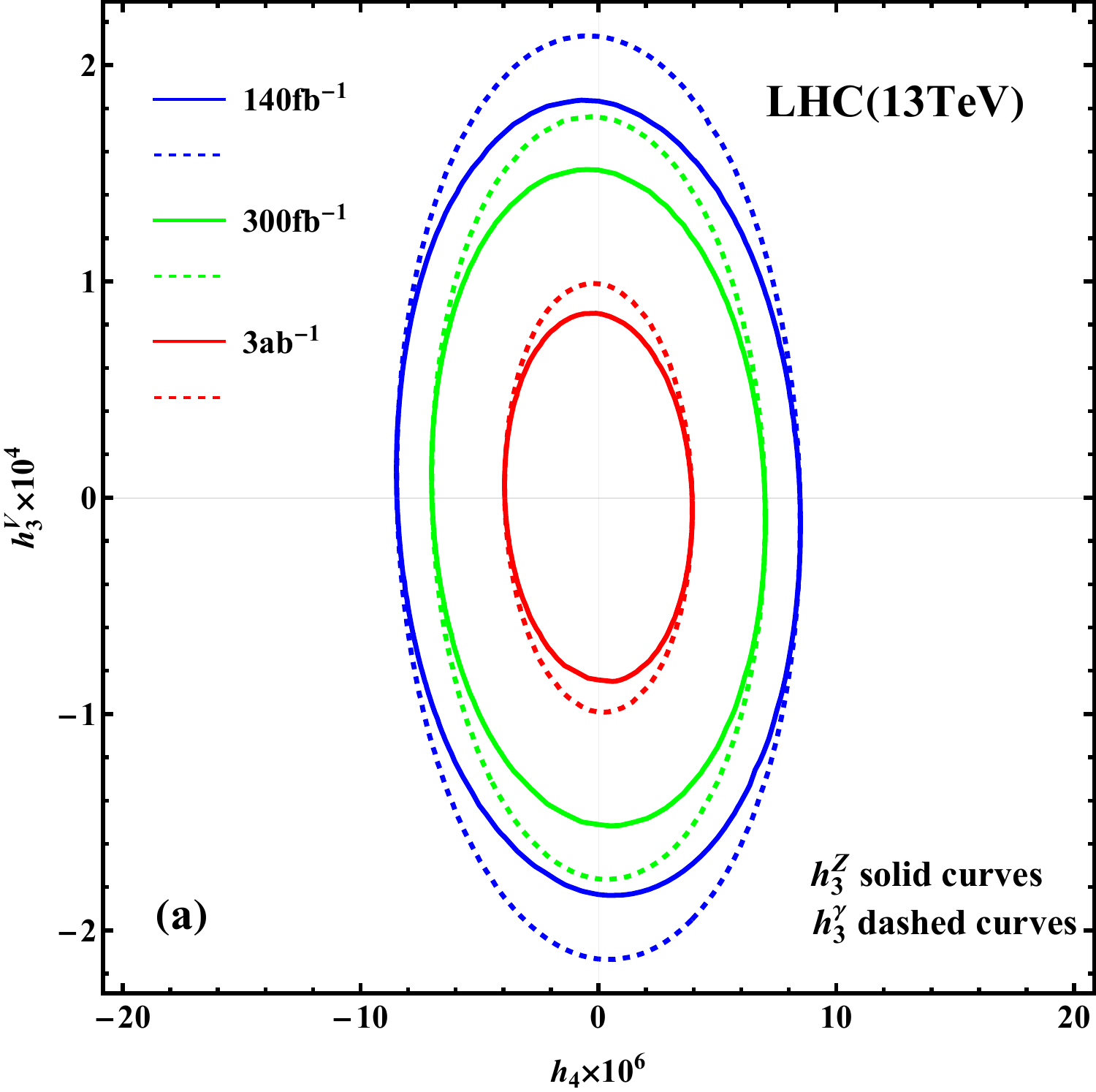}
\hspace*{1mm}
\includegraphics[height=7cm,width=7.5cm]{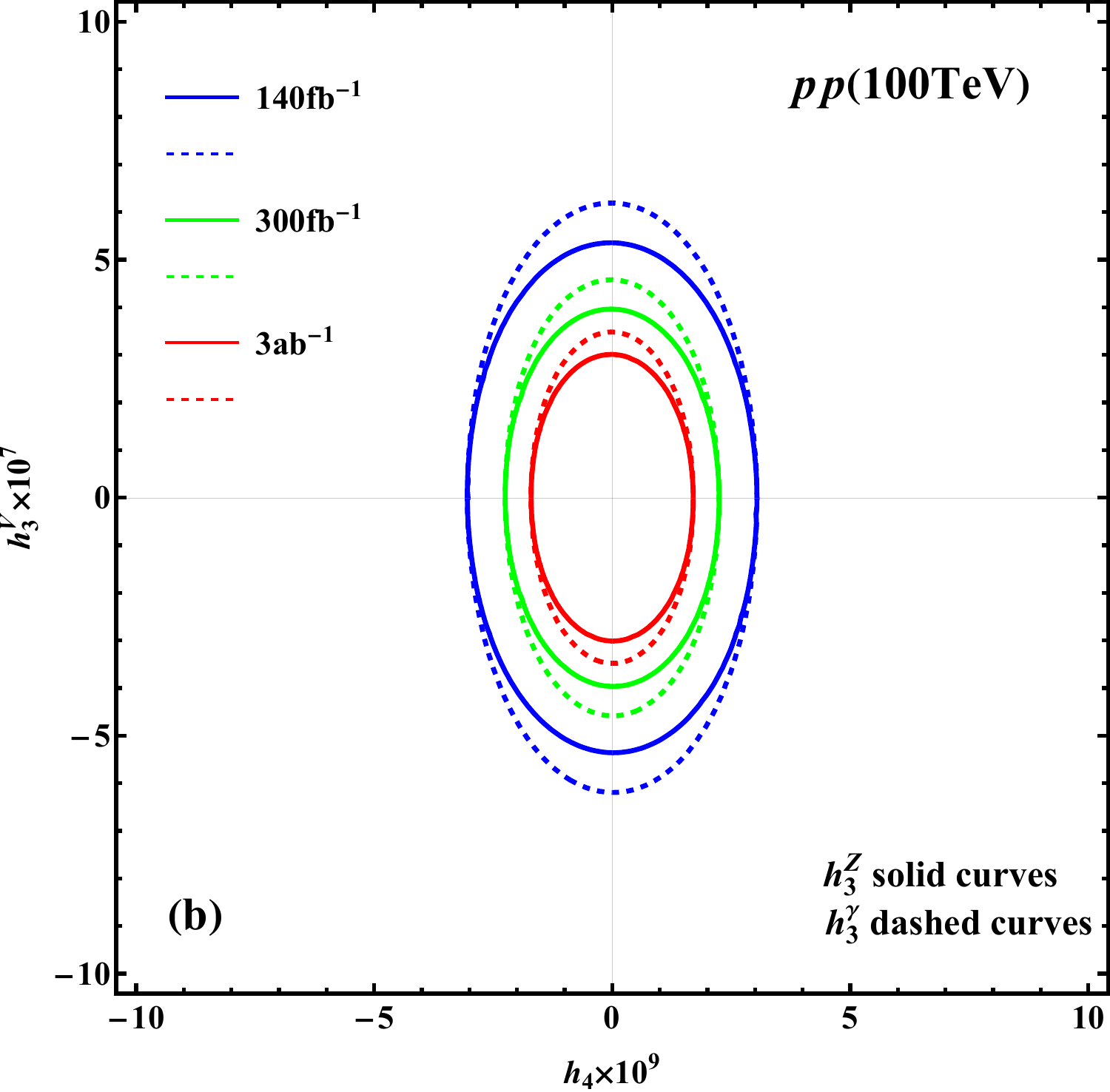}
\\[3mm]
\includegraphics[height=7cm,width=7.5cm]{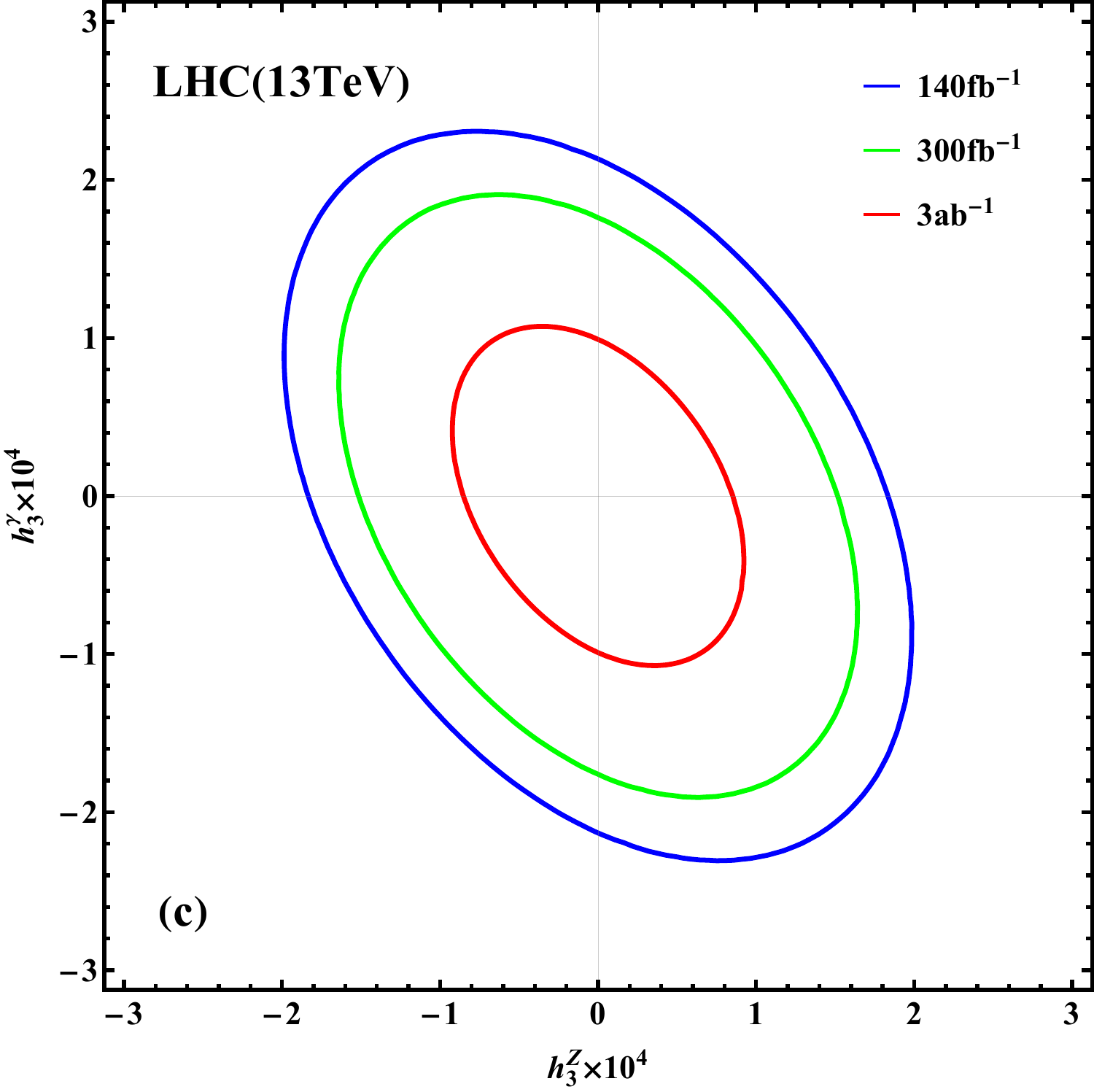}
\hspace*{1mm}
\includegraphics[height=7cm,width=7.5cm]{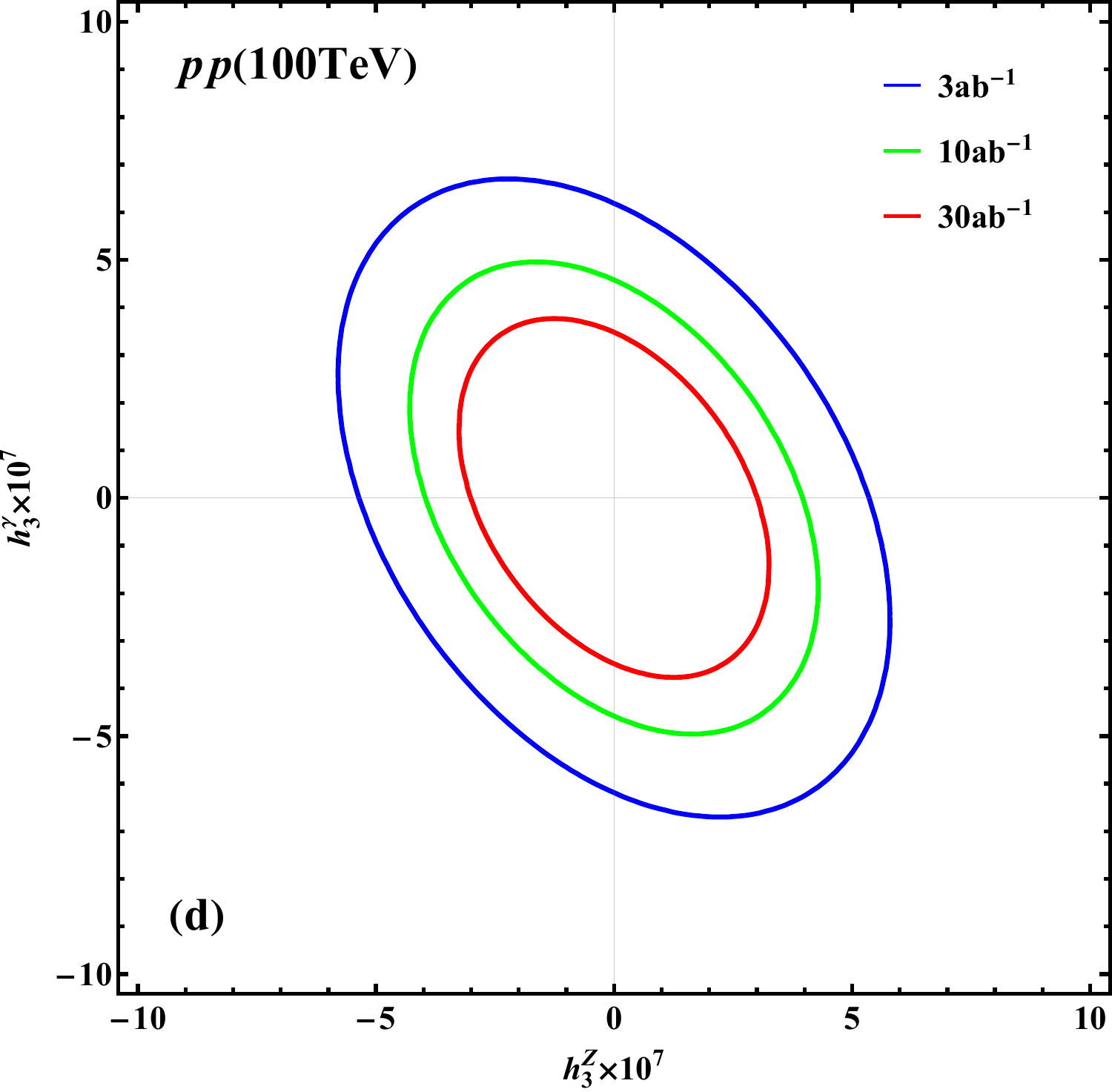}
\caption{\small\hspace*{-2mm}
{\it Correlation contours 
	of the sensitivity reaches (95\%\,C.L.) 
	for the indicated pairs of nTGC form factors at the 
	LHC}\,(13\,TeV) {\it [\,panels\,$(a)$ and $(c)$]
	and a} 100\,TeV {\it $pp$ collider
	[\,panels\,$(b)$ and $(d)$].\
	Panels\,(a) and (b) show the correlation contours of 
	$(h_4^{},\,h_3^Z)$ (solid curves) and 
	$(h_4^{},\,h_3^\gamma)$ (dashed curves),
	and panels\,(c) and (d) depict 
	the correlation contours of $(h_3^Z,\,h_3^\gamma)$.\ 
}}
\label{fig:4}
\vspace*{1mm}
\end{figure}

\vspace*{1mm}

We present our findings in Fig.\,\ref{fig:4}.\
Panels\,(a) and (b) show the correlation contours 
of the form factors $(h_4^{},\,h_3^Z)$ (solid curve) and
$(h_4^{},\,h_3^\gamma)$ (dashed curve)
at the 95\%\,C.L.,
and panels\,(c) and (d) depict the correlation contours 
of the form factors $(h_3^Z,\,h_3^\gamma)$ 
at the 95\%\,C.L.\ 
Panels\,(a) and (c) show the correlation contours  
for the LHC with different integrated luminosities
$\mL=(140,\hs 300,\hs 3000)\hs$fb$^{-1}$ 
(marked by the blue, green,  and red colors, respectively),
and panels\,(b) and (d) depict 
the correlation contours for the 100\,TeV 
$pp$ collider with different integrated luminosities
$\mL=(3,\hs 10,\hs 30)\hs$ab$^{-1}$
(marked by the blue, green,  and red colors, respectively).

\vspace*{1mm}

Inspecting Figs.\ref{fig:4}(a) and (b), 
we see that each elliptical contour has its axes 
nearly aligned with the frame axes, which shows that 
the form factors $(h_4^{},\hs h_3^V)$ 
have rather weak correlation.\   
This feature can be understood by examining the structure of the 
$\chi^2$ function \eqref{eq:chi2}.\ 
For a qualitative understanding of such correlation features, 
here we simplify Eq.\eqref{eq:chi2} 
by considering a single bin analysis.\ Since the squared term
$\sigma_2^{}$ in Eq.\eqref{eq:CS2-h4h3-size} 
dominates over the interference term $\sigma_1^{}$, 
from Eq.\eqref{eq:chi2} we have 
$\hs\chi^2\!\sim\! [(\sigma_2^{})^2/\sigma_0^{}]
(\mL\hsm\times\!\epsilon)
\hsm\!\propto\hsm\!(\sigma_2^{})^2\hs$,
where the SM cross section 
$\hs\sigma_0^{}\hs$  
does not contain any new physics parameter and is thus
irrelevant to the correlation issue.\  
Since each elliptical contour has a fixed value of $\chi^2$, 
the cross section $\hs\sigma_2^{}\hs$ is given
by $\hs\sigma_2^{}\!\sim\!
\sqrt{\chi^2\hs\sigma_0^{}/(\mL\hsm\times\!\epsilon)\,}\hs$.\ 
We note that $\hs\sigma_2^{}\hs$ is a quadratic function 
of the form factors, so we can use the usual 
statistical method\,\cite{statistics}\cite{PDG2022} to analyze the 
quadratic function of $\hs\sigma_2^{}\hs$, which suffices
for examining the correlation property 
of each elliptical contour.\

Using Eqs.\eqref{eq:CS2-qq-Zgamma-H34} 
and \eqref{eq:CS2-h4h3-size},  
we express the quadratic form of $\hs\sigma_2^{}\hs$ as follows,
exhibiting explicitly the energy-scaling behavior of each term:
\beq
\label{eq:sigma2-new} 
\sigma_2^{} \,=\, 
\sbar^3\hs\bar{\sigma}_2^{44}\hs (h_4^{}\hsm )^2 +
\sbar^2\hs\bar{\sigma}_{2V}^{33}\hs (h_3^V\hsm )^2 + 
\sbar^2\hs\bar{\sigma}_{2V}^{43}\hs (h_4^{}h_3^V\hsm ) + 
\sbar^2\hs\bar{\sigma}_{2ZA}^{33}\hs (h_3^Zh_3^\gamma\hsm )\,,
\eeq 
where $\,\sbar \hsm =\hsm \shat /M_Z^2\,$ 
is a scaled dimensionless energy factor and  
$\hs\bar{\sigma}_{2}^{ij}\hs$ denotes the 
coefficient of each leading cross-section term in Eq.\eqref{eq:CS2-qq-Zgamma-H34} in the high-energy expansion.\

\vspace*{1mm}

To examine the correlations between $h_3^V$ and $h_4^{}$, only the
first three terms of Eq.\eqref{eq:sigma2-new} are relevant.\
Denoting the form factors 
$\hs (h_3^V,\hs h_4^{})\hsm =\hsm (x,\hs y) \hsm \equiv\hsm\XX
\hs$, 
we can express the relevant terms of 
Eq.\eqref{eq:sigma2-new} in the following quadratic form: 
\beqs 
\label{eq:sigma2-xy-Vinv}
\begin{align}
\label{eq:sigma2-xy-ABC}
\sigma_2^{}(x,\hs y) &\hs =\hs  
A\hs x^2\! +\hsm B\hs y^2\! +\hsm 2\hs C\hs x\hs y
\,=\,\XX\hs V^{-1}\hs\XX^T ,
\\[0.5mm]
\label{eq:Vinv-ABC}
V^{-1} &\hs =
\left(\!\!\hsm\ba{cc} A & C 
\\[0.3mm] C & B\ea\!\hsm\right)\!,
\end{align}
\eeqs 
where the coefficients $\,(A,\hs B,\hs C)\! \equiv \!
(\sbar^2\hs\bar{\sigma}_{2V}^{33},\hs
\sbar^3\hs\bar{\sigma}_2^{44},\hs
\fr{1}{2}\sbar^2\hs\bar{\sigma}_{2V}^{43})\hs$.\
The correlation contour of $(x,\hs y)$
is clearly an elliptical curve.\ 
For the above quadratic form
$\,\sigma_2^{}(x,\hs y)\!=\!\XX\hs V^{-1}\hs\XX^T$
with two parameters $\hs\XX\!=\!(x,\hs y)\hs$,
we express the covariance matrix 
as follows\,\cite{statistics}:
\begin{eqnarray}
\label{eq:V}
V=\(\!\hsm 
\begin{array}{cc}
\hat{\sigma}_{\!x}^2 & {\rho}\hsx\hat{\sigma}_{\!x}^{}\hs\hat{\sigma}_{\!y}^{}
\\[1mm]
{\rho}\hsx\hat{\sigma}_{\!x}^{}\hs\hat{\sigma}_{\!y}^{} & \hat{\sigma}_{\!y}^2 
\end{array}
\!\!\)\!,
\end{eqnarray}
where $(\sigmaX ,\hs\sigmaY)$ are related to the errors 
in the parameters $(x,\hs y)$.\ 
The inverse of the covariance matrix $V$ is derived as
\begin{eqnarray}
V^{-1}=\(\!\!\!
\begin{array}{cc}
\dis\frac{1}{~(1\!-\hsm\rho^2)\hs\hat{\sigma}_{\!x}^2~} & 
\dis -\frac{\rho}{~(1\!-\rho^2)\hs\sigmaX\hs\sigmaY~} 
\\[4mm]
\dis -\frac{\rho}{~(1\!-\rho^2)\hs\sigmaX\hs\sigmaY~}
& \dis\frac{1}{~(1\!-\hsm\rho^2)\hs\hat{\sigma}_{\!y}^2~} 
\end{array}
\!\!\!\) =
\(\!\!
\begin{array}{cc}
A & C \\[1.5mm]
C & B 
\end{array}
\hsm\!\)\!,
\end{eqnarray}
with the correlation parameter $\,\rho\,$ given by 
\beqa
\label{eq:rho}
\rho \,=\, -\hs C/\hsm\sqrt{AB\,}\,,
\eeqa 
where $(\sigmaX ,\hs\sigmaY)$ are connected to 
$(A,B,C)\hs$ through the relations,
%
$\sigmaX \!=\! [(1\!-\!\rho^2)A\hs ]^{-\frac{1}{2}}\hs$ and
$\sigmaY \!=\! [(1\!-\!\rho^2)B\hs ]^{-\frac{1}{2}}\hs$.\ 
%
Thus, using Eq.\eqref{eq:sigma2-xy-Vinv} 
we compute the correlation parameter \eqref{eq:rho} 
for the $(h_3^V\hsmx ,\hs h_4^{})$ contour as follows:
\beqa
\label{eq:rho-h3h4}
\rho (h_3^V\hsmx , h_4^{})\,=\, -\frac{\bar{\sigma}_{2V}^{43}}
{~2\sqrt{\bar{\sigma}_{2V}^{33}\hs\bar{\sigma}_{2}^{44}\,}~}
\,\sbar^{-\frac{1}{2}}\hs .
\eeqa 
In the above, 
$(\bar{\sigma}_{2}^{44},\hs
\bar{\sigma}_{2V}^{33},\hs 
\bar{\sigma}_{2V}^{43})$
correspond to the leading-energy terms of the cross sections
\eqref{eq:sigma2-44}-\eqref{eq:sigma2A-43}.\  
We see from Eqs.\eqref{eq:sigma2-44}-\eqref{eq:sigma2A-33}
and 
Eqs.\eqref{eq:sigma2Z-43}-\eqref{eq:sigma2A-43}
that the cross section coefficients
$(\bar{\sigma}_{2}^{44},\hs\bar{\sigma}_{2V}^{33})$ 
of the leading energy terms
are always positive and the cross section coefficients
$\,\bar{\sigma}_{2V}^{43}\,$ of the leading energy terms
are positive for any quark flavor.\
Hence, we deduce that the correlation parameter 
$\hsx\rho (h_3^V\hsmx , h_4^{})\!\hsm <\! 0\,$
in Eq.\eqref{eq:rho-h3h4}, but it is suppressed by a large
energy factor $\hs 1/\!\sqrt{\sbar\,}\,$.\
This means that the apex of the contour 
(where the slope $\hs y'\!=\hsm 0\hs$) 
must lie on the left-hand side (LHS) of the $y$ axis.\ 
These features explain why the orientations of the contours  
in Figs.\ref{fig:4}(a) and (b)
are not only nearly vertical, but also are 
aligned slightly towards the upper-left direction.\ 
Moreover, the deviation of the orientation of each 
contour from the vertical axis of Fig.\ref{fig:4}(b) 
is almost invisible because
of the more severe suppression by the energy factor
$\hs 1/\!\sqrt{\sbar\,}\,$ at the 100\,TeV
$pp$ collider than at the LHC.

\vspace*{1mm}

Then, we use Eq.\eqref{eq:chi2} to perform
the exact $\chi^2$ analysis 
for the form factors $(h_3^Z,\hs h_3^\gamma)$.\
The $(h_3^Z,\hs h_3^\gamma)$ contours are plotted  
in Figs.\ref{fig:4}(c) and (d) for the LHC and
the 100\,TeV $pp$ collider respectively,
which show strong correlations and are oriented towards
the upper-left quadrant, very different from the contours 
in Figs.\,\ref{fig:4}(a) and (b).\
To understand the correlation features of
Figs.\ref{fig:4}(c) and (d), 
we examine the relevant leading energy terms
in the cross section \eqref{eq:sigma2-new}
that include the form factors $(h_3^Z,\hs h_3^\gamma)$ and their
products.\
From Eq.\eqref{eq:sigma2-new},
we find that the cross section $\hs\sigma_2^{}\hs$ contains 
the following leading energy-dependent contributions:
%
\beqs 
\label{eq:chi2-h3Z-h3A-xy}  
\begin{align}
\label{eq:chi2-Ecount-h3Z-h3A}  
\sigma_2^{}(h_3^Z\hsm , h_3^\ga) &\,=\, 
\sbar^2\hs\bar{\sigma}_{2Z}^{33}\hs (h_3^Z\hsm )^2 + 
\sbar^2\hs\bar{\sigma}_{2A}^{33}\hs (h_3^\gamma\hsm )^2 + 
\sbar^2\hs\bar{\sigma}_{2ZA}^{33}\hs (h_3^Zh_3^\gamma )
\nn\\
&\,=\, A\hs x^2+B\hs y^2+2\hs C\hs x\hs y
\,=\,\XX\hs V^{-1}\hs\XX^T \,, 
\\[1mm]
(A,\hs B,\hs C) &\,\equiv\,
(\sbar^2\hs\bar{\sigma}_{2Z}^{33},\hs
\sbar^2\hs\bar{\sigma}_{2A}^{33},\hs
\fr{1}{2}\sbar^2\hs\bar{\sigma}_{2ZA}^{33})\hs,
\end{align}	
\eeqs 
where we denote the form factors 
$\hs (h_3^Z,\hs h_3^\ga )\!\equiv\! (x,\hs y)\!\equiv\!\XX\,$
and the matrix $V^{-1}$ takes the form of Eq.\eqref{eq:Vinv-ABC}.\
Thus, using $\hs\sigma_2^{}\hs$ formula in 
Eq.\eqref{eq:chi2-h3Z-h3A-xy}, 
we compute the correlation parameter \eqref{eq:rho} 
for the $(h_3^Z\hsmx ,\hs h_3^\ga)$ contour as follows:
\beqa
\label{eq:rho-h3Z-h3A}
\rho(h_3^Z\hsmx , h_3^\ga) \hs =\hs -\frac{\bar{\sigma}_{2ZA}^{33}}
{~2\sqrt{\bar{\sigma}_{2Z}^{33}\hs\bar{\sigma}_{2A}^{33}\,}~}
\hs\sbar^0 \hs.
\eeqa 
This shows that the correlation parameter $\hs\rho\hs$ 
is of $O(\sbar^0)$ and not suppressed by any energy factor,
unlike the case of Eq.\eqref{eq:rho-h3h4} which is suppressed
by $\hs 1/\!\sqrt{\sbar\,}\,$.\ 
From Eqs.\eqref{eq:sigma2Z-33}-\eqref{eq:sigma2A-33} and 
Eq.\eqref{eq:sigma2ZA-43}, we deduce that 
$\hs\bar{\sigma}_{2Z}^{33}\hs\bar{\sigma}_{2A}^{33}\!>\!0\hs$
and $\hs\bar{\sigma}_{2ZA}^{33}\!>\!0\hs$
always holds which lead to 
$\hs\rho(h_3^Z\hsmx , h_3^\ga)\!<\!0\hs$.\ 
These facts explain why the correlation between
$\hs (h_3^Z,\hs h_3^\gamma)\hs$ is large and all the contours
of Figs.\,\ref{fig:4}(c) and (d) are oriented towards the 
upper-left quadrant.

\begin{figure}[t]
\vspace*{-5mm}
\centering
\includegraphics[height=7.4cm,width=7.8cm]{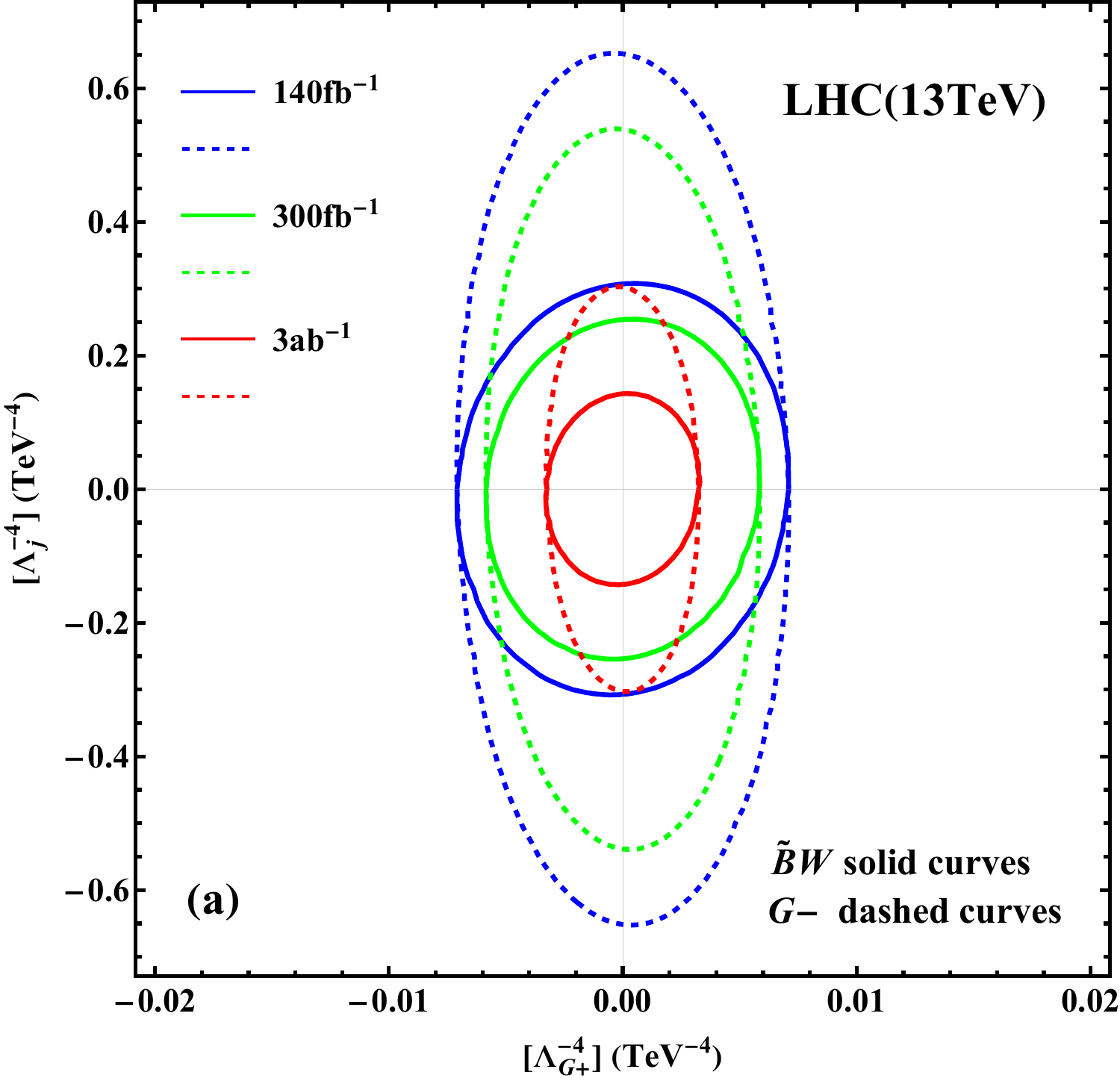}
\includegraphics[height=7.4cm,width=7.8cm]{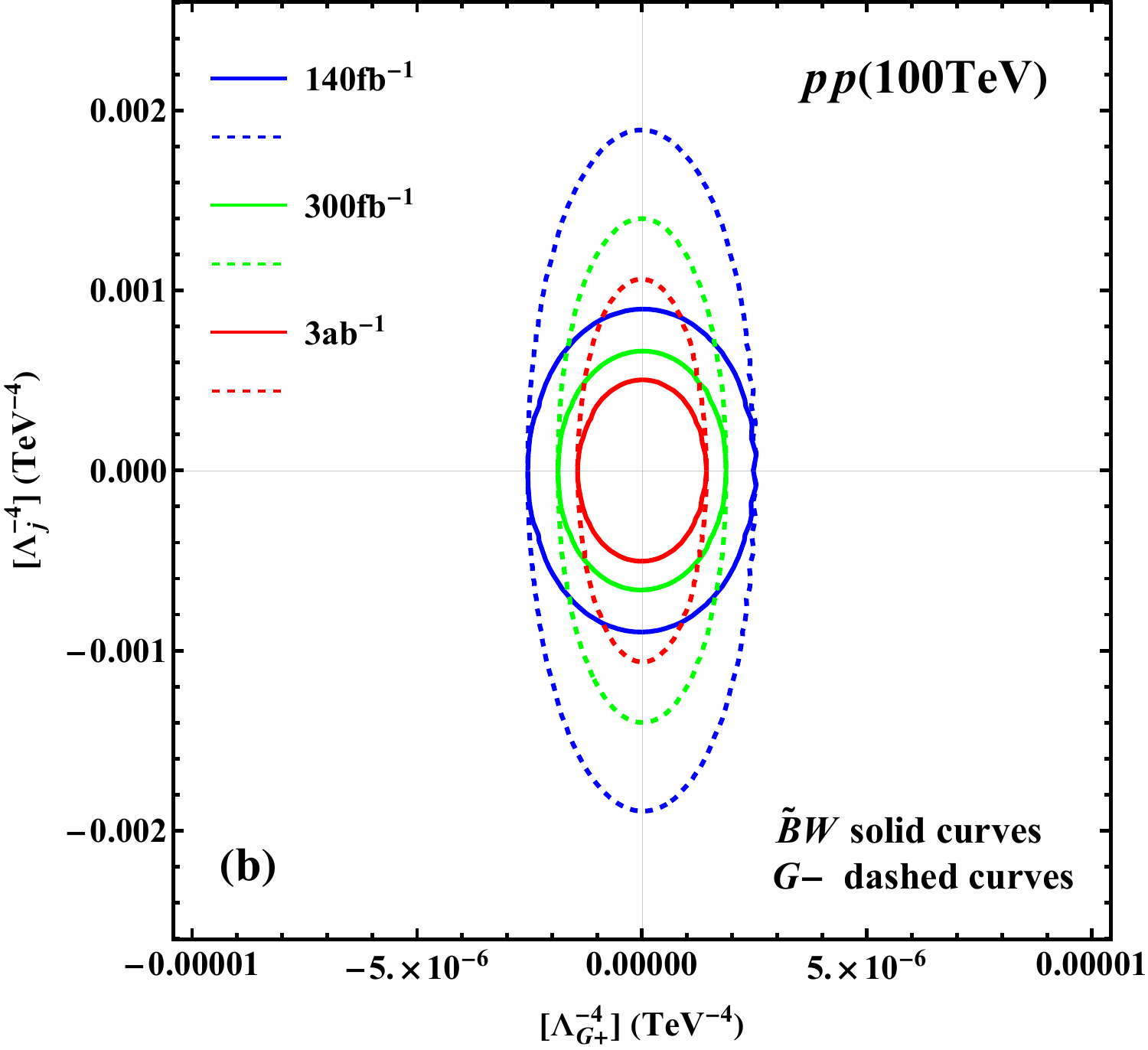}
\caption{\small\hspace*{-2mm}
{\it Correlation contours of the sensitivity reaches 
	(95\%\,C.L.) for the indicated pairs of nTGC operators at the 
	LHC}\,(13\,TeV) {\it [\,panel $(a)$] and a} 100\,TeV {\it $pp$
	{collider} [\,panel $(b)$].\
	Panels\,(a) and (b) show the correlation contours of 
	$(\OGP,\,\OBW)$ (solid curves) and 
	$(\OGP,\,\OGM)$ (dashed curves).\ 
}}
\label{fig:5}
\end{figure}

\vspace*{1mm}

We then consider the nTGC formulation using the 
dimension-8 SMEFT operators as given in Section\,\ref{sec:2}
and study correlations of the sensitivity reaches 
between each pair of the nTGC operators.\ 
We first study the correlations between the pairs of nTGC
operators $(\OGP,\hs\OBW)$ and $(\OGP,\hs\OGM)$.\ 
We perform the $\chi^2$ analysis using Eq.\eqref{eq:chi2}
and present the findings in Fig.\,\ref{fig:5}
for the LHC\,(13\,TeV) [panel (a)] and the 100\,TeV $pp$ collider
[panel (b)] for a set of sample integrated luminosities, 
respectively.\ In each panel, the $(\OGP,\hs\OBW)$ correlations
are shown by the contours in solid curves, whereas the
$(\OGP,\hs\OGM)$ correlations are depicted by the contours 
in dashed curves.\ We see that the correlations  
of the operators $(\OGP,\hs\OBW)$ and $(\OGP,\hs\OGM)$ are rather
weak, similar to the case of the $(h_4^{},\hs h_3^V)$ contours 
in Figs.\,\ref{fig:4}(a) and (b).

\vspace*{1mm}

The correlation features of the contours in Fig.\,\ref{fig:5}
can be understood in the following way.\ 
Using the relations in Eq.\eqref{eq:h-dim8}, we here denote 
$\hs (x,\hs y)\!=\!(h_4^{},\hs h_3^V)
\!=\!(r_4^{}\bar x,\hs r_3^V\bar y)\hs$ and 
$\hs (\bar x,\hs \bar y)=
([\cut_{V}^{-4}],\hs[\cut_{G+}^{-4}])$,
where $V\!=\!Z,A$ and 
$(\cut_{Z}^{-4},\hs\cut_{A}^{-4})\!\equiv\! 
(\cut_{\widetilde BW}^{-4},\hs \cut_{G-}^{-4})\hs$.\ 
With these, we express the leading cross section 
$\hs\sigma_2^{}\hs$ in Eqs.\eqref{eq:sigma2-new}   
and \eqref{eq:sigma2-xy-ABC} as follows:
\vspace*{-2mm}
\beqs 
\label{eq:sigma2-ABCV-d8}
\begin{align}
\label{eq:sigma2-xy-d8}
\hspace*{-7mm}
\sigma_2^{}(\xb,\hs \yb) &= 
A\hs \xb^2\hsm +\hsm B\hs \yb^2\hsm +\hsm 2\hs C\hs \xb\hs \yb
\,=\hs\over{\XX}\hs V^{-1}\hs\over{\XX}^T \hs,
\\[1mm]
\label{eq:sigma2-ABC-d8}
(A,\hs B,\hs C) & \equiv
\(\sbar^2\hs (r_3^V)^2\bar{\sigma}_{2V}^{33},\hs
\sbar^3\hs r_4^2\bar{\sigma}_2^{44},\hs
\fr{1}{2}\sbar^2\hs r_3^Vr_4^{}\bar{\sigma}_{2V}^{43}\) 
\hs,
\end{align}	
\eeqs 
where $\hs\over{\XX} \equiv (\xb,\hs \yb)\hs$ 
and the matrix $V^{-1}$ takes the form in 
Eq.\eqref{eq:Vinv-ABC}.\  
Thus, using Eq.\eqref{eq:sigma2-ABCV-d8}, 
we compute the correlation parameter \eqref{eq:rho} for 
$(\xb,\hs\yb)\hsm =\hsm 
\([\cut_{V}^{-4}],\hs [\cut_{G+}^{-4}]\)$ 
as follows:
\beqa
\label{eq:rho-OG+OV}
\rho\!\([\cut_{V}^{-4}], [\cut_{G+}^{-4}]\)
\,=\, -\text{sign}(r_3^Vr_4^{})
\frac{\bar{\sigma}_{2V}^{43}\,\sbar^{-\frac{1}{2}}}
{~2\sqrt{\bar{\sigma}_{2V}^{33}\hs\bar{\sigma}_{2}^{44}\,}~}
\,=\, \text{sign}(r_3^Vr_4^{})\hs\rho (h_3^V\hsmx , h_4^{})
\hs ,
\eeqa 
where the correlation parameter
$\hs\rho (h_3^V\hsmx , h_4^{})\!<\! 0\,$ is derived in
Eq.\eqref{eq:rho-h3h4}.\ 
According to Eq.\eqref{eq:r4-r3},  we have 
$\hs\text{sign}(r_3^Zr_4^{}) \!<\! 0\hs$ and
$\hs\text{sign}(r_3^\ga r_4^{}) \!>\! 0\hs$.\
Thus, we can infer the signs of the corresponding 
correlation parameters:
\beq
\rho\big([\cut_{\widetilde BW}^{-4}], [\cut_{G+}^{-4}]\big)
> 0\,, \hspace*{5mm}
\rho\!\([\cut_{G-}^{-4}], [\cut_{G+}^{-4}]\) < 0\,.
\eeq 
These nicely explain why in Fig.\,\ref{fig:5} the orientations of
the correlation contours (solid curves) of 
the operators $(\OGP,\,\OBW)$ are slightly aligned towards to
the right-hand-side of the vertical axis, whereas the
orientations of the correlation contours (dashed curves) 
of the operators $(\OGP,\,\OGM)$ 
are slightly aligned towards to the left-hand-side of the vertical axis.\ 
Their deviations from the vertical axis are rather small
because of the energy suppression factor 
$\hs 1/\!\sqrt{\sbar\,}\,$ in Eq.\eqref{eq:rho-OG+OV}, 
and they become even smaller 
for the contours of Fig.\,\ref{fig:5}(b)
at a 100\,TeV $pp$ collider, as expected.

\begin{figure}[t]
\vspace*{-6mm}
\centering
\includegraphics[height=7cm,width=7.5cm]{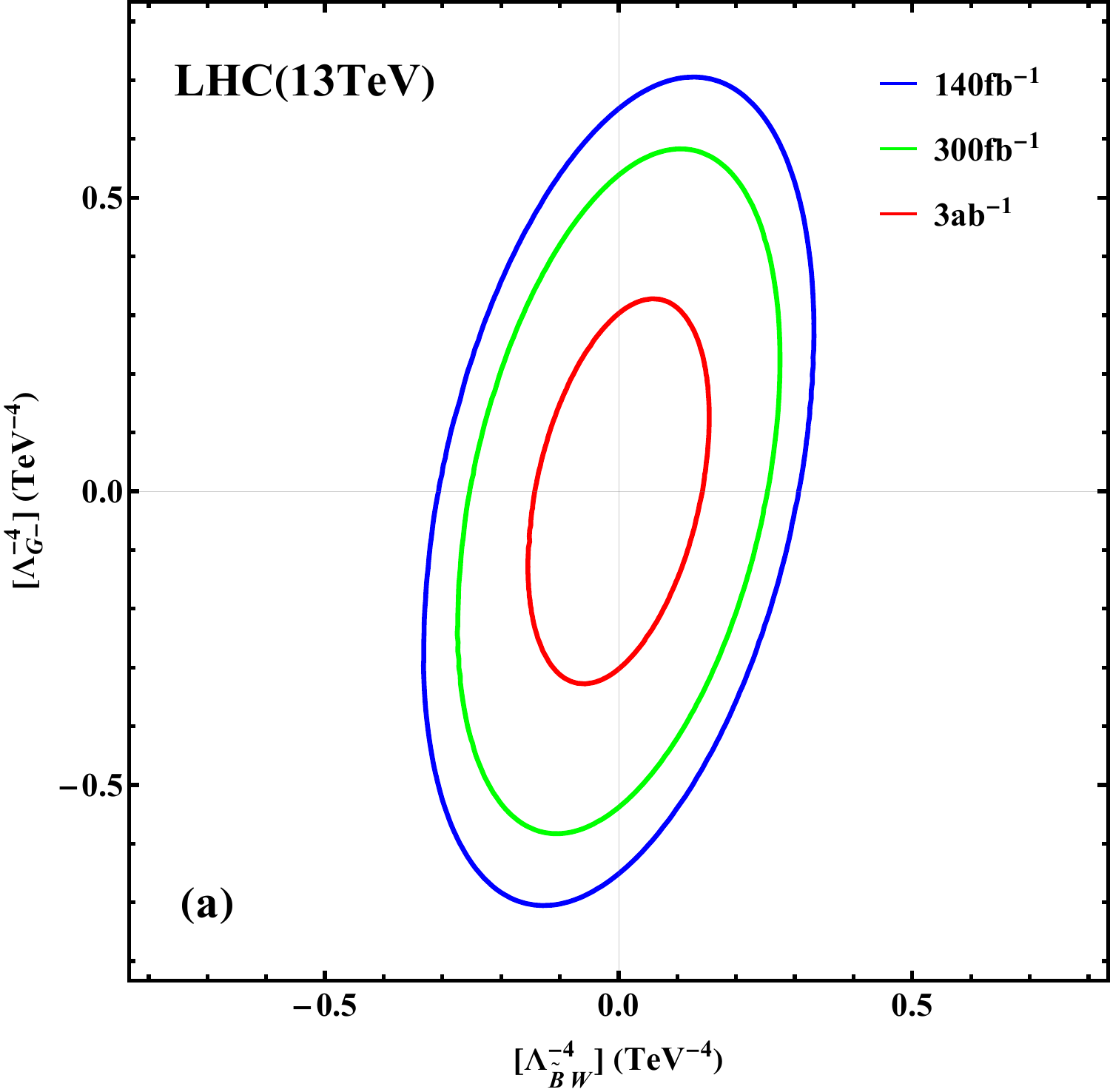}
\includegraphics[height=7cm,width=7.5cm]{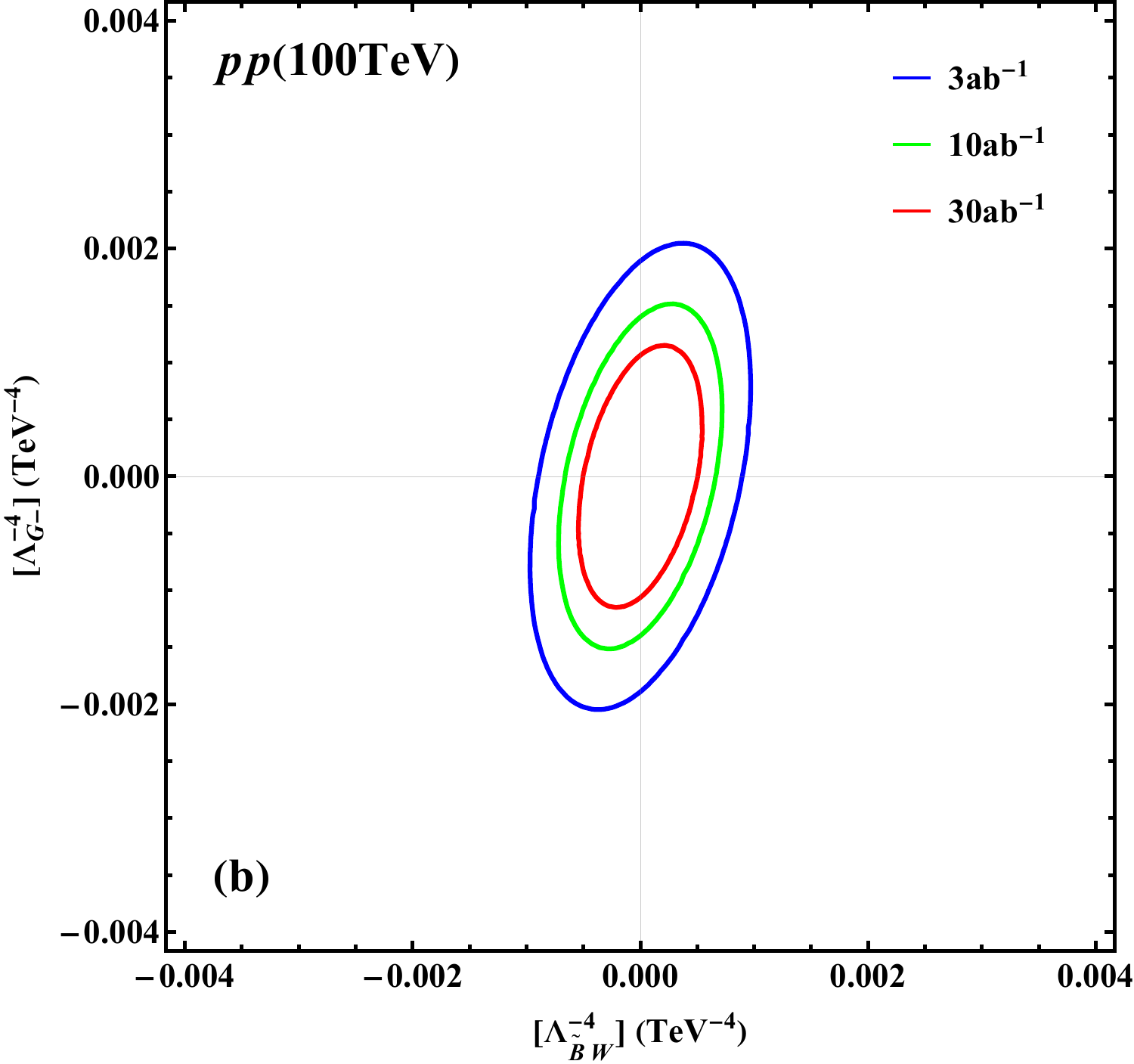}
\\[3mm]
\includegraphics[height=7cm,width=7.5cm]{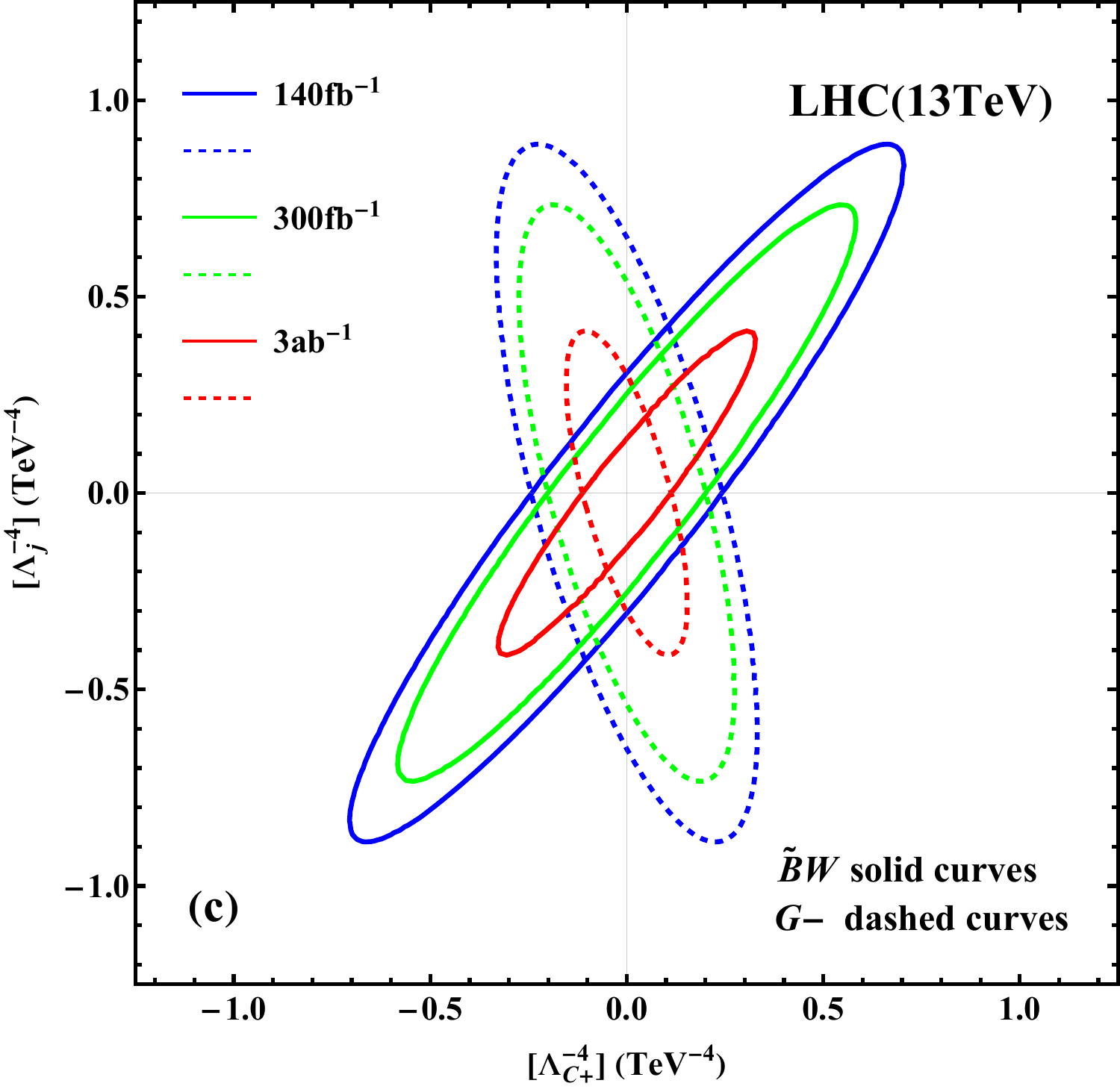}
\includegraphics[height=7cm,width=7.5cm]{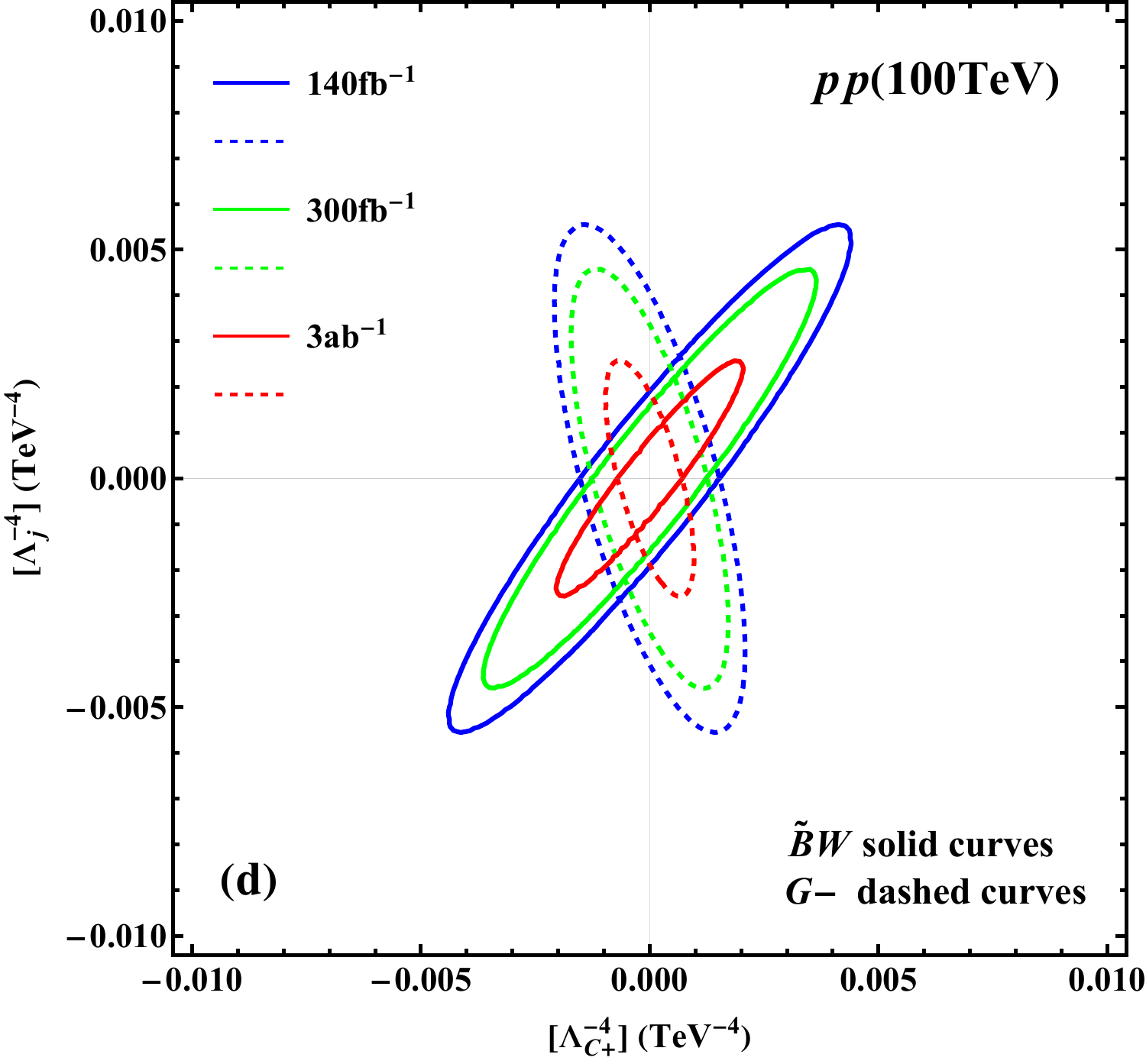}
\caption{\small\hspace*{-2mm}
{\it Correlation contours of sensitivity reaches (95\%\,C.L.)
	for the indicated pairs of nTGC operators 
	at the LHC}\,(13\,TeV) {\it [\,panels $(a)$ and $(c)$] and the} 100\,TeV {\it $pp$ collider [\,panels $(b)$ and $(d)$].\
	Panels\,(a) and (b) show the correlation contours of 
	$(\OBW,\,\OGM)$, whereas panels\,(c) and (d) depict 
	the correlation contours of $(\OCP,\,\OBW)$ (solid curves) and 
	$(\OCP,\,\OGM)$ (dashed curves). 
}}
\label{fig:6}
\vspace*{2mm}
\end{figure}
%


Next, we study the correlations between the nTGC operators
$(\OBW,\hs\OGM)$ and $(\OCP,\,\OGM)$.\  
We perform a $\chi^2$ analysis using Eq.\eqref{eq:chi2} and present the findings in Fig.\,\ref{fig:6}.\
Using the relations \eqref{eq:h-OBW}-\eqref{eq:h-OGM}
we find $\hs[\cut^{-4}_{\widetilde BW}]\!\propto\!h_3^Z\hs$
and $\hs[\cut^{-4}_{G-}]\!\propto\!h_3^\ga\hs$.\ 
So we expect that 
the $(\OBW,\hs\OGM)$ contour should be related to the 
$(h_3^Z\hsm ,\hs h_3^\ga)$ contour.\ 
Inspecting the contours in Figs.\,\ref{fig:4}(c)-(d)
and Figs.\,\ref{fig:6}(a)-(b), we see that they all
exhibit significant correlations, but in Figs.\,\ref{fig:6}(a)-(b)
the contours are aligned along different directions
from those of Figs.\,\ref{fig:4}(c)-(d).\ 
We can understand this difference in the following way.\ 
For convenience, we define 
$\hs(x,\hs y)\!=\!(r_3^Z\xt,\hs r_3^{\ga}\hs\yt)\hs$ with 
$\widetilde{X}\hsm\equiv\hsm (\xt,\hs \yt)
\hsm =\hsm ([\cut^{-4}_{\widetilde BW}],\hs 
[\cut^{-4}_{G-}])$.\ 
With these and using Eq.\eqref{eq:chi2-h3Z-h3A-xy},
we express the leading terms of the cross section 
$\hs\sigma_2^{}\hs$ as follows:
%
\beqs 
\label{eq:sigma2-xyt-d8}  
\begin{align}
\sigma_2^{}(\xt,\hs \yt) &\hs =\hs  
A\hs \xt^2\! +\hsm B\hs\yt^2\! +\hsm 2\hs C\hs \xt\hs \yt
\hs =\hs\widetilde{X}\hs V^{-1}\widetilde{X}^T , 
\\[1mm]
(A,\hs B,\hs C) &\hs =\hs 
\(\sbar^2(r_3^Z)^2\hs\bar{\sigma}_{2Z}^{33},\,
\sbar^2(r_3^\ga)^2\hs\bar{\sigma}_{2A}^{33},\,
\fr{1}{2}\sbar^2r_3^Zr_3^\ga\hs\bar{\sigma}_{2ZA}^{33}\)\!,
\end{align}	
\eeqs 
where the matrix $V^{-1}$ takes the form of 
Eq.\eqref{eq:Vinv-ABC}.\  From the above, 
we compute the correlation parameter \eqref{eq:rho} 
for the operators $(\OBW,\hs\OGM)$ as follows:
\beqa
\label{eq:rho-h3Z-h3A}
\rho\big([\cut^{-4}_{\widetilde BW}],
[\cut^{-4}_{G-}]\big) \hs =\hs -\text{sign}(r_3^Zr_3^\ga)\hs
\frac{\bar{\sigma}_{2ZA}^{33}\hs\sbar^0}
{~2\sqrt{\bar{\sigma}_{2Z}^{33}\hs\bar{\sigma}_{2A}^{33}\,}~}
=\, \text{sign}(r_3^Zr_3^\ga)\hs\rho(h_3^Z\hsmx , h_3^\ga)
\hs .
\eeqa 
Because Eq.\eqref{eq:r4-r3} gives 
$\hs\text{sign}(r_3^Zr_3^\ga) \!<\! 0\hs$,
we deduce 
$\hs\rho\big([\cut^{-4}_{\widetilde BW}],
[\cut^{-4}_{G-}]\big)\hsm\!=\!
-\rho(h_3^Z\hsmx , h_3^\ga)\!=\!O(\sbar^0)\!\!>\!0\hsx$.\ 
This explains why the contours of 
$\hs (\OBW,\hs\OGM)\hs$ in Figs.\,\ref{fig:6}(a) and (b) 
exhibit strong correlations [similar to those in 
Figs.\,\ref{fig:4}(c) and (d)], but have their orientations 
aligned towards the upper-right quadrant [unlike Figs.\,\ref{fig:4}(c) and (d), in which all the contours
are oriented towards the upper-left quadrant].

\vspace*{1mm}

Finally, we examine the correlations of the
fermionic contact operator $\OCP$ with the nTGC operators
$\OBW$ and $\OGM$.\ 
Since
$\OCP$ is a combination of two other operators
$\hs\OCP\!\!=\hsm\OGM\!\hsm -\OBW\hs$
viathe equation of motions \eqref{eq:OCP=OGM-OBW},
it is connected to both of the form factors $(h_3^Z,\hs h_3^\gamma)$, which
would complicate the correlation analysis in the form factor formulation
\eqref{eq:CS2-qq-Zgamma-H34}.\ 
Instead, we analyze directly the contributions of
the operators $(\OCP,\hs\OGM,\hs\OBW)$ to
the helicity amplitudes 
\eqref{eq:T8-OG-OBW-OC+}-\eqref{eq:Coup-OG-OBW-OC+}.\
As shown by Eq.\eqref{eq:Coup-OG-OBW-OC+}, the operator
$\OCP$  has a nonzero left-handed coupling
$\hs c_{L(C+)}'\!\!=\!-T_3^{}\hsx$ only.\ 
So for examining its correlations
with $\OGM$ and $\OBW$, the contributions
of $\OGM$ and $\OBW$ from the left-handed-quark couplings 
$\hs c_{L(G-)}'\hs$ and $\hs c_{L(\widetilde BW)}'\hs$
play key roles.\ 
Thus, we can express as follows the relevant helicity amplitudes
\eqref{eq:T8-OG-OBW-OC+}-\eqref{eq:Coup-OG-OBW-OC+}
containing left-handed (right-handed) 
initial-state quarks:
\beqs 
\label{eq:T8LR}
\begin{align}
\label{eq:T8L}
\TT_{8L}^{} &\,=\, \over{\TT}_{\!8L}^{}\!\times\!
\big\{c_{L(C+)}'[\cut_{C+}^{-4}]
+c_{L(\!\widetilde BW)}'
[\cut_{\widetilde BW}^{-4}]
+c_{L(G-)}' [\cut_{G-}^{-4}]\big\}
\nn\\[0.5mm]
&\,=\, \over{\TT}_{\!8L}^{}\hs 
(f_{L0}^{}\hs x + f_{L1}^{}\hs y_1^{} 
+f_{L2}^{}\hs y_2^{})\,,
\\[1mm]
\label{eq:T8R}
\TT_{8R}^{} &\,=\, \over{\TT}_{\!8R}^{}\!\times\!
\big\{ c_{R(\!\widetilde BW)}'
[\cut_{\widetilde BW}^{-4}]
+c_{R(G-)}' [\cut_{G-}^{-4}]\big\}
\hs\equiv\hs \over{\TT}_{\!8R}^{}\, (f_{R1}^{}\hs y_1^{} 
\!+\!f_{R2}^{}\hs y_2^{})\,,
\end{align}
\eeqs 
where $\over{\TT}_{8L}^{}$ (or $\over{\TT}_{8R}^{}$)
is the remaining common part of the helicity amplitudes
\eqref{eq:T8-OG-OBW-OC+}-\eqref{eq:Coup-OG-OBW-OC+} 
after separating out the coupling $\hs c_{Lj}'$ 
(or $\hs c_{Rj}'$) and
the cutoff factor $[\cut_j^{-4}]$.\ 
In the above, we have defined
$\hs (x,\hs y_1^{},\hs y_2^{}) \! \equiv \!
([\cut_{C+}^{-4}],\hs [\cut_{\widetilde BW}^{-4}],\hs
[\cut_{G-}^{-4}])\hs$ and
\beqs
\label{eq:fL-fR}
\begin{align}
f_{L0}^{} &=c_{L(C+)}'\!=-T_3^{}
\hs,~~~
f_{L1}^{} \!=c_{\!L(\widetilde BW)}'\!
=T_3^{}\!-\hsm Qs_W^2 \hs,~~~
f_{L2}^{} \!=c_{L(G-)}' \!= -Qs_W^2 \hs ,
\\[1mm]
f_{R1}^{} &=c_{\!R(\widetilde BW)}'\!
=-Qs_W^2 \hs,~~~
f_{R2}^{} =c_{R(G-)}'\!= c_{L(G-)}' \!=f_{L2}^{}
\equiv f_{2}^{}\, .
\end{align}
\eeqs
%
With the above, we perform a $\chi^2$ analysis based upon 
Eq.\eqref{eq:chi2}.\ We present the correlation contours
of $(\OCP,\hs \OBW)$ and $(\OCP,\hs \OGM)$ in 
Figs.\,\ref{fig:6}(c) and (d) for the LHC and 
the 100\,TeV $pp$ collider, respectively.\
We find that all these contours exhibit strong correlations.\ 
In particular, the $(\OCP,\hs \OBW)$ contours (solid curves)
are oriented towards the upper-right quadrant, whereas the
$(\OCP,\hs \OGM)$ contours (dashed curves)
are oriented towards the upper-left quadrant.\ 

\vspace*{1mm}

To understand the qualitative features of the correlation contours in 
Figs.\,\ref{fig:6}(c) and (d), we examine the cross section
$\sigma_2^{}$, which contains the squared part of the dimension-8 contributions and dominates the $\chi^2$ function.\ 
From Eq.\eqref{eq:T8LR}, we derive the cross section
$\sigma_2^{}$ as follows:
\vspace*{-1.5mm}
\begin{align}
\sigma_2^{}(x,y_1^{},y_2^{}) &\,=\,
(f_{L0}^{}\hs x \hsm +\! f_{L1}^{}\hs y_1^{} \!+\! f_2^{}\hs y_2^{})^2\!\left<|\over{\TT}_{\!8L}^{}|^{\,2}\right>
+(f_{R1}^{}\hs y_1^{} \!+\! f_{2}^{}\hs y_2^{})^2\! 
\left<|\over{\TT}_{\!8R}^{}|^{\,2}\right> 
\nn\\
&\,=\, \left[(f_{L0}^{}\hs x \hsm +\! f_{L1}^{}\hs y_1^{} \!+\! f_2^{}\hs y_2^{})^2\!
+\!(f_{R1}^{}\hs y_1^{} \!+\! f_{2}^{}\hs y_2^{})^2\right]\! 
\left<|\over{\TT}_{\!8}^{}|^{\,2}\right> ,
\label{eq:sigma2-C+BWG-}
\end{align}
where we have defined the notations 
$\left<|\over{\TT}_{\!8L}^{}|^{\,2}\right>\!=\!\int_{\rm{PS}}^{}
|\over{\TT}_{\!8L}^{}|^{\,2}$ and
$\left<|\over{\TT}_{\!8R}^{}|^{\,2}\right>\!=\!\int_{\rm{PS}}^{}
|\over{\TT}_{\!8R}^{}|^{\,2}$ 
with $\int_{\rm{PS}}^{}$
denoting the phase space integration
for the final state.\ 
From the squared term of the cross section
\eqref{eq:CS-qqZA-d8other},
we can further deduce the equality
$\left<|\over{\TT}_{\!8L}^{}|^{\,2}\right>\!=\!
\left<|\over{\TT}_{\!8R}^{}|^{\,2}\right>
\!\equiv\!\left<|\over{\TT}_{\!8}^{}|^{\,2}\right>$,
which is used in the last step of Eq.\eqref{eq:sigma2-C+BWG-}.

\vspace*{1mm}

For analyzing the correlations, the overall factor 
$\left<|\over{\TT}_{\!8}^{}|^{\,2}\right>$
is irrelevant.\ So we define the following rescaled cross sections
for the convenience of analyzing the two-parameter
correlations:
\beq
\bar\sigma_2^{}(x,y_1^{}) \equiv  
\sigma_2^{}(x,y_1^{},0)/\!\left<|\over{\TT}_{\!8}^{}|^{\,2}\right>
\hsm,
~~~~
\bar\sigma_2^{}(x,y_2^{}) \equiv 
\sigma_2^{}(x,0,y_2^{})/\!\left<|\over{\TT}_{\!8}^{}|^{\,2}\right>
\hsm.
\eeq 
Thus, $\hs\bar\sigma_2^{}(x,y_1^{})\hs$ and 
$\hs\bar\sigma_2^{}(x,y_2^{})\hs$ are expressed in the following
quadratic form:
\beqs 
\begin{align}
\bar\sigma_2^{}(x,y_1^{}) &=\,
A\hs x^2 + B_1^{}y_1^2 + 2\hs C_1^{}\hs x\hs y_1^{}
\equiv\hs \XX_1^{}V_1^{-1}\hs\XX_1^T\,,
\\
\bar\sigma_2^{}(x,y_2^{}) &=\,
A\hs x^2 + B_2^{}y_2^2 + 2\hs C_2^{}\hs x\hs y_2^{}
\equiv\hs \XX_2^{}V_2^{-1}\hs\XX_2^T\,,
\end{align}
\eeqs
where we have defined
$\hs\XX_1^{}\! \equiv \! (x,\hs y_1^{})\hs$ and
$\hs\XX_2^{}\! \equiv \! (x,\hs y_2^{})\hs$ 
as well as the following notations,
\vspace*{-2mm}
\beqs 
\label{eq:AB1C1-AB2C2-V1V2}
\begin{align} 
(A,\hs B_1^{},\hs C_1^{}) & \equiv (f_{L0}^2,\,
f_{L1}^2\!+\!f_{R1}^2,\, f_{L0}^{}f_{L1}^{})\hs,  
\\
(A,\hs B_2^{},\hs C_2^{}) & \equiv (f_{L0}^2,\,
2\hs f_{2}^2,\, f_{L0}^{}f_{2}^{})\hs,  
\\[1mm]
V_1^{-1} &=\!
\(\!\!\!\ba{cc} A\, & C_1^{} \\ \hs C_1^{} & B_1^{}\ea\hsm\!\!\)\!,
~~~~
V_2^{-1} =\!
\(\!\!\!\ba{cc} A\, & C_2^{} \\ \hs C_2^{} & B_2^{}\ea\hsm\!\!\)\!.
\end{align}
\eeqs
Thus, we can deduce the following correlation parameter for the two cases:
\vspace*{-1mm}
\beqs 
\label{eq:rho-xy1-xy2}
\begin{align} 
\label{eq:rho-xy1}
\rho_1^{}(x,y_1^{}) &\,=\hs 
\frac{-\hs C_1^{}}{\,\sqrt{\hsm AB_1^{}}\,}
=\frac{~-\hs\text{sign}(f_{L0}^{}f_{L1}^{})~}
{\,\sqrt{1\!+\!f_{R1}^2/\!f_{L1}^2\,}\,}> 0\,,
\\[1mm]
\label{eq:rho-xy2}
\rho_2^{}(x,y_2^{}) &\,=\hs 
\frac{-\hs C_2^{}}{\,\sqrt{\hsm AB_2^{}}\,}
=-\frac{1}{\sqrt{2\,}\,}\hs\text{sign}(f_{L0}^{}f_{2}^{})
< 0\,,
\end{align}
\eeqs 
where  
$\hs (x,\hs y_1^{})\hsm \equiv \hsm ([\cut_{C+}^{-4}],\hs 
[\cut_{\widetilde BW}^{-4}])\hs$ and 
$(x,\hs y_2^{})\hsm \equiv \hsm
([\cut_{C+}^{-4}],\hs [\cut_{G-}^{-4}])$.\ 
Using the coupling formula \eqref{eq:fL-fR}, we derive
$\hs f_{L0}^{}\hs f_{L1}^{}\!=\!-T_3^{}\hs (T_3^{}- Qs_W^2)
\!<\hsm 0\,$ and 
$\hs f_{L0}^{}f_2^{}\!=\!T_3^{}Q\hs s_W^2
\!>\hsm 0\,$, where each inequality holds for both
up-type and down-type quarks.\
From these, we deduce that the operators
$(\OCP,\hs \OBW)$ are correlated positively, whereas the
operators $(\OCP,\hs \OGM)$ are correlated negatively.\ 
Moreover, Eq.\eqref{eq:rho-xy1-xy2} shows that 
both correlation parameters are of $\hs O(\sbar^0)\hs$ 
and not suppressed by any energy factor.\ 
This predicts strong correlations for the operators 
$(\OCP,\hs \OBW)$ and $(\OCP,\hs \OGM)$, respectively.\ 
These features are indeed reflected in  
Figs.\,\ref{fig:6}(c) and (d).\ 
We see that the correlation contours of $\,(\OCP,\,\OBW)$ 
(solid curves) are oriented towards the upper-right quadrant
due to the positive correlation parameter
$\hs\rho_1^{}(x,y_1^{})\!>\!0\,$ given by Eq.\eqref{eq:rho-xy1},
whereas the correlation contours of $(\OCP,\,\OGM)$ (dashed curves) 
are aligned towards the upper-left quadrant
due to the negative correlation parameter
$\hs\rho_2^{}(x,y_2^{})\!<\!0\,$ 
given by Eq.\eqref{eq:rho-xy2}.\

\vspace*{1mm}
\subsection{\hspace*{-2.5mm}Comparison with the Existing LHC Bounds on nTGCs}
\label{sec:4.7}
\vspace*{1mm}

In this subsection, we make direct comparison with the
published LHC measurements of nTGCs 
through the reaction 
$\hs p{\hs}p{\hs}(q{\hs}\bar{q})\hsm\ito\hsm Z\gamma\hs$
with $Z\!\ito\nu{\hs}\bar{\nu}{\hs}\hs$ 
by the ATLAS\,\cite{Atlas2018nTGC-FF} and 
CMS\,\cite{CMS2016nTGC-FF} collaborations
using the conventional nTGC form factor formula 
\eqref{eq:FF0-nTGC}.\ 
The CMS collaboration analyzed 19.6\,fb$^{-1}$ of Run-1 data 
at $\sqrt{s\,}\! =\! 8$\,TeV
\cite{CMS2016nTGC-FF},
whereas the ATLAS collaboration analyzed 36.1\,fb$^{-1}$ of Run-2 data
at $\sqrt{s\,}\hsm =\hsm 13$\,TeV
\cite{Atlas2018nTGC-FF}. They obtained the following
sensitivity bounds (95\%\,C.L.) on the form factors:
\beqs 
\label{eq:LHC-exp-h3h4}
\begin{align} 
\hspace*{-15mm}
\label{eq:CMS-h3h4}
\hspace*{-15mm}
\text{CMS:} \hspace*{6mm}
& h_3^Z\in (-1.5,\,1.6)\!\times\! 10^{-3},
\hspace*{-15mm}
& h_3^\ga\in (-1.1,\,0.9)\!\times\! 10^{-3},
\nn\\
\hspace*{-15mm}
& h_4^Z\in (-3.9,\,4.5)\!\times\! 10^{-6},
\hspace*{-15mm}
& h_4^\ga\in (-3.8,\,4.3)\!\times\! 10^{-6};
\\[0.5mm]
\label{eq:Atlas-h3h4}
\text{ATLAS:} \hspace*{6mm}
& h_3^Z\in (-3.2,\,3.3)\!\times\! 10^{-4},
\hspace*{-15mm}
& h_3^\ga\in (-3.7,\,3.7)\!\times\! 10^{-4},
\nn\\
\hspace*{-15mm}
& h_4^Z\in (-4.5,\,4.4)\!\times\! 10^{-7},
\hspace*{-15mm}
& h_4^\ga\in (-4.4,\,4.3)\!\times\! 10^{-7}.
\end{align}
\eeqs
We see that the CMS and ATLAS  analyses both obtained 
much stronger bounds on $(h_4^Z,\,h_4^\ga)$ 
than on  $(h_3^Z,\,h_3^\ga)$, i.e., by factors 
$\sim\!(210-380)$ at CMS (Run-1) 
and $\sim\!(710\hsm -\hsm 860)$ at ATLAS (Run-2).\  
In comparison, we see in Table\,\ref{tab:5}
using our SMEFT form factor formulation
\eqref{eq:FF2-nTGC} that the LHC
sensitivity bounds on $h_4^V$ are stronger than those
on $\hs h_3^V$ only by factors of about $20\hs$.
Our Table\,\ref{tab:6} further demonstrates that using
the conventional form factor formulation \eqref{eq:FF0-nTGC}
would generate spuriously stronger $h_4^V$ bounds 
(marked in blue) at the LHC\,(13{\hs}TeV)
than the SMEFT bounds (marked in red) 
by a factor of about $20$, 
and thus much stronger than
the $h_3^V$ bounds by a large factor of 
$\sim\!20\!\times\!20=400\hs$, 
which agrees with the ATLAS results in Eq.\eqref{eq:Atlas-h3h4}
within a factor of $2\hsx$.\footnote{%
Since our analyses 
in Tables\,\ref{tab:5}-\ref{tab:6} have used as input the full
Run-2 integrated luminosity of $140\,$fb$^{-1}$ as well as
different kinematic cuts for each bin, unlike 
the experimental analyses
of ATLAS\,\cite{Atlas2018nTGC-FF} and CMS\,\cite{CMS2016nTGC-FF},
such a minor difference in
the bounds could be expected.}\ 
Unfortunately, this means that the strong experimental bounds 
\eqref{eq:LHC-exp-h3h4} on $(h_4^Z,\,h_4^\ga)$
are {\it unreliable} because they were obtained
by using the conventional form factor formulation
\eqref{eq:FF0-nTGC}, which does not respect the SM electroweak
gauge symmetry of 
SU(2)$_{\rm L}^{}\otimes\hs$U(1)$_{\rm Y}^{}$
as incorporated in the SMEFT.\

\vspace*{1mm}

To study quantitatively the conventional parametrization \eqref{eq:FF0-nTGC} 
including the nTGC form factors $(h_3^V,\hs h_4^V)$ only, 
we denote their contributions to the total cross section by
$\,\widetilde{\sigma}(Z\ga )\!
=\sigma_0^{}+\widetilde{\sigma}_1^{}+\widetilde{\sigma}_2^{}\,$,
where  $\widetilde{\sigma}_1^{}$ is the interference term
and $\widetilde{\sigma}_2^{}$ is the squared contribution.\ 
This is similar to what we did around Eq.\eqref{eq:Xsigma12-dim8}.\ 
We find that $\widetilde{\sigma}_2^{}$ always dominates over  
$\widetilde{\sigma}_1^{}$ for both the LHC and the 100{\,}TeV $pp$ collider.\ 
Using the conventional form factor formula \eqref{eq:FF0-nTGC}, 
we derive the squared contribution $\widetilde{\sigma}_2^{}$
as follows:
\begin{equation}
\widetilde\sigma_2=\frac{~e^4 (X_{LR}^V)^2 
(\hat s -M_Z^2)^3 \!\hsm\left[4 (h_3^V)^2\! 
\(\!\frac{M_Z^2}{\hat s}\hsm +\hsm 1\!\)\!+\hsm 4\hs h_3^V h_4^V  \!\(\!1\!-\!\frac{\hat{s}}{M_Z^2}\!\)\!+\!
(h_4^V)^2\!\(\!1\!-\!\frac{\hat{s}}{M_Z^2}\!\)^{\!\!2}\right]~}
{768\hs\pi\hs s_W^2c_W^2 M_Z^{6}\,\hat{s} },
\end{equation}
where the coupling factor $(X_{LR}^V)^2$ is defined as
\beqs 
\label{eq:XLR-V}
\begin{align}
& (X_{LR}^V)^2 \equiv (X_L^V)^2\!+\!(X_R^V)^2,~~~~
X_{L,R}^Z \equiv x_{L,R}^Z\,,~~~~
X_{L,R}^\gamma \equiv -\frac{c_W^{}}{s_W^{}}x_{L,R}^A\,,
\hspace*{10mm}
\\
& (x_L^Z,\hs x_R^Z)=(T_3^{}\!-\! Qs_W^2,\hs -Qs_W^2)\hs,~~~~
(x_L^A,\hs x_R^A)=-Qs_W^2(1,\hs 1)\hs.
\end{align}
\eeqs  
Defining a scaled dimensionless energy parameter 
$\,\sbar = \shat /M_Z^2\,$ 
and making the high-energy expansion for $\,\sbar\!\gg\! 1\,$,
we can compare the leading energy-dependence of each term of  $\,\widetilde{\sigma}_2^{}$
with that of $\,\sigma_2^{}\hs$, as follows:
\beqs
\label{eq:EXPD-sigma2T-sigma2}
\begin{align}
\label{eq:EXPD-sigma2T}
\widetilde\sigma_2^{} &\,\approx\,
\frac{~e^4 (X_{LR}^V)^2 \!
\left[(h_4^V)^2\hs\sbar^4
\!-\hsm 4\hs h_4^Vh_3^V\sbar^3
\!+\hsm 4(h_3^V)^2\hs\sbar^2\hs \right]~}
{768\hs\pi\hs s_W^2\hs c_W^2M_Z^2} \hs ,
\\[2mm]
\label{eq:EXPD-sigma2}
\sigma_2^{} &\,\approx\,
\frac{~e^4 \!\left[T_3^2(h_4^Z)^2\hs\sbar^3
\!+8\hs T_3^{}X_L^V h_4^Zh_3^V\hs\sbar^2
\!+\!4\hs (X_{LR}^V)^2(h_3^V)^2\hs\sbar^2\hs 
\right]~}
{~768\hs\pi s_W^2\hs c_W^2M_Z^2\,}\hs ,
\end{align}
\eeqs
where the cross section $\,\sigma_2^{}\hs$ is given by 
our SMEFT form factor formula \eqref{eq:FF2-nTGC}.\ 
We note that the form factors $(h_4^Z,\,h_4^\ga)$
in the above cross section $\hs\sigma_2^{}\hs$ should
obey the condition \eqref{eq:h4Z-h4A} due to the underlying
electroweak gauge symmetry of the SM that is respected by the
corresponding dimension-8 nTGC operators.\ 
We have used the relation \eqref{eq:h4Z-h4A} to 
combine the $\hs h_4^\ga\hs$ contribution with that of 
$\hs h_4^Z\hs$.\ 
To examine the correlation of $(h_3^\ga,\,h_4^\ga)$
from Eq.\eqref{eq:EXPD-sigma2},
we can use Eq.\eqref{eq:h4Z-h4A} to replace 
$\hs h_4^Z\hs$ by $\hs h_4^\ga\hs$.\ 
Inspecting Eq.\eqref{eq:EXPD-sigma2T-sigma2}, we see that  
both the $(h_4^V)^2$ and $(h_4^Vh_3^V)$ terms 
in $\hs\widetilde\sigma_2^{}$
have higher energy dependences than those of
$\sigma_2^{}$ by an extra factor $\hs\sbar^1$,
which leads erroneously to much stronger bounds on 
$h_4^V\hs$.\

\vspace*{1mm}

We first make a one-parameter analysis and derive the
bound on each form factor coefficient
$\,h_j^V$ individually (where $j\!=\!3,4\hs$ and $\hs V\!=\!Z,\ga\,$)  
using the conventional form factor parametrization
\eqref{eq:FF0-nTGC}.\ 
To make a more precise comparison with the ATLAS bounds 
\eqref{eq:Atlas-h3h4}, 
we adopt the same kinematic cut on the transverse momentum
of the final-state photon, $P_T^\gamma$\,>\,600\,GeV,
and the same integrated luminosity 
$\hs\mL\hsm =\hsm 36.1$\,fb$^{-1}$
as in the ATLAS analysis\,\cite{Atlas2018nTGC-FF}.\
For illustration, 
we ignore the other detector-level cuts 
and the systematic errors,
and choose a typical detection efficiency 
$\,\ep\!=\!75\%\hsx$.\footnote{%
We thank our ATLAS colleague Shu Li for discussing the typical
detection efficiency of the ATLAS 
detector\,\cite{Atlas2018nTGC-FF}.}\ 
\,With these, we derive the following bounds on the nTGCs 
(95\%\,C.L.) when using the conventional form factor
parametrization \eqref{eq:FF0-nTGC}:
%
\beq
\label{eq:Atlas-Xh34-theory}
|h_3^Z|< 3.0\!\times\! 10^{-4},
\hspace*{5mm}
|h_3^\ga|<3.4\!\times\! 10^{-4},
\hspace*{5mm}
|h_4^Z|<4.4\!\times\! 10^{-7},
\hspace*{5mm}
|h_4^\ga|< 4.9\!\times\! 10^{-7},
\eeq 
%
and note that the squared 
nTGC contributions dominate the sensitivity.\ 
Comparing the above estimated bounds \eqref{eq:Atlas-Xh34-theory}
with the ATLAS experimental bounds \eqref{eq:Atlas-h3h4},
we see that they agree well with each other:
the agreements for $h_4^Z$ are within about 
$2\%\hsx$ and
the agreements for $(h_4^\ga,h_3^Z,\hs h_3^\ga)$ 
are within about $(8\!-\!13)\%\hsx$.\   
This means that by making plausible simplifications  
we can reproduce quite accurately the experimental bounds
\eqref{eq:Atlas-h3h4}
established by the ATLAS collaboration\,\cite{Atlas2018nTGC-FF} using the conventional form factor 
formulation in Eq.\eqref{eq:FF0-nTGC}.

\vspace*{1mm}

\begin{figure}[]
\centering
\includegraphics[height=7.2cm,width=7.2cm]{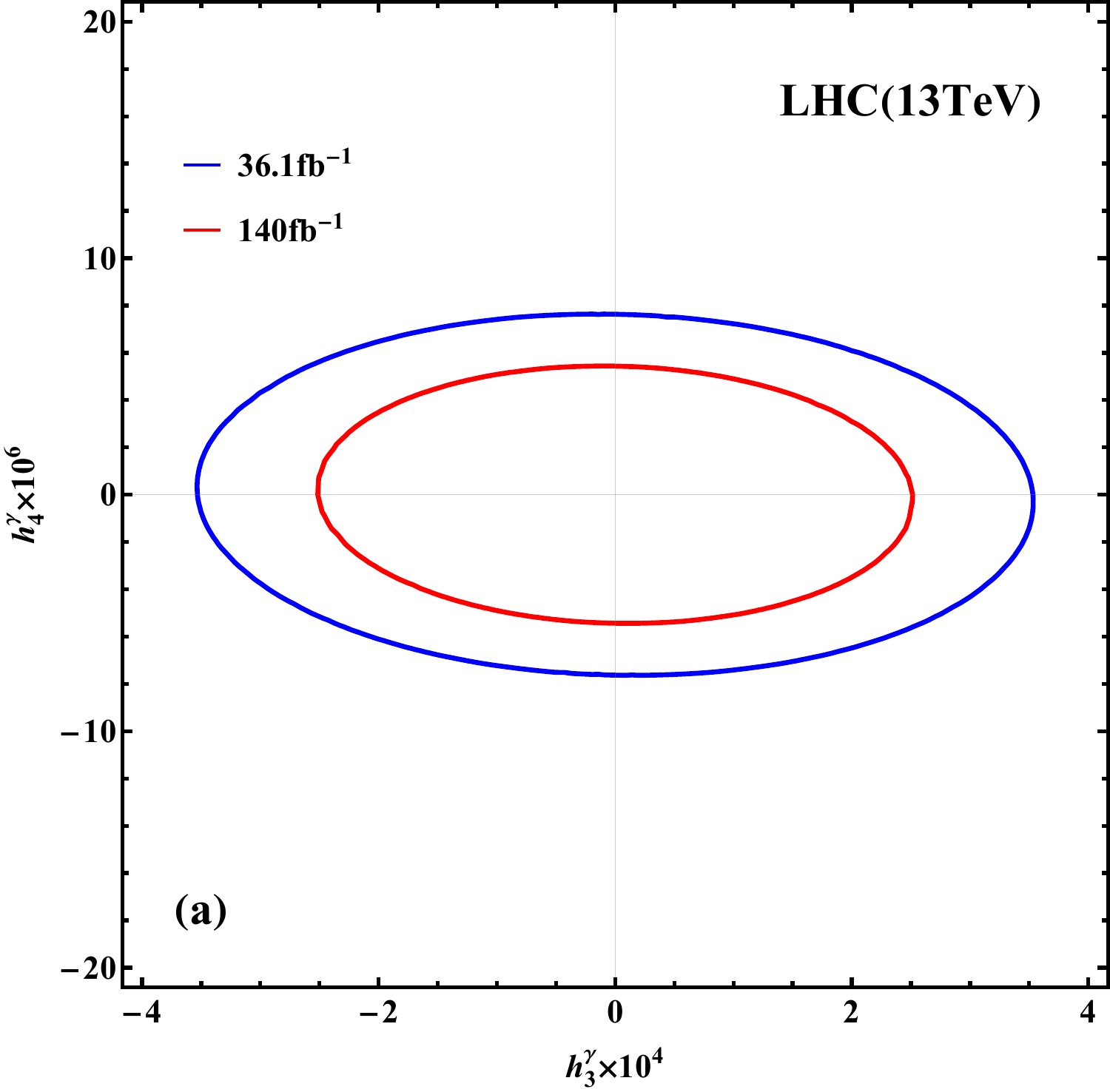}
\hspace*{2mm}
\includegraphics[height=7.2cm,width=7.2cm]{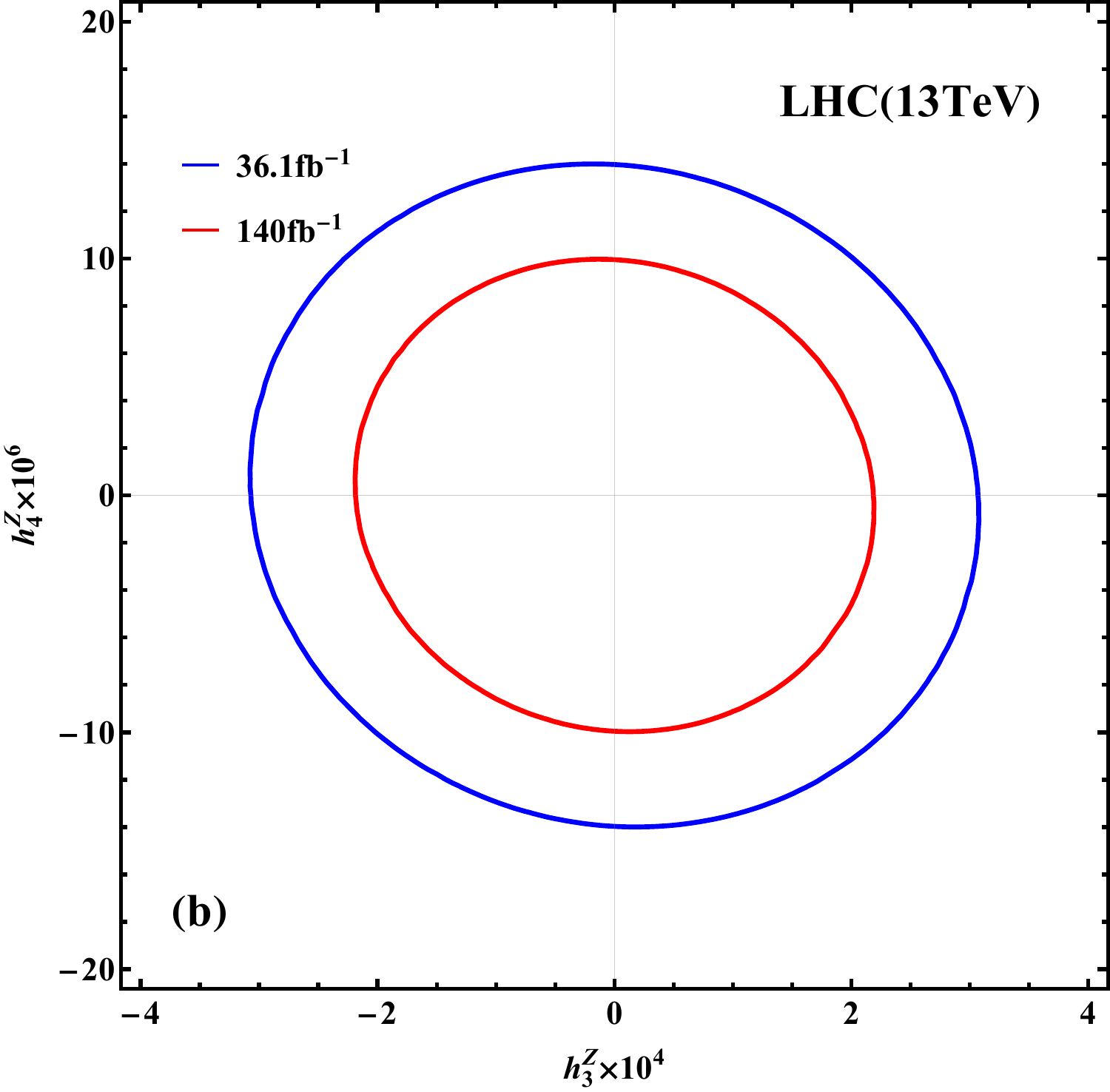}
\\[1mm]
\includegraphics[height=7.2cm,width=7.2cm]{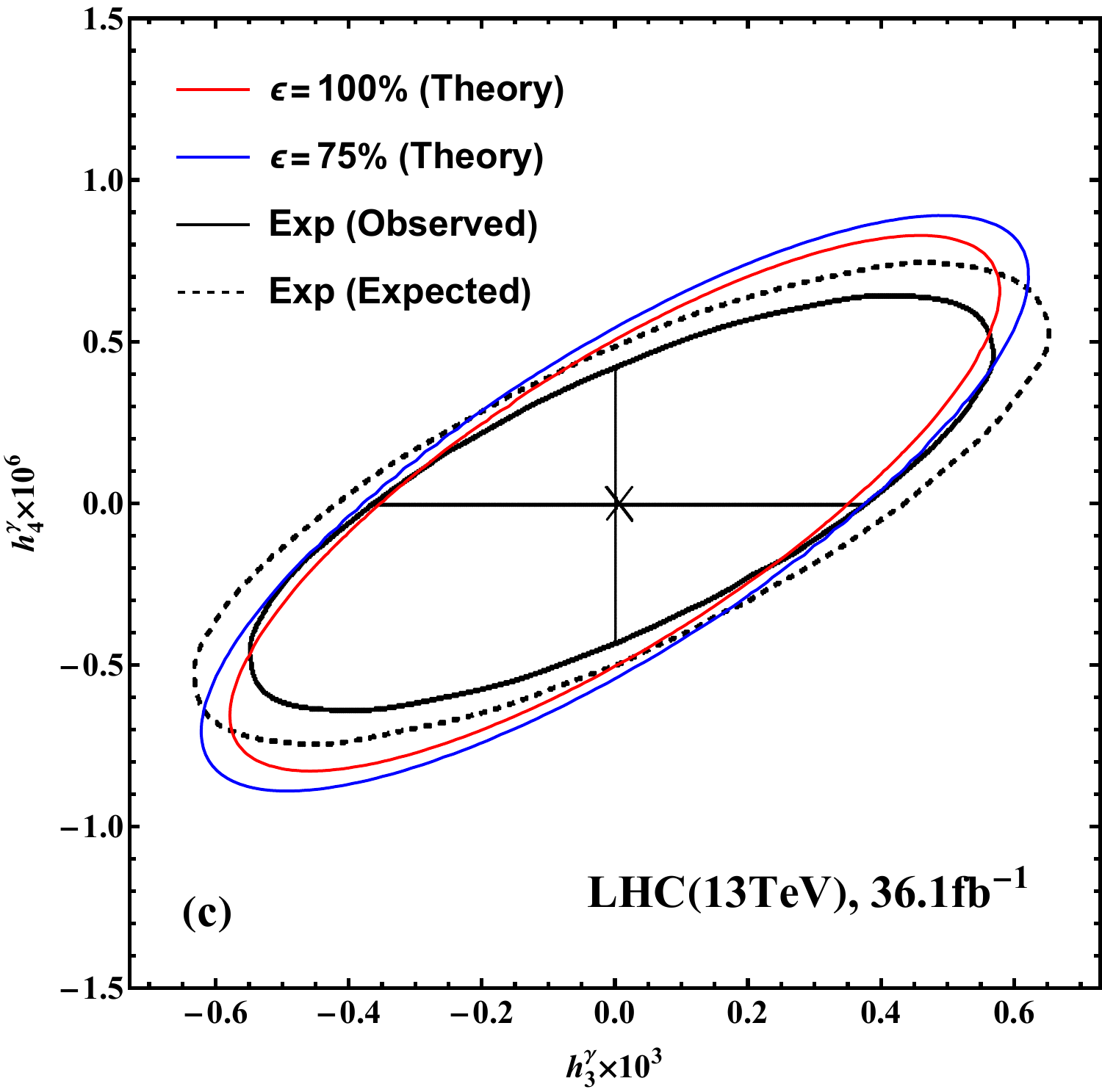}
\hspace*{2mm}
\includegraphics[height=7.2cm,width=7.2cm]{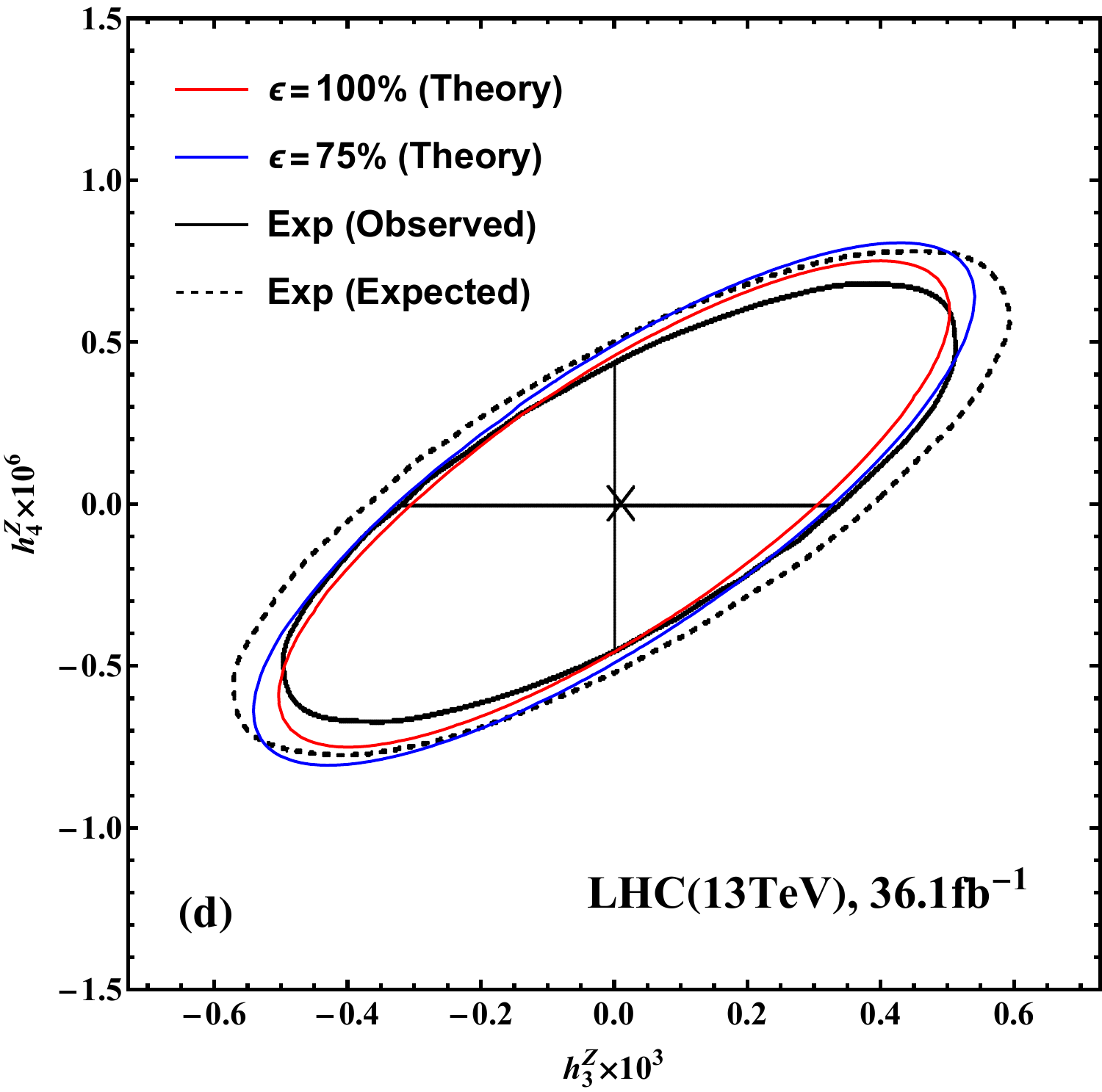}
\caption{\small\hspace*{-2mm}
{\it Correlation contours of the sensitivity reaches} 
(95\%\,C.L.) {\it for the indicated pairs of 
	nTGC form factors at the LHC}\,(13\,TeV).\ 
{\it Panels\,(a) and (b) present  the 
	correlation contours for $(h_3^\ga ,\,h_4^\ga)$ and
	$(h_3^Z,\,h_4^Z)$ respectively, by using our SMEFT form factor formula \eqref{eq:FF2-nTGC}, where in each panel the red contour 
	inputs the full integrated luminosity}  
140\,fb$^{-1}\!\!$ {\it of Run-2 and the blue contour inputs 
	a partial integrated luminosity} 36.1\,fb$^{-1}\!$ 
{\it as in the ATLAS analysis\,\cite{Atlas2018nTGC-FF}.}\ 
{\it Panels\,(c) and (d) compare the theoretical correlation 
	contours (red and blue colors) with the experimental contours 
	(black color) from the ATLAS analysis\,\cite{Atlas2018nTGC-FF},
	where we derived the red and blue contours  
	by using the conventional form factor formula \eqref{eq:FF0-nTGC}
	and by assuming an ideal detection efficiency $\ep\!=\!100\%$ 
	(for red contours) or a reduced detection efficiency 
	$\hs\ep\!=\!75\%$ (for blue contours).\ 
	The ATLAS contours are shown by the black solid curves (observed) 
	and the black dashed curves (expected). 
}}
\label{fig:10}
\end{figure}

Next, we analyze the correlation contours for
$(h_3^\ga,\hs h_4^\ga)$ and $(h_3^Z,\hs h_4^Z)$,
respectively, using the conventional form factor 
parametrization \eqref{eq:FF0-nTGC},
which can be compared to the correlation contours
obtained by using our SMEFT form factor formulation
\eqref{eq:FF2-nTGC}.\ 
Fig.\,\ref{fig:10} displays the correlation contours at 95\%\,C.L.\ 
for LHC Run-2.\  
Panels (a) and (b) show the correlation contours 
based on the SMEFT form factor formula \eqref{eq:FF2-nTGC}, 
where the blue (red) contours correspond to inputting 
integrated LHC luminosities of $36.1\hs$fb$^{-1}$\,($140\hs$fb$^{-1}$).\ 
Panels (c) and (d) present the correlation contours 
based on the conventional form factor 
parametrization \eqref{eq:FF0-nTGC},
where the red and blue contours are given by our theoretical
analysis with the assumed detection efficiencies 
$\hs\ep \!=\!100\%\hs$ and $\hs\ep \!=\!75\%\hs$ respectively.
For comparison, we show in panels (c) and (d) the experimental contours 
as extracted from 
the ATLAS results\,\cite{Atlas2018nTGC-FF}
based on the conventional form factor formula \eqref{eq:FF0-nTGC},
where the black solid curves depict the observed
bounds and the black dashed curves show the 
expected limits.\ 
It is impressive to see in panels (c) and (d)
that our theoretical contours agree well with 
the experimental contours
obtained by using the conventional form factor 
parametrization \eqref{eq:FF0-nTGC}.

\vspace*{1mm}  

We note that the correlation contours 
of panels (a) and (b) in Fig.\,\ref{fig:10} 
have very different features from 
those of panels (c) and (d), which can be understood as follows.\  
For convenience, we denote 
$\hs \XX\!=\!(x,\,y)\!\equiv\!
(h_3^V\hsmx ,\hs h_4^Z)$.\ 
Thus, we can express the cross sections of
Eqs.\eqref{eq:EXPD-sigma2T} and \eqref{eq:EXPD-sigma2} as follows:
\vspace*{-2mm}
\beqs 
\label{eq:sigma2T-sigma2-xy}
\begin{align} 
\label{eq:sigma2T-xy}
\widetilde\sigma_2^{} &\,\propto\,
\widetilde{A}{x}^2\!+ \widetilde{B} {y}^2\!
+2\hs\widetilde{C}{x}\hs{y}
\hs\equiv\hs \XX\hs \widetilde{V}^{-1}\hs\XX^T ,
\\
\label{eq:sigma2-xy}
\sigma_2^{} &\,\propto\,
Ax^2 \!+ By^2
\!+ 2\hs Cx\hs y
\hs\equiv\hs \XX\hs V^{-1}\hs\XX^T ,
\end{align}
\eeqs
where we have defined the following notations,
\beqs 
\label{eq:ABCt-ABC-VtV}
\begin{align} 
(\widetilde{A},\hs \widetilde{B},\hs \widetilde{C}) 
&\,=\hs 
(4\hs\sbar^2\hsm,\hs\sbar^4\hsm,\hs -2\sbar^3)\hs,
\\
(A,\hs B,\hs C) &\,=\hs 
\(4\hs\sbar^2(X_{LR}^V)^2\hsm,\hs\sbar^3 T_3^2\hsm,\hs 
4\hs\sbar^2 X_L^VT_3^{}\)\hsmx ,
\\[0.5mm]
\widetilde{V}^{-1} &=\!
\(\!\!\!\ba{cc} \widetilde{A}\, & \widetilde{C} 
\\[0.4mm] 
\hs \widetilde{C} & \widetilde{B} \ea\hsm\hsm\!\)\!,
~~~~
V^{-1} =\!
\(\!\!\!\ba{cc} A\, & C 
\\[0.3mm]  
\hs C & B \ea\hsm\hsm\!\)\!.
\end{align}
\eeqs
With these we can compute the correlation parameter 
of the form factors in each case:
\vspace*{-1mm}
\beqa 
\label{eq:rho-xy-compare}
\widetilde{\rho}^{} \hs =
\frac{-\hs \widetilde{C}}
{\,\sqrt{\hsm \widetilde{A}\hs\widetilde{B}\,}\,}
= 1\! >\hsm 0\,,~~~~~
\rho \hs = 
\frac{-\hs C}{\,\sqrt{\hsm AB\,}\,}
=-\frac{\,2\hsx\text{sign}(T_3^{}) X_L^V\,}
{|X_{LR}^V|\,}\hs\sbar^{-\frac{1}{2}}
\!<\hsm 0\,.
\eeqa
%
The fact of $\hs\widetilde{\rho}^{}\!=\!O(\sbar^0)\!\!>\!0\hs$
explains why the $\hs(h_3^V\hsmx ,\hs h_4^V)\,$ contours
in Figs.\,\ref{fig:10}(c) and (d) exhibit strong 
correlations and have their orientations aligned towards the 
upper-right quadrant.\
On the other hand, from Eq.\eqref{eq:XLR-V} we find that 
$\hs\text{sign}(T_3^{})X_L^V\hsm\!>\!0\,$ 
holds for the initial-state quarks 
being either up-type or down-type, and thus
Eq.\eqref{eq:rho-xy-compare} gives $\hs\rho\!<\! 0\hsx$.\ 
This means that the $\hs(h_3^V\hsmx ,\hs h_4^V)\,$ contours
in Figs.\,\ref{fig:10}(a) and (b)
should have their orientations towards the upper-left quadrant, 
but this correlation is almost invisible because 
$\hs\rho\! =\! O(\sbar^{-\frac{1}{2}})\hs$
receives a large energy-suppression factor 
at the LHC.\ Thus, the correlation features of the 
$\hs(h_3^V\hsmx ,\hs h_4^V)\,$ contours are well understood 
both for Figs.\,\ref{fig:10}(a)-(b) [based on the SMEFT  
form factor formula \eqref{eq:FF2-nTGC}] 
and for Figs.\,\ref{fig:10}(c)-(d)
[based on the conventional form factor formula \eqref{eq:FF0-nTGC}].

\vspace*{1mm}

Our quantitative comparisons in Figs.\,\ref{fig:10}
are instructive and encouraging.\  
We suggest that the ATLAS and CMS colleagues 
perform a systematic nTGC analysis 
based on the new SMEFT form factor formula \eqref{eq:FF2-nTGC},
using the full Run-2 data set.\
Moreover, we note that in 
Refs.\,\cite{CMS2016nTGC-FF}-\cite{Atlas2018nTGC-FF}
the CMS and ATLAS collaborations analyzed the
correlations between the form factors
$(h_3^V,\,h_4^V)$ and found strong correlations.\ 
We have reproduced this feature in 
Figs.\,\ref{fig:10}(c)-(d), 
but we note that those correlation contours 
differ substantially from our new correlation contours 
in Figs.\,\ref{fig:10}(a)-(b).\ 
Based upon the above analysis, 
we suggest that the CMS and ATLAS collaborations should 
make updated analyses on the $(h_3^V,\,h_4^V)$ correlations
using our new SMEFT form factor formulation with their 
full Run-2 data sets.\  We anticipate that such new 
analyses should yield results similar to the 
theoretical predictions for LHC Run-2 given in
Table\,\ref{tab:5} and Figs.\,\ref{fig:10}(a)-(b).

\vspace*{1mm}
\section{\hspace*{-2.5mm}Comparison with Probes of nTGCs at Lepton Colliders}
\label{sec:5}
\vspace*{1mm}

In this Section we first summarize the sensitivity reaches
of nTGC new physics scales 
at high-energy $e^+e^-$ colliders 
found in our previous work\,\cite{Ellis:2020ljj}.\
Then we analyze the sensitivity reaches of the
nTGC form factors at these $e^+e^-$ colliders.\ 
Finally, we compare these sensitivity limits with those
obtained for the hadron colliders as given in
Section\,\ref{sec:4} of the present study.\ 
%

\vspace*{1mm}

At high-energy $e^+e^-$ colliders, 
we found in Ref.\,\cite{Ellis:2020ljj} that the reaction
$\,e^+e^-\!\ito Z\ga\,$ with hadronic decays 
$\hs Z\!\ito q\hs\bar{q}\,$ gives greater sensitivity reach
than the leptonic and invisible decays 
$\hs Z\ito \ell\hs\bar{\ell},\nu\hs\bar{\nu}\hs$.
Therefore we choose for comparison the sensitivity reaches obtained using 
hadronic $Z$ decays, and consider the  $e^+e^-$ collision energies
$\sqrt{s}=(0.25,\hs 0.5,\hs 1,\hs 3,\hs 5)$~TeV with a benchmark
integrated luminosities $\mL=5\hs$ab$^{-1}$.\ 
These results are summarized in the upper half of Table\,\ref{tab:7}
for the new physics scale $\,\cut\,$ 
of each dimension-8 nTGC operator or related contact operator $(\OGP,\,\OGM,\,\OBW,\,\OCP )$ at the $2{\hs}\sigma$ level,
where each entry has two limits which correspond to the 
(unpolarized,\,polarized) $e^\mp$ beams.\ 
For the polarized $e^\mp$ beams, we choose the benchmark 
polarizations $(P_L^e,\, P_R^{\bar e})=(0.9,\,0.65)$.\
For comparison, we summarize in the lower half of Table\,\ref{tab:7}
the sensitivity reaches of $\,\cut\,$ via the reaction
$\,pp(q{\hs}\bar{q})\!\to\!Z\ga\,$ with 
$\hs Z\ito \ell\hs\bar{\ell},\nu\hs\bar{\nu}\hs$
at the LHC\,(13\,TeV) and the 100\,TeV $\!pp$ collider,
based on Tables\,\ref{tab:3} and \ref{tab:4} of Section\,\ref{sec:3}.

\vspace*{1mm}

From the comparison in Table\,\ref{tab:7}, we see that
the the sensitivity reaches for the nTGC operator $\OGP$ 
(and also the contact operator $\OCM$)
at the LHC\,(13\,TeV) with integrated luminosities 
$\mL\!=\!(0.14,\,0.3,\,3)$\,ab$^{-1}$ 
are higher than those of
$e^+e^-$ colliders with collision energies 
$\sqrt{s}\hsm =\hsm (250,\hs 500)$\,GeV, and are comparable
to those of an $e^+e^-$ collider of energy $\sqrt{s}=1$\,TeV,
but much lower than that of the CLIC with $\sqrt{s}=(3-5)$\,TeV.
On the other hand, the sensitivity reaches of 
the 100\,TeV $pp$ collider
with an integrated luminosity 
$\mL=3\hsx$ab$^{-1}$ can surpass those of all the $e^+e^-$ colliders
with collision energies up to $(3\hsm -\hsm 5)$\,TeV.

{\small
\tabcolsep 1pt
\begin{table}[t]
\begin{center}
\begin{tabular}{c|c||c|c|c|c}
\hline\hline
&&&&&\\[-3.5mm]
$\sqrt{s\,}$\,(TeV)\, & ~$\mL$\,(ab$^{\hsm -1}$)
& $\Lambda_{G+}^{}$ 
& $\Lambda^{}_{G-}$
& $\Lambda^{}_{\!\widetilde{B}W}$
& $\Lambda^{}_{C+}$\\
&&&&& \\[-3.7mm]
\hline\hline
&&&&& \\[-3.8mm]
~$e^+e^-${\hs}(0.25)~ & 5 & 
  ({1.3},\,{1.6}) & ({0.90},\,{1.2})
& ({1.2},\,{1.3}) & ({1.2},\,{1.6})
\\
&&&&& \\[-3.8mm]
\hline
&&&&& \\[-3.8mm]
~$e^+e^-${\hs}(0.5)~ &5 &	
({2.3},\,{2.7})  & ({1.4},\,{1.7})
&({1.8},\,{1.9}) & ({1.8},\,{2.2}) 
\\
&&&&& \\[-3.8mm]
\hline
&&&&& \\[-3.8mm]
~$e^+e^-${\hs}(1)~ & 5  & ({3.9},\,{4.7}) 
& ({1.9},\,{2.5})  & ({2.5},\,{2.6})
& ({2.6},\,{2.9})
\\
&&&&& \\[-3.8mm]
\hline
&&&&& \\[-3.8mm]
~$e^+e^-${\hs}(3)~ & 5  
& ({9.2},\,{11.0}) & ({3.4},\,{4.3})
& ({4.3},\,{4.5})  & ({4.4},\,{5.2}) 
\\
&&&&& \\[-3.8mm]
\hline
&&&&& \\[-3.8mm]
~$e^+e^-${\hs}(5)~ & 5  
&\,	({13.4},\,{15.9}) \,	
&\, ({4.4},\,{5.6}) \,		
&\, ({5.7},\,{5.9})\,
&\, ({5.7},\,{6.8}) \,
\\
&&&&& \\[-4.4mm]
\hline\hline
&&&&& \\[-4mm]
& 0.14 & \,3.3\, &\,1.1\,&\,1.3\,&\,1.4\\
&&&&& \\[-4.4mm]
LHC{\hs}(13) & 0.3 & \,3.6\, &\,1.2\,&\,1.4\,&\,1.5\\
&&&&& \\[-4.4mm]
&3& \,4.2\, &\,1.4\,&\,1.7\,&\,1.7\\
\hline
&&&&& \\[-4.4mm]
&3& \,23\, &\,4.6\,&\,5.6\,&\,5.9\\
&&&&& \\[-4.4mm]
$pp${\hs}(100)  & 10 & \,26\, &\,5.1\,&\,6.1\,&\,6.5\\
&&&&& \\[-4.4mm]
& 30 & \,28\, &\,5.5\,&\,6.7\,&\,7.1\\
\hline\hline
\end{tabular}
\end{center}
\vspace*{-3mm}
\caption{{\small \it
Comparisons of $\,2{\hs}\sigma$ sensitivities to 
the new physics scale $\,\cut\,$} (in TeV)
{\it for each dimension-8 nTGC operator or related contact operator $(\OGP,\,\OGM,\,\OBW,\,\OCP )$,
at $e^+e^-\!$ colliders of different collision energies, and at
the LHC and the}  $pp$\,(100\,TeV) {\it collider.\
The reactions $\,e^-e^+\!\!\to\!Z\ga\!\ito\! q\bar{q}\ga$\, and 
$\,pp\hs (q{\hs}\bar{q})\!\ito\!Z\ga\!\ito\!  
 \ell\bar{\ell}\ga,\hs\nu\bar{\nu}\ga$\,
are analyzed for the lepton and hadron colliders respectively.\
For the $\hs e^+e^-\!$ colliders, each entry corresponds to 
(unpolarized,\,polarized) $e^\mp$ beams, where we choose the  
benchmark $e^\mp$ beam polarizations as 
$(P_L^e,\hs P_R^{\bar e})\hsm =\hsm (0.9,\,0.65)$.
}}
\label{tab:7}
\end{table}
}

\vspace*{1mm}

We consider next the other three dimension-8 operators
$(\OGM,\,\OBW,\,\OCP )$.\ 
Table\,\ref{tab:7} shows that the LHC has sensitivities
to $\cut$ that are comparable to those of $e^+e^-$ colliders with
$\sqrt{s\,}\hsm =\hsm (250,\hs 500)$\,GeV, but are clearly lower than
those of $e^+e^-$ colliders with collision energies
$\sqrt{s\,}\!\geqq\! 1$\,TeV.
On the other hand, we find that the sensitivities of 
the 100\,TeV $\!pp$ collider
with an integrated luminosity $\mL\!=\!3\hs$ab$^{-1}$
are significantly greater than those of the $e^+e^-$ colliders
with energy $\sqrt{s\,}\!\leqq\! 3$\,TeV. 
Moreover, a 100\,TeV $\!pp$ collider 
with an integrated luminosity 
$\hs\mL\hsm =\hsm (10\hsm -\hsm 30)\hs$ab$^{-1}$
has sensitivities comparable to those 
of an $e^+e^-$ collider with $\sqrt{s\,}=5\hs$~TeV,
while a 100\,TeV $\!pp$ collider with an
integrated luminosity of $30\hs$ab$^{-1}$ 
would have higher sensitivities
than an $e^+e^-$ collider with $\sqrt{s\,}=5$\,TeV.\
In passing, we find that our collider limits given in Table\,\ref{tab:7}
are much stronger than the unitarity limits
of Table\,\ref{tab:0} and Fig.\hs\ref{fig:0}.\
This shows that the perturbation expansion 
in the SMEFT formulation is well justified for 
the present collider analyses of probing the nTGCs.\

{\small
\tabcolsep 1pt
\begin{table}[]
{\small
\begin{center}
\begin{tabular}{c|c||c|c|c}
\hline\hline
&&&\\[-3.5mm]				
$\sqrt{s\,}$\,(TeV)\, & ~$\mL$\,(ab$^{\hsm -1}$)~
& $|h_4|$ & $|h_3^Z|$ & $|h_3^\gamma|$
\\
&&& \\[-3.7mm]
\hline\hline
&&& \\[-3.8mm]
~$e^+e^-${\hs}(0.25)~  & 5 & ({3.9},\,{2.0})$\times10^{-4}$
& ({2.7},\,{2.3})$\times10^{-4}$
& ({4.9},\,{1.6})$\times10^{-4}$					
\\
&&& \\[-3.8mm]
\hline
&&& \\[-3.8mm]
~$e^+e^-${\hs}(0.5)~ &5 &	({3.8},\,{1.9})$\times10^{-5}$
& ({6.2},\,{5.2})$\times10^{-5}$ & ({10},\,{3.7})$\times10^{-5}$
\\
				&&& \\[-3.8mm]
				\hline
				&&& \\[-3.8mm]
				~$e^+e^-${\hs}(1)~  &5  &	({4.5},\,{2.3})$\times10^{-6}$
				& ({1.5},\,{1.2})$\times10^{-5}$& ({2.3},\,{1.0})$\times10^{-5}$
				\\
				&&& \\[-3.8mm]
				\hline
				&&& \\[-3.8mm]
				~$e^+e^-${\hs}(3)~  &5 & ({1.6},\,{0.84})$\times10^{-7}$
				&({1.7},\,{1.4})$\times10^{-6}$ & ({2.5},\,{1.0})$\times10^{-6}$
				\\
				&&&\\[-3.8mm]
				\hline
				&&& \\[-3.8mm]
				~$e^+e^-${\hs}(5)~  &5 &\,	({3.6},\,{1.8})$\times10^{-8}$
				\,&\, ({5.8},\,{4.9})$\times10^{-7}$
				\,&\, ({8.9},\,{3.4})$\times10^{-7}$ \,
				\\
				\hline\hline
				&&&& \\[-4mm]
				& 0.14 & \,9.6$\times10^{-6}$\,
				& \,1.9$\times10^{-4}$\,&\,2.2$\times10^{-4}$\,\\
				&&&& \\[-4.4mm]
				LHC{\hs}(13) &0.3& \,7.5$\times10^{-6}$\, &\,1.5$\times10^{-4}$\,&\,1.8$\times10^{-4}$\,\\
				&&&& \\[-4.4mm]
				& 3 & \,3.8$\times10^{-6}$\, &\,0.80$\times10^{-4}$\,&\,0.97$\times10^{-4}$\,\\
				\hline
				&&&& \\[-4mm]
				&3& \,4.0$\times10^{-9}$\, &\,6.1$\times10^{-7}$\,&\,7.2$\times10^{-7}$\,\\
				&&&& \\[-4.4mm]
				$pp${\hs}(100) & 10 & \,2.6$\times10^{-9}$\,  &\,4.2$\times10^{-7}$\,&4.9$\times10^{-7}$\\
				&&&& \\[-4.4mm]
				& 30 & \,1.9$\times10^{-9}$\, &\,3.0$\times10^{-7}$\,&\,3.5$\times10^{-7}$\,\\
				\hline\hline
			\end{tabular}
		\end{center}
		\vspace*{-3mm}
		\caption{{\small\it\hspace*{-2mm}
				Sensitivity reaches on the nTGC form factors 				
				at the $2\hs\sigma$ level
				of \hs$e^+e^-\!\!$ colliders with different collision energies,
				compared with those of the LHC}
			{\it and the} $pp$\,(100\,TeV)$\!$ {\it collider.\
				The reactions $\,e^-e^+\!\!\to\!Z\ga\!\to\! q\bar{q}\ga$\,
				and $\,pp(q{\hs}\bar{q})\!\to\!Z\ga\!\to\!
				\ell\bar{\ell}\ga,\hs\nu\bar{\nu}\ga\hs$
				are considered for the lepton and hadron colliders respectively.
				For the $e^+e^-$ colliders, each entry corresponds to
				(unpolarized,\,polarized) $e^\mp$ beams.\
				As benchmarks for the $e^\mp$ beam polarizations we choose
				$(P_L^e,\, P_R^{\bar e})=(0.9,\,0.65)$.}}
\label{tab:8}
}
\end{table}
}

\vspace*{1mm}

Next, we analyze the probes of nTGCs at $e^+e^-$ colliders 
using the form factor formulation we described in Section\,\ref{sec:3}.
According to the relations we derived in Eq.\eqref{eq:h-dim8},
can translate our sensitivity reaches on the new physics scale
$\cut_j^{}$  of each dimension-8 operator $\mO_j^{}$ to that 
of the related form factor $h_j^V$. 
The corresponding sensitivities on the form factors
$(h_4^{},\hs h_3^Z,\hs h_3^\gamma)$ are presented
in the upper half of Table\,\ref{tab:8}.
For comparison, we also show the 
sensitivities of the LHC\,(13\,TeV) and a 100\,TeV $\!pp$ collider 
in the lower half of Table\,\ref{tab:8}.

\vspace*{1mm}

We see from Table\,\ref{tab:8} that the LHC has sensitivities 
for the form factor 
$|h_4^{}|$ that are higher than those of the
$e^+e^-$ colliders with
$\sqrt{s\,}\hsm =\hsm (250,\hs 500)$\,GeV by a factor of
${O}(10\!-\!10^2)$, but has comparable sensitivities to that
of an $e^+e^-$ collider with
$\sqrt{s\,}\hsm =\hsm 1$\,TeV, 
whereas the LHC sensitivities are lower than those of
the $e^+e^-$ colliders with
$\sqrt{s\,}\hsm =\hsm (3\hsm -\hsm 5)$\,TeV
by a factor of $O(10-10^2)$.
On the other hand, a 100\,TeV $pp$ collider would
have much higher sensitivities than all the 
$e^+e^-$ colliders with
$\sqrt{s}\hsm \leqq\! 5$\,TeV,
by factors ranging from $O(10\hsm -\! 10^5)$.
We also see that a 100\,TeV $\!pp$ collider
has a sensitivity for probing the
form factor $h_4^{}$
that is better than that of the LHC
by a factor ${O}(10^{3})\hs$.

\vspace*{1mm}

Similar features
hold for the form factors
$(h_3^Z,\hs h_3^\gamma)$, as can be seen by inspecting Table\,\ref{tab:8}.
We find that an $e^+e^-$ collider of any given
collision energy $\sqrt{s\,}\,$ 
has comparable sensitivities for probes of $(h_3^Z,\hs h_3^\gamma)$,
with the differences being less than a factor of 2{\hs}.
We see also that the sensitivities improve from $O(10^{-4})$
to $O(10^{-7})$ when the collider energy increases
from $\sqrt{s\,}\hsm =\hsm 0.25$\,TeV to 5\,TeV.
We further note that 
the LHC and 100\,TeV $\!pp$ colliders
have comparable sensitivities 
to $(h_3^Z,\hs h_3^\gamma)$ for any given integrated luminosity.\
When the integrated luminosity of the LHC (or  
the 100\,TeV $pp$ collider) increases over the range from
$\,\mL\!=\!(0.14\hsm -\hsm 3)\hs$ab$^{-1}$ 
[or $\hs\mL\!=\!(3\hsm -\hsm 30)\hs$ab$^{-1}$], 
we see that the sensitivities to the form factors 
$(h_3^Z,\hs h_3^\gamma)$ increase by about a factor of $2\hs$.
Comparing the sensitivity reaches of
the $e^+e^-$ and hadron colliders in Table\,\ref{tab:8},
we find that the sensitivities of the LHC are comparable
to those of a 0.25\,TeV $e^+e^-$ collider, but lower than
those of $e^+e^-$ colliders with 
$\sqrt{s\,}\!=\!(0.5\hsmx -\hsmx 1)$\,TeV 
by a factor of $O(10)$, and lower than those
of \hs$e^+e^-$ colliders with 
$\sqrt{s}\!=\!(3\hsm -\hsm 5)\hs$TeV by factors of 
${O}(10^{2}\hsmx -\hsmx 10^{3})$.\ 
On the other hand, the sensitivities of the 
$pp\hs$(100{\hs}TeV)
collider for probing $(h_3^Z,\hs h_3^\gamma)$ are generally higher
than those of the 250\,GeV $e^+e^-$
collider by a factor
of ${O}(10^{3})$, higher than those of the 
0.5{\hs}TeV to 1{\hs}TeV $e^+e^-$ colliders 
by a factor of ${O}(10^{2})$, 
and higher than those of the 
3{\hs}TeV $e^+e^-$ collider by a factor
of ${O}(10)$, while they are comparable to those of
a 5{\hs}TeV $e^+e^-$ collider.

\vspace*{1mm}

\begin{figure}[t]
\vspace*{-6mm}
\includegraphics[width=8cm,height=5.9cm]{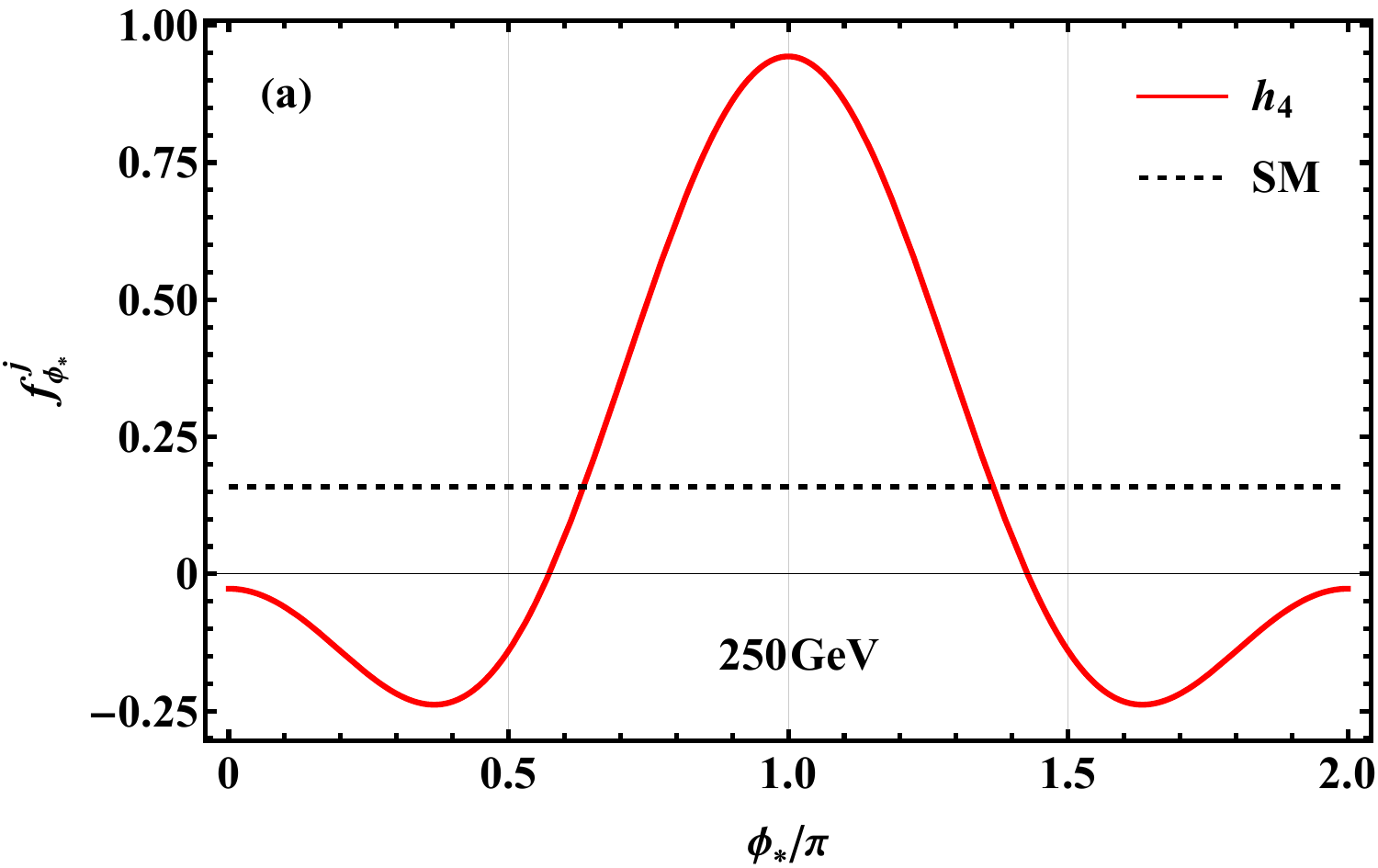}
\includegraphics[width=7.8cm,height=5.8cm]{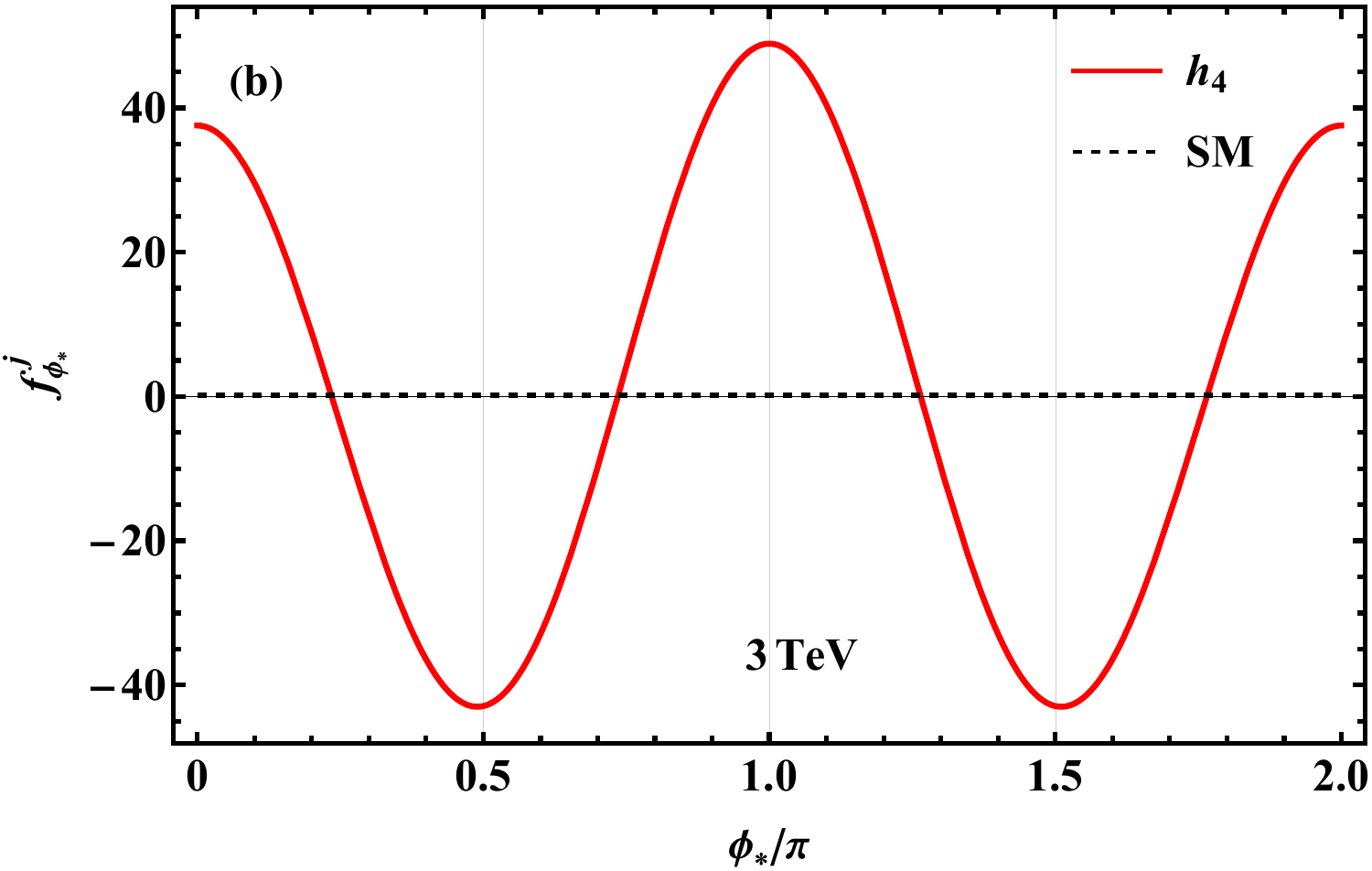}
\\[2mm]
\hspace*{1mm}
\includegraphics[width=7.8cm,height=5.8cm]{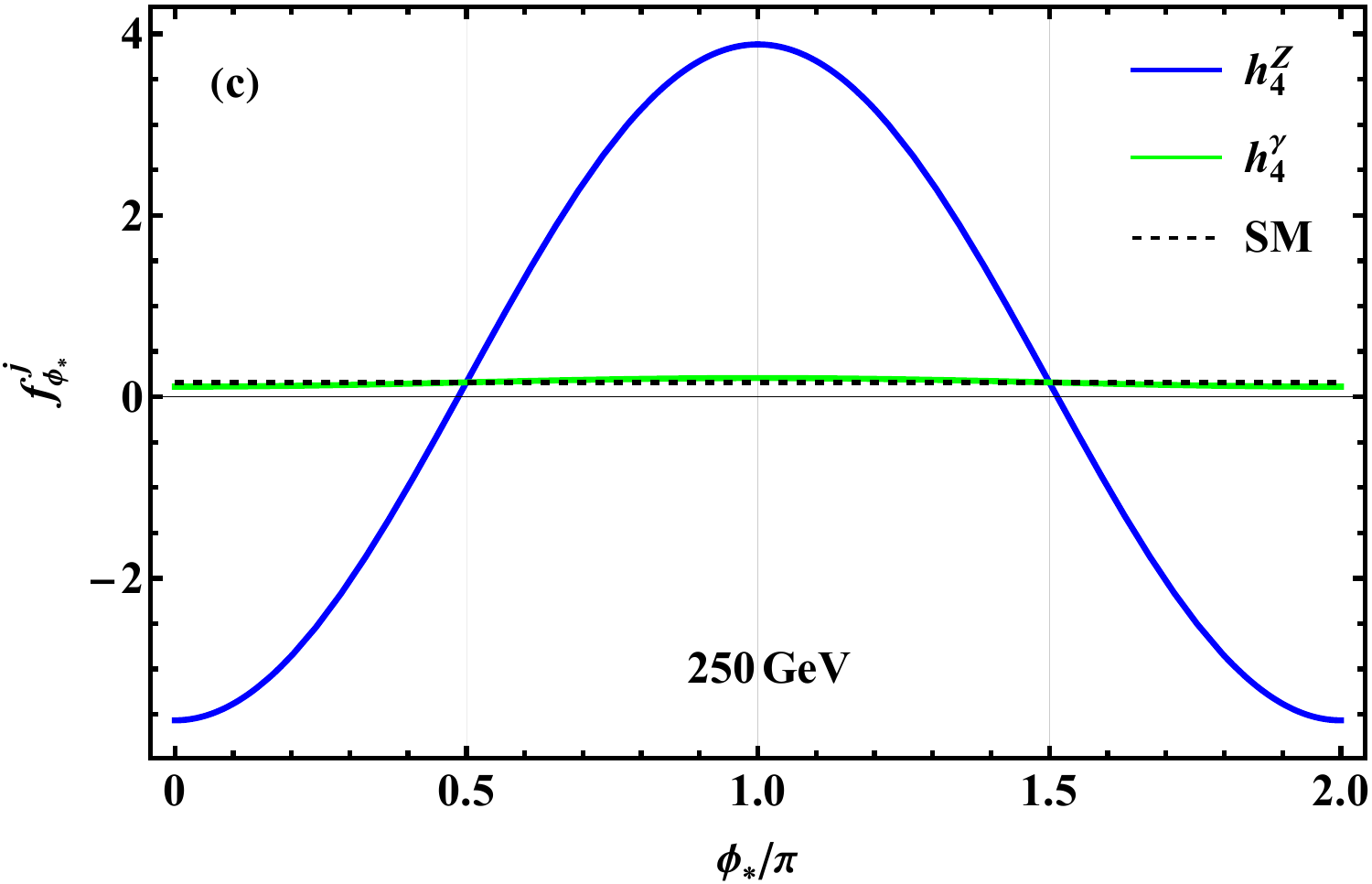}
\includegraphics[width=7.8cm,height=5.8cm]{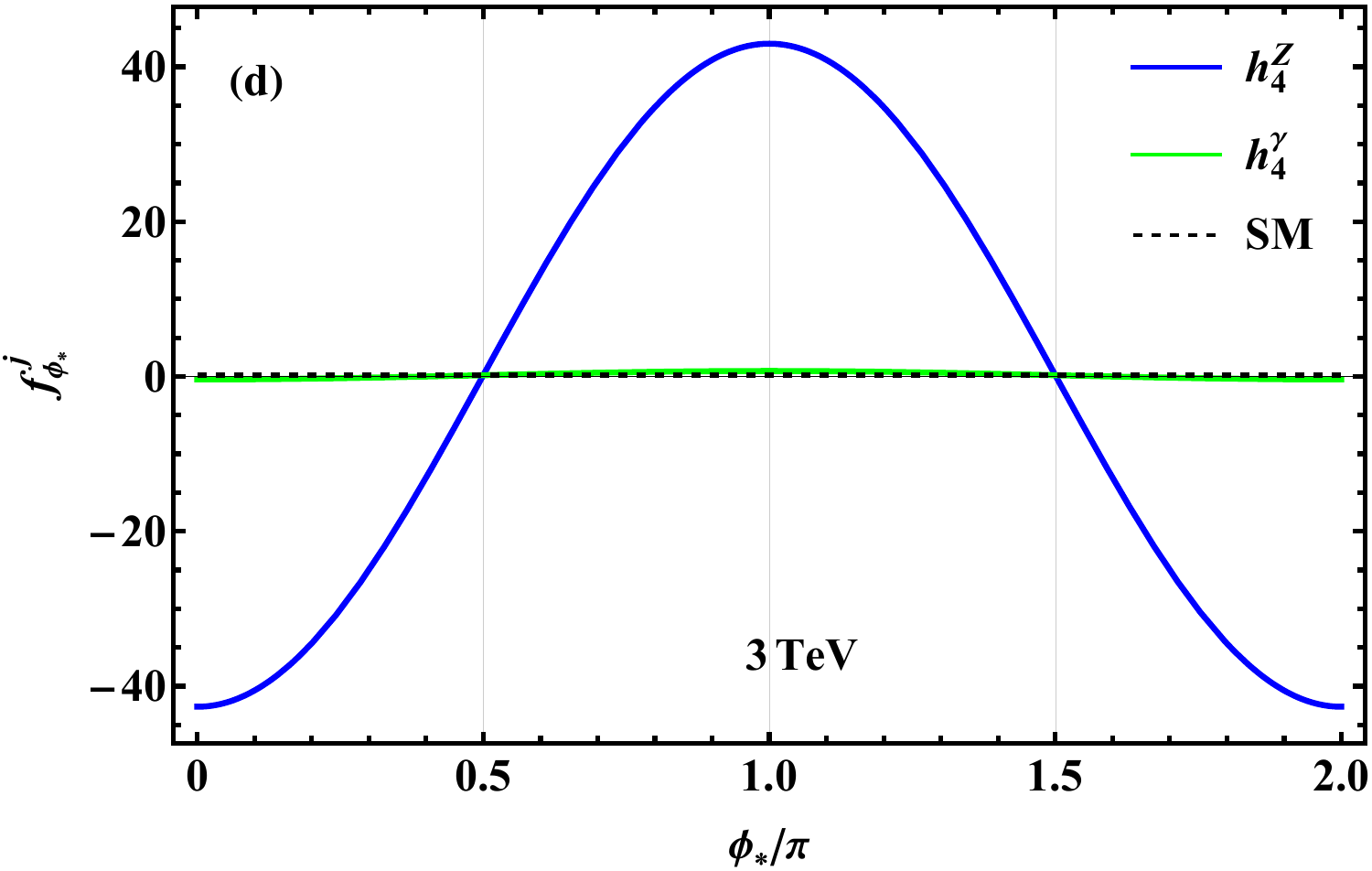}
\vspace*{-2mm}
\caption{\small{\hspace*{-1mm}%
\it Normalized angular distributions in 
$\phi_*^{}$ for $\hs e^+\hs e^{-}\!\ito Z\ga\hs$ 
with $Z\!\ito d\hs\bar{d}\hs$, as generated by  $h_4^{}$ 
in our form factor formulation \eqref{eq:FF2-nTGC}  
in panels $(a)$ and $(b)$, 
and as generated by $(h_4^Z,\hs h_4^\gamma)$ in the conventional
form factor formulation \eqref{eq:FF0-nTGC} with 
$h_5^V\hsmx\!=\hsm 0\hs$ in panels $(c)$ and $(d)$.\ 
The panels $(a)$ and $(c)$ correspond to the
$e^+e^-$ colliders with} $\sqrt{s}\!=\!250\,$GeV {\it and 
the panels $(b)$ and $(d)$ correspond to}  $\sqrt{s}=3\,$TeV. 
}
\label{fig:7}
\end{figure}

Finally, it is instructive to present the $\phi_*^{}$
angular distributions for the form factor
$h_4^V$. In the gauge-invariant form factor formulation given in 
Eq.\eqref{eq:FF2-nTGC}, we have imposed the constraints \eqref{eq:h4-h5}-\eqref{eq:h4Z-h4A}. Hence, the form factor
$\hs h_5^V\hs$ is not independent, and should be replaced by
$\hs h_5^V\!=-h_4^V/2\hs$, according to Eq.\eqref{eq:h4-h5}.\
Moreover, Eq.\eqref{eq:h4Z-h4A} shows that $\hs h_4^\gamma \hs$ 
is not independent, so the form factors 
$(h_4^V,\,h_5^V)$ reduce to a single parameter
$\hs h_4^{}\hs (\equiv h_4^Z)$
as shown below Eq.\eqref{eq:h4h5-h4ZA}. 
We can then derive the interference cross section $\sigma_1^{}$
contributed by $\hs h_4^{}\hs$ and the normalized angular
distribution $\hs f^1_{\phi_*}\hs$ as follows:
\beqs 
\begin{align}
\label{eq:sigma1-h4} 
\sigma_1^{} &\,=\,
\frac{~e^2\!\(-\fr{1}{2}\hsm +\hsm s_W^2\)\!\hsm (s\!-\!M_Z^2)~}
{8\pi s_W^{}c_W^{} v^2\,s}h_4^{} \,,
\\[1.5mm]
\label{eq:f1-h4}
f_{\phi_*^{}}^{1} &\,=\,
\frac{1}{\,2\pi\,} \hsm -\hsm 
\frac{\,3\pi (f_L^2\!-\!f_R^2)(M_Z^2+5\hs s)
	\cos\phi_*^{}\,}{256(f_L^2\!+\!f_R^2)M_Z\sqrt s}+
\frac{\,s\cos2\phi_*^{}\,}{8\pi M_Z^2} \,.
\end{align}
\eeqs 

\noindent 
We see that the interference cross section scales as 
$\,\sigma_1^{}\!\propto\!E^0\,$, while the angular
distribution $\hs f^1_{\phi_*}\hs$ has 
the leading term $\cos2\phi_*^{}$ 
enhanced by $E^2$ and the subleading term $\cos\phi_*^{}$
enhanced by $E^1$ for large energy 
$\sqrt{s\,}\!=\!E\hs$.
We plot the angular distribution $f_{\phi_*^{}}^{1}$ 
in Fig.\,\ref{fig:7}(a) and \ref{fig:7}(b) 
for the $e^+e^-$ collider energies
$\sqrt{s\,}=250\,$GeV and $3\,$TeV, respectively.
In each panel, the $\hs h_4^{}\hs$ contribution is depicted by
the red solid curve, and the SM contribution is shown as the
black dashed curve which is almost flat.\ 
We also observe that $\cos\phi_*^{}$ and $\cos2\phi_*^{}$ terms in the function $f_{\phi_*^{}}^{1}$ in Eq.\eqref{eq:f1-h4}
have opposite signs. They are comparable for lower collision energy
$\sqrt{s\,}=250\,$GeV, but $\cos(2\phi_*^{})$ becomes dominant
for a large collision energy $\sqrt{s\,}=3\,$TeV.
We can evaluate the numerical coefficients of 
$f_{\phi_*^{}}^{1}$, as follows:
\beqs 
\begin{align}
\label{eq:f1-h4-250GeV}
f_{\phi_*^{}}^{1} &=\,
\frac{1}{\,2\pi\,} \hsm -\hsm 0.485\cos\phi_*^{}+
0.299\cos2\phi_*^{} \hs , 
&\hspace*{-5mm}& 
\text{for}~\sqrt{s\,}\hsm =\hsm 250\hs\text{GeV},
\\[1.5mm]
\label{eq:f1-h4-3TeV}
f_{\phi_*^{}}^{1}&=\,
\frac{1}{\,2\pi\,} \hsm -\hsm 5.67\cos\phi_*^{}+
43.1\cos2\phi_*^{} \hs ,   
&\hspace*{-5mm}& 
\text{for}~\sqrt{s\,}\hsm =\hsm 3\hs\text{TeV}.
\end{align}
\eeqs 
This explains why panel\,(a) of  Fig.\,\ref{fig:7}
exhibits a significant cancellation between the 
$\cos\phi_*^{}$ and $\cos(2\phi_*^{})$ terms, whereas in
panel\,(b) the  $\cos(2\phi_*^{})$ term dominates and thus
the red curve exhibits
interesting $\cos(2\phi_*^{})$ behavior.

\begin{table}[t]
\begin{center}
	\begin{tabular}{c||c|c|c|c|c}
		\hline\hline	
		& & & & &  
		\\[-4mm]
		$\sqrt{s\,}$\,(TeV) & 0.25&0.5 & 1&3 &5
		\\
		\hline\hline
		& & & & &  
		\\[-4.3mm]
		\red$|h_4|^{}$ 
		&\red $3.9\!\times\!10^{-4}$&\red $3.8\!\times\!10^{-5}$ &\red$4.5\!\times\!10^{-6}$ & \red$1.6\!\times\!10^{-7}$ &\red $3.6\!\times\!10^{-8}$ \\
		\hline
		& & & & &  
		\\[-4.3mm]
		\blu$|h_4^Z|$ 
		& \blu $8.9\!\times\!10^{-5}$ &\blu $4.2\!\times\!10^{-6}$ &\blu $2.5\!\times\!10^{-7}$ &\blu $3.0\!\times\!10^{-9}$ &\blu $3.9\!\times\!10^{-10}$ \\
		\hline
		& & & & & 
		\\[-4.3mm]
		\blu$|h_4^\gamma|$ 
		& \blu $6.7\!\times\!10^{-4}$ &\blu $3.2\!\times\!10^{-5}$ &\blu $1.9\!\times\!10^{-6}$ &\blu $2.3\!\times\!10^{-8}$ &\blu $2.9\!\times\!10^{-9}$ \\
		\hline\hline
	\end{tabular}
\end{center}
\vspace*{-4mm}
\caption{\small%
{\it Comparisons of the $2\hs\sigma$ sensitivities
to probing the form factor $\hs h^{}_4\hs$ of our SMEFT formulation 
\eqref{eq:FF2-nTGC} (marked in red color) 
and the conventional form factors 
$(h_4^Z,\,h_4^\gamma)$ that take into account only U(1)$_{em}$ gauge invariance (marked in blue color),
as derived by analyzing the reaction
$\,e^+\hs e^-\!\ito Z\ga \ito q\hs\bar{q}\ga$\,		
at various \,$e^+e^-\!$ colliders with} $\mL\!=5\,$ab$^{-1}\!$
{\it and unpolarized $e^\mp$ beams.\
As discussed in the text, the conventional form-factor limits 
(blue color) are included for illustration only, 
as they do not respect the full SM gauge symmetry, 
and hence are invalid.}
}
\label{tab:9}
\end{table}
%

\begin{figure}[t]
\vspace*{-5mm}
\centering
\includegraphics[width=12.5cm,height=7cm]{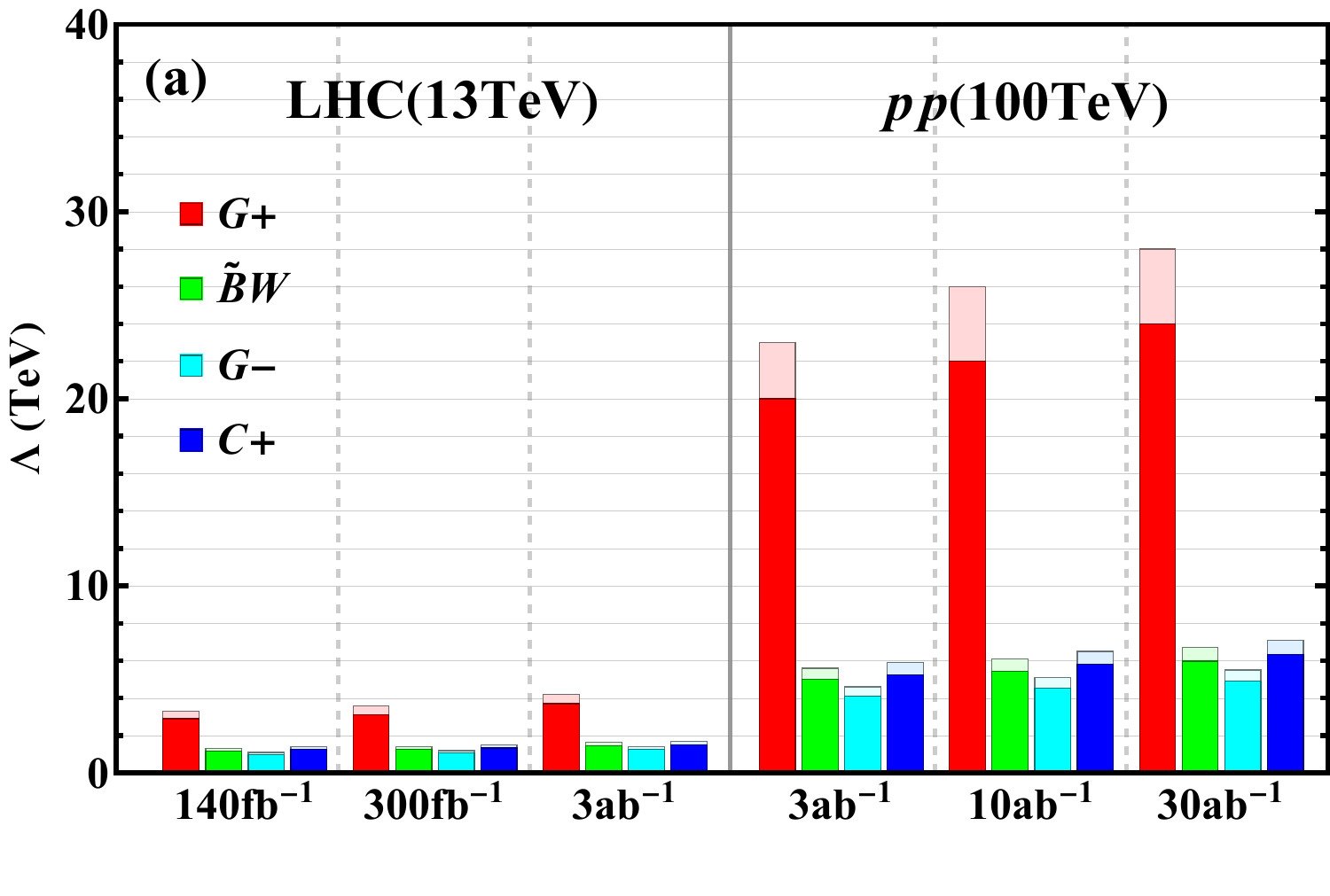}
\\[-1mm]
\includegraphics[width=12.5cm,height=7cm]{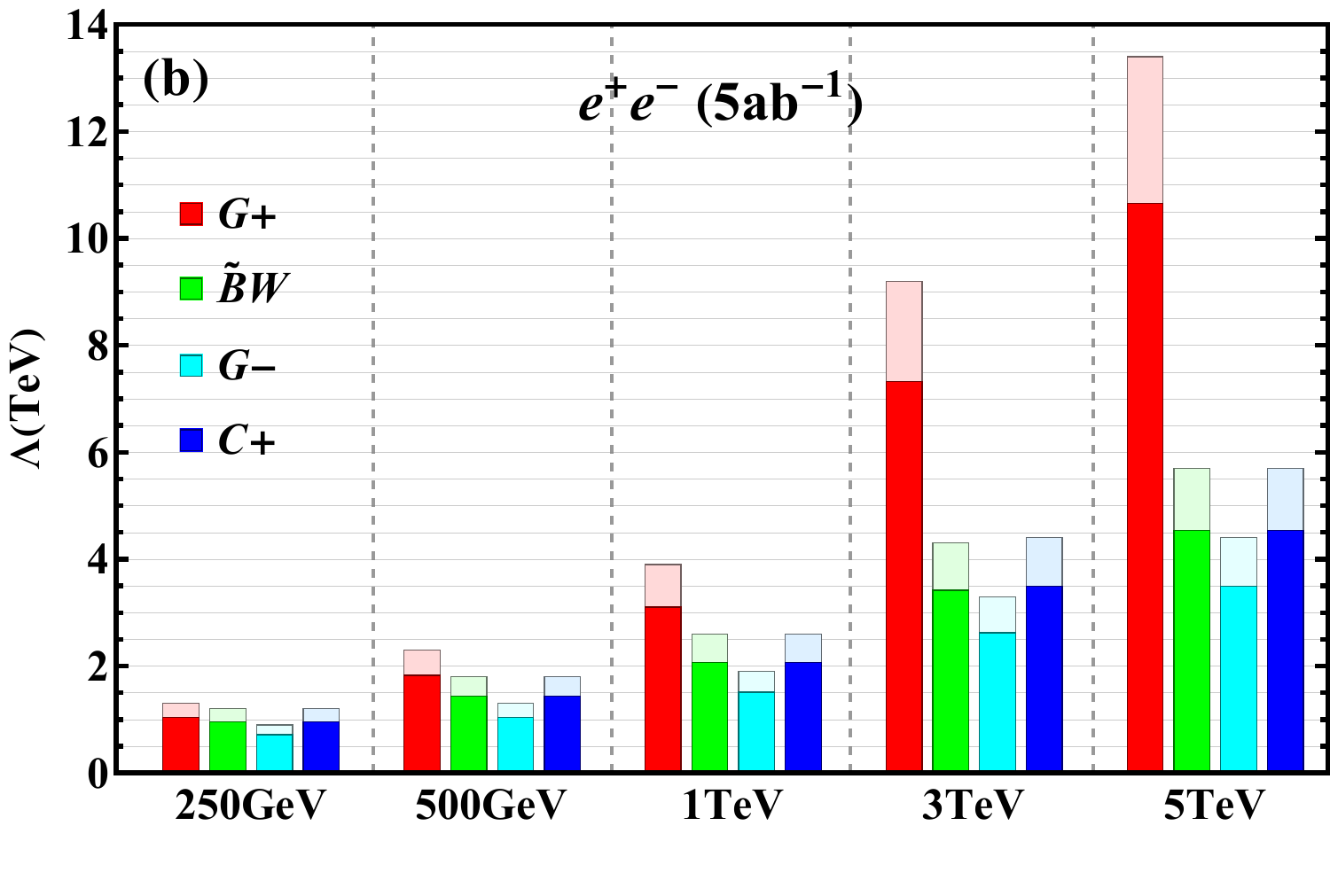}
\vspace*{-5mm}
\caption{\hspace*{-1mm}\small{\it  
Sensitivity reaches for the new physics scale $\cut$ 
of the nTGC operators at the hadron colliders
LHC}\,(13\,TeV) {\it and} $pp$\,(100\,TeV)
{\it in plot (a) and $e^+e^-$\! colliders 
with collision energies 
$\sqrt{s\,}\!=\!(0.25,\,0.5,\,1,\,3,\,5)$}\,TeV {\it in plot (b).
In each plot, the (${\hsx}2{\hs}\sigma,\,5{\hs}\sigma$) 
sensitivities are shown in (light,\,heavy) colors, respectively.}}
\label{fig:8}
\vspace*{1mm}
\end{figure}
%

\vspace*{1mm}

For comparison, we consider the conventional form factor
formulation \eqref{eq:FF0-nTGC} with $h_5^V\hsmx\!=\hsm 0\hs$,
where $(h_4^Z,\hs h_4^\gamma)$ 
are treated as two independent parameters.\ 
In this case we derive the following interference cross sections 
$(\tilde\sigma_1^Z,\hs \tilde\sigma_1^A)$ 
contributed by $(h_4^Z,\hs h_4^\gamma)$ and their normalized angular
distributions 
$\hs (\tilde{f}^{1Z}_{\phi_*},\hs \tilde{f}^{1A}_{\phi_*})\hs$: 
\beqs 
\label{eq:sigma1-f1-ZA}
\begin{align}
\label{eq:sigma1-ZA}
(\tilde\sigma_1^Z,\,\tilde\sigma_1^A) &\,=\, 
\frac{~e^4(s\!-\!M_Z^2)^2~}{128\pi M_Z^4\,s}
\(\!\frac{\,1\!-\!4s_W^2\,}{s_W^2c_W^2}h_4^Z,~
\frac{2}{\,s_W^{}c_W^{}\,}h_4^\gamma\!\) \!,
\\[1.5mm]
\tilde f_{\phi_*}^{1Z}  &\,=\,  \frac{1}{\,2\pi\,}-
\frac{~3\pi(f_L^2\!-\!f_R^2)(3s\!+\!M_Z^2)~}
{~128M_Z^{}(f_L^2\!+\!f_R^2)\sqrt{s}~}
\frac{~1\!-\!4s_W^2\!+\!8s_W^4~}{1\!-\!4s_W^2}\cos\hsm\phi_*^{}\hs ,
\\[1.5mm]
\label{eq:f1-h4ZA}
\tilde f_{\phi_*}^{1A}  &\,=\,  \frac{1}{2\pi} - 
\frac{~3\pi(f_L^2\!-\!f_R^2)(3s\!+\!M_Z^2)~}
{~128M_Z^{}(f_L^2+f_R^2)\sqrt{s}~} (1\!-\!4s_W^2)
\cos\hsm\phi_*^{}\hs .
\end{align}
\eeqs 
We see that the interference cross sections in (\ref{eq:sigma1-f1-ZA}) scale as 
$\,(\tilde{\sigma}_1^{Z},\hs\tilde{\sigma}_1^{A})
\!\propto\!E^2\,$, while the angular
distributions $\hs (f^{1Z}_{\phi_*}\hsm ,\hs f^{1A}_{\phi_*})\hs$ 
have leading terms $\propto \cos\hsm\phi_*^{}$ 
enhanced by $E^1$ for large energy 
$\sqrt{s\,}\!=\!E^1\hs$.
We also note that the distribution $f^{1A}_{\phi_*}$ is much
suppressed relative to $f^{1Z}_{\phi_*}$ due to the small factor
$\,(1\hsm - 4s_W^2)\!\ll\!1\hs$.\ 
We plot the angular
distributions $\hs (f^{1Z}_{\phi_*}\hsm,\hsx f^{1A}_{\phi_*})\hs$ 
of Eq.\eqref{eq:f1-h4ZA} as the blue solid curves
in Fig.\,\ref{fig:7}(c)-(d), while the
squared distributions 
$\hs (f^{2Z}_{\phi_*}\hsm ,\hs f^{2A}_{\phi_*})\hs$
and the SM distribution $\hs f^{0}_{\phi_*}\hs$ are
plotted as the green solid curves and 
black dashed curves, respectively. 
As expected, the distributions
$\hs (f^{2Z}_{\phi_*}\hsm ,\hs f^{2A}_{\phi_*})\hs$ and  
$\hs f^{0}_{\phi_*}\hs$ are dominated by the constant term and
thus nearly flat.

\begin{figure}[t]
\vspace*{-5mm}
\centering
\includegraphics[width=12.5cm,height=7cm]{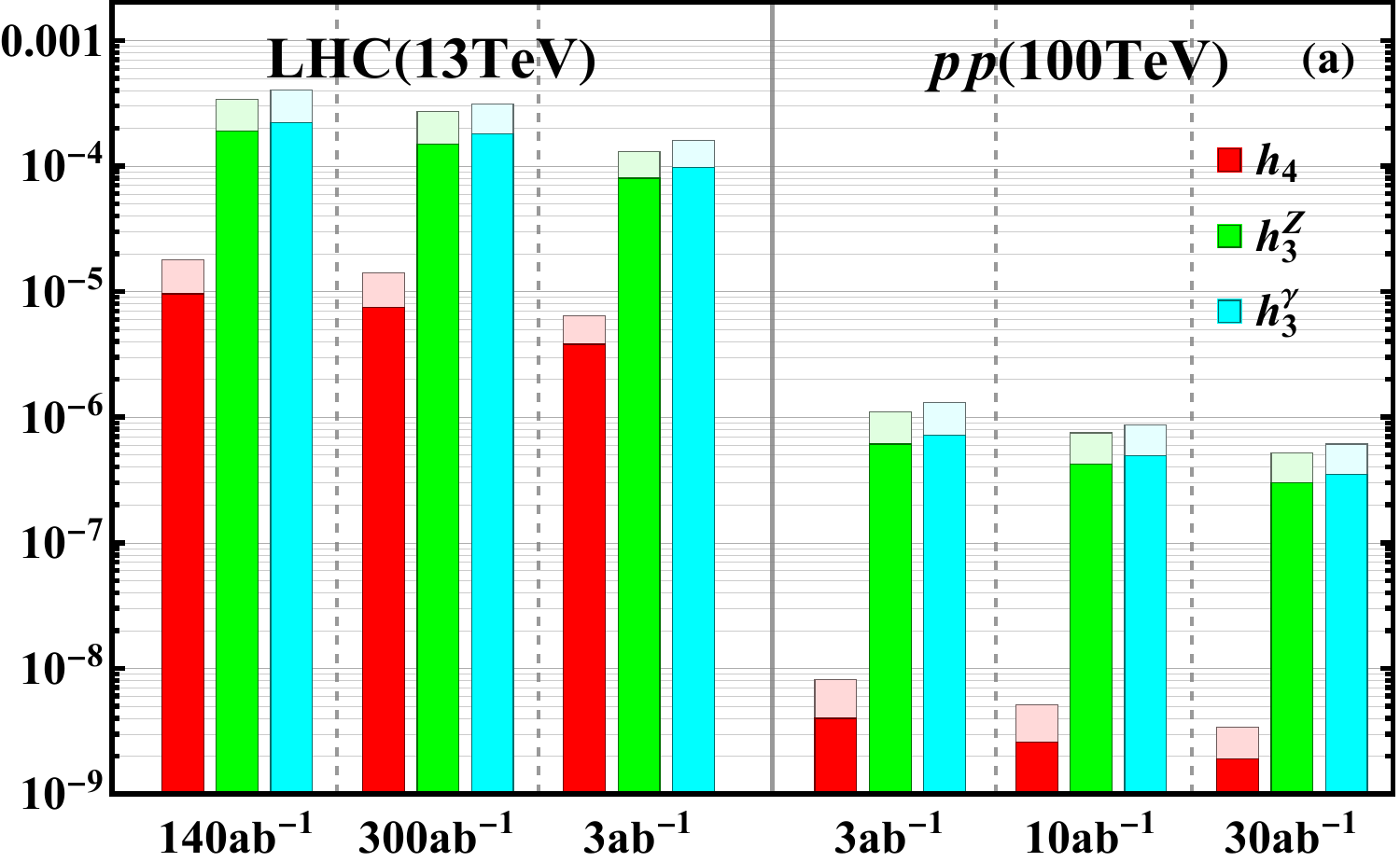}
\\[3mm]
\includegraphics[width=12.5cm,height=7cm]{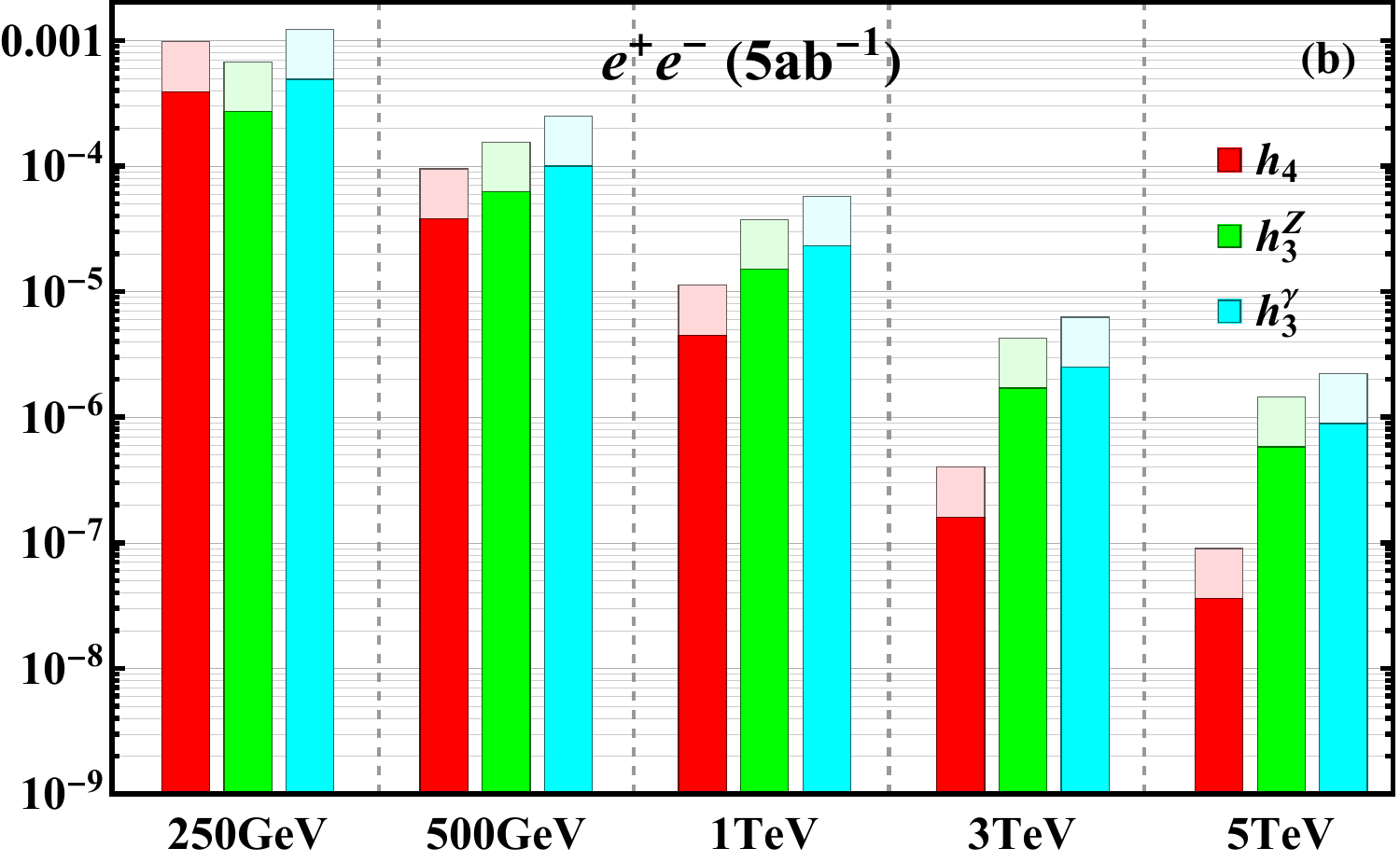}
\caption{\small\hspace*{-3mm}
	{\it Sensitivity reaches for the nTGC form factors 
		$(h_4^{},\,h_3^Z,\,h_3^\gamma)$
		at the hadron colliders 
		LHC} (13\,TeV) {\it and} $pp$\,(100\,TeV) {\it 
		in plot\,(a) and at $e^+e^-$\! colliders with collision energies} 
	$\sqrt{s\,}\!=\!(0.25,\,0.5,$ $\,1,\,3,\,5)$\,TeV {\it in plot\,(b).\
		In each plot, the {($\,2{\hs}\sigma,\,5{\hs}\sigma$)} 
		sensitivities are shown in (heavy,\,light) colors, respectively.}
}
\label{fig:9}
\vspace*{1mm}
\end{figure}

\vspace*{1mm}

We stress that Eq.\eqref{eq:sigma1-f1-ZA}
and the corresponding Figs.\,\ref{fig:7}(c)-(d)
{\it are incorrect} because the conventional form factor formulation \eqref{eq:FF0-nTGC} with 
$\,h_5^V\hsmx\!=\hsm 0\,$ does not
obey the consistency conditions \eqref{eq:h4h5-h4ZA} imposed by the
spontaneous breaking of the electroweak gauge symmetry
SU(2)$_L^{}\otimes$U(1)$_Y^{}$ of the SMEFT. 
In the following we further show that the 
conventional form factor formulation also leads to 
erroneously strong sensitivity limits on the form factors
$(h_4^Z,\,h_4^\gamma)$.

\vspace*{1mm}

Following the steps leading to Eq.\eqref{eq:O1-G+},
we construct the following observables 
$(\mathbb{O}_1^{},\,\mathbb{\widetilde{O}}_1^{})$
for probing the form factors:
\beqs 
\label{eq:FF-O1-O1T}
\begin{align}
\mathbb{O}_1^{} &\,=\, 
\left|\hs\sigma_1^{}\!\int\!\!\di\phi_*^{}\,
f_{\phi_*^{}}^1\!\hsm\times\hsm\text{sign}(\cos\hsm 2\phi_*^{})
\right|,
\label{eq:O1-h4}
\\
\mathbb{\widetilde{O}}_1^{} &\,=\, 
\left|\hs\tilde\sigma_1^{}\!\int\!\!\di\phi_*^{}\,
\tilde f_{\phi_*^{}}^1\!\hsm\times\hsm\text{sign}(\cos\hsm \phi_*^{})
\right|.
\label{eq:O1T-h4ZA}
\end{align}
\eeqs 
From the above formulae \eqref{eq:sigma1-h4} and 
\eqref{eq:sigma1-f1-ZA}, we deduce that energy-dependences
of the observables are
$\hs\mathbb{O}_1^{}(h_4^{})\hsmx\!\propto\hsmx\!E^2\hs$ and
$\hs\mathbb{\widetilde{O}}_1^{}(h_4^V)\hsmx\!\propto\hsmx\! E^3\hs$.\ 
This shows that in the conventional form factor formulation  
$\mathbb{\widetilde{O}}_1^c$ has an erroneously large
energy-dependence ($E^3$ instead of $E^2$),
leading to incorrectly strong sensitivities
to the form factors $(h_4^Z,\,h_4^\gamma)$.\ 
We present these incorrect sensitivities in blue color in Table\,\ref{tab:9}.\ 
For comparison, we also show in this Table 
the correct sensitivities (red color) to the form factors 
$h_4^{}$, as derived within our consistent 
form factor formulation \eqref{eq:FF2-nTGC} with the constraints \eqref{eq:h4-h5} and \eqref{eq:h4Z-h4A}.
From this comparison, we see that for the form factor 
$\hs h_4^Z\hs$ the conventional sensitivities (blue color) 
are erroneously stronger than our new sensitivities (red color)
by a factor of $5$ for the 
collider energy $\sqrt{s\,}=250$\,GeV,
by a factor of $\,{O}(10\hsmx -\hsmx 20)$ for the collider energies 
$\sqrt{s\,}=(0.5\hsmx -\hsmx 1)$\,TeV,
and by a factor of ${O}(10^2)$ for the collider energies
$\sqrt{s\,}=(3\hsmx -\hsmx 5)$\,TeV.

\vspace*{1mm}

Finally, for an intuitive comparison and overview, we 
summarize in Fig.\,\ref{fig:8} the sensitivity reaches
for the new physics scales of the dimension-8 nTGC operators
at the hadron colliders [panel\,(a)] and the $e^+e^-$ colliders
[panel\,(b)] from Tables\,\ref{tab:3}-\ref{tab:4} and 
Table\,\ref{tab:7}.\ We present these limits at both
the $(2{\hs}\sigma,\hs 5{\hs}\sigma)$ levels,
which are indicated by the (light,\,heavy) colors respectively.\ 
In Fig.\,\ref{fig:8}(b) we only plot the sensitivity reaches
for $e^-e^+$ collisions with unpolarized electron/positron beams.\ 
We note that according to Table\,\ref{tab:7},
adding the $e^-/e^+$ beam polarizations can increase
the sensitivity reaches on the new physics scale 
by about $20\%$ for $\OGP$ and $\OCP$, 
and by about $5\%$ for $\OBW$.\ 
Also, we summarize in Fig.\,\ref{fig:9} the sensitivity reaches
for probing the nTGC form factors
$(h_4^{},\hs h_3^Z,\hs h_3^\gamma)$ 
at the hadron colliders [panel\,(a)] and the $e^+e^-$ colliders
[panel\,(b)] from Table\,\ref{tab:5} and Table\,\ref{tab:8},
where the $(2{\hs}\sigma,\hs 5{\hs}\sigma)$ limits 
are marked by the (light,\,heavy) colors, respectively.\ 
In Fig.\,\ref{fig:8}(b) we present only the sensitivity reaches
for $e^-e^+$ collisions with unpolarized electron/positron beams.\ 
We note that according to Table\,\ref{tab:8},
adding the $e^-/e^+$ beam polarizations can increase
the sensitivity reaches for the nTGC form factors 
by about $100\%$ for $h_4^{}$,  
by about $20\%$ for $h_3^Z$, and
by about $160\%$ for $h_3^{\ga}$.\

\vspace*{1mm}

The reason that the effects of beam polarization 
for probing the nTGC form factors in $e^-e^+$ collisions
appear much stronger than those for probing the new physics cutoff scales
of the dimension-8 nTGC operators can be understood as follows.\
We note that the relation between the polarized and unpolarized cross sections
of the SM backgrounds is given by\,\cite{Ellis:2020ljj}:
\begin{eqnarray}
	\sigma_0(P_L^e,P_R^{\bar e}) &\!\!=\!\!&
	4\frac{\,P_L^e P_R^{\bar e}c_L^2\!+\!
		(1\!-\!P_L^e)(1\!-\!P_R^{\bar e})c_R^2\,}{c_L^2\!+c_R^2}
	\,\sigma_0(0.5,0.5) \,,
\end{eqnarray}
where $P_L^e$ $(P_R^{\bar e})$ denotes the fraction
of left-handed (right-handed) electrons (positrons) 
in the $e^-$\,($e^+$) beam and
$\hs P_L^{}\hsm +\!P_R^{}\hsm =\!1\hs$
holds for both $e^-$ and $e^+$ beams.\footnote{%
Note that the degree of longitudinal beam polarization 
for $e^-$ or $e^+$ is defined as
$\,\widehat{P}\!=\!P_R^{}\!-\!P_L^{}$ \cite{Ellis:2020ljj}.\ 
Thus, the left-handed and right-handed fractions of
$e^-$ and $e^+$ in the beam can be expressed as
$\,P_{L,R}^e\!=\!\fr{1}{2}(1\mp\widehat{P}^e)$\, and
$P_{L,R}^{\bar{e}}\!=\!\fr{1}{2}(1\mp\widehat{P}^{\bar{e}})$,\,
respectively. For instance, unpolarized $e^-$ and $e^+$ beams have
vanishing degrees of polarization 
$\,(\widehat{P}^e\!,\hs\widehat{P}^{\bar{e}})\hsm =\hsm 0$\,,
whereas a polarized $e^-$ beam with fraction $P_L^e=90\%$
has $\,\widehat{P}^e\!=\!-0.8$\,
and a polarized $e^+$ beam with fraction $P_R^{\bar{e}}=65\%$
has $\,\widehat{P}^{\bar{e}}\!=\!0.3$\,.
}

\vspace{1mm}

According to Ref.\,\cite{Ellis:2020ljj}, we construct 
the following three kinds of
$\mathbb{O}_1$ observables and extract the different signal terms
of $O(\cut^{-4})$ in the differential cross section:
%
\beqs	
\label{eq:O-ABC}
\begin{eqnarray}
\mathbb{O}_A &\!\!=\!\!&
		\int\!\! \di\theta \di\theta_*^{}\di\phi_*^{} \,
		\frac{d^3\sigma_1}{\di\theta \di\theta_*^{}\di\phi_*^{}}
		\text{sign}(\cos\!\phi_*^{})\,,
		\\
		\mathbb{O}_B &\!\!=\!\!&
		\int\!\! \di\theta \di\theta_*^{}\di\phi_*^{} \,
		\frac{d^3\sigma_1}{\di\theta \di\theta_*^{}\di\phi_*^{}}
		\text{sign}(\cos\!\phi_*^{})\text{sign}(\cos\!\theta^{})
		\text{sign}(\cos\!\theta_*^{})\,,
		\\
		\mathbb{O}_C &\!\!=\!\!&
		\int\!\! \di\theta \di\theta_*^{}\di\phi_*^{} \,
		\frac{d^3\sigma_1}{\di\theta \di\theta_*^{}\di\phi_*^{}}
		\text{sign}(\cos\!2\phi_*^{})\,. 
\end{eqnarray}
\eeqs	
For these observables, we can deduce the following:
\beqs
\vspace*{-2mm}
\begin{eqnarray}
	\mathbb{O}_A(G+) &\!\!=\!\!&
	\mathbb{A}\times \fr{1}{2}c_L^{}P_L^eP_R^{\bar{e}}
	(5s\!+\!M_Z^2)\Lambda^{-4}_{G+}\,,
	\\[1mm]
	\mathbb{O}_A(j) &\!\!=\!\!&
	\mathbb{A}\times
	3\!\left[c_L x_L P_L^eP_R^{\bar{e}}\!+\!
	c_R x_R(1\!-\!P_L^e)(1\!-\!P_R^{\bar{e}})\right]\!
	(s\!+\!M_Z^2)\Lambda^{-4}_{j}\,,
	\\[1mm]
	\mathbb{O}_B(G+) &\!\!=\!\!&
	\mathbb{B}\times \fr 1 2c_L^{}P_L^eP_R^{\bar{e}}
	(5s\!+\!M_Z^2)\Lambda^{-4}_{G+}\,,
	\\[1mm]
	\mathbb{O}_B(j) &\!\!=\!\!&
	\mathbb{B}\times 3\!\left[c_Lx_LP_L^eP_R^{\bar{e}}\!-\!
	c_Rx_R(1\!-\!P_L^e)(1\!-\!P_R^{\bar{e}})\right]\!
	(s\!+\!M_Z^2)\Lambda^{-4}_{j}\,,
	\\[1mm]
		\mathbb{O}_C(G+) &\!\!=\!\!&
	\mathbb{C}\times \fr 1 2c_L^{}P_L^eP_R^{\bar{e}}\,s\,\Lambda^{-4}_{G+}\,,
	\\[1mm]
	\mathbb{O}_C(j) &\!\!=\!\!&
	\mathbb{C}\times \!\left[c_Lx_LP_L^eP_R^{\bar{e}}\!-\!
	c_Rx_R(1\!-\!P_L^e)(1\!-\!P_R^{\bar{e}})\right]\!
	M_Z^2\Lambda^{-4}_{j}\,,
\end{eqnarray}
\eeqs
where the index $j$ denotes the operators 
$(\OGM,\hs\OBW,\hs\OCP)$, respectively.
The values of the coefficients 
$(\mathbb{A},\hs\mathbb{B},\hs\mathbb{C})$ in these formulae
are given by the numerical results for the observables in Eq.\eqref{eq:O-ABC}.\ 
The dependence of each sensitivity limit on the polarization choice 
is determined by the relation between the left- and right-handed couplings.\ The most sensitive observable for probing $\OGP\,(h_4)$ is 
$\mathbb{O}_C^{}$, while the most sensitive observable for probing $\OBW\,(h_3^Z)$ and $\OCP$ 
is $\mathbb{O}_A$.\ 
For probing $\OGM\,(h_3^\gamma)$, the most sensitive observable is $\mathbb{O}_B^{}$ in the case of unpolarized beams, 
and is $\mathbb{O}_A^{}$ in the case of polarized beams
[for the choice $(P_L^e,\hs P_R^{\bar{e}})\!=\!(0.9,\hs 0.65)$].\ 
We note that the sensitivity limits for the nTGC form factors 
scale as $\,(h_3^V\hsm,\hs h_4^V)\!\propto\!\mathbb O_X^{}$ 
(where $X\hsm\!=\!A,B,C$), 
and that the new physics cutoff reaches of the dimension-8 nTGC operators
behave like $\Lambda\!\propto\! \mathbb O_X^{1/4}$.\  
Hence, the improvements from the beam polarizations can be significant  
for the form factors, but become rather mild for the cutoff scales of
the dimension-8 operators.

\vspace*{1mm}

For convenience, we express a given observable 
$\,\mathbb{O}_X^{}\!\equiv\hsm\over{\mathbb{O}}_X^{}/\cut_j^{4}\,$
for the dimension-8 operator formulation and
$\,\mathbb{O}_X^{}\!\equiv\hsm\over{\mathbb{O}}_X^{}h_i^V\,$
for the form factor formulation.\ 
We may estimate the significance by 
$\,\SZZ \!\simeq\! S/\!\sqrt{B\,}$.$\hs$\
If we require the significances of 
the polarized and unpolarized cases to be equal,
$\SZZ_{\rm{pol}}^{}\!=\!\SZZ_{\rm{unpol}}^{}\hs$,
we can derive the following ratio of the polarized/unpolarized
limits on the dimension-8 cutoff scales and on the form factors, respectively:  
%
\begin{align}
\label{eq:Rcut-Rh}
\mathcal{R}_{\cut_j}^{}
=\frac{\cut_j^{}(\rm{pol})}{\,\cut_j^{}(\rm{unpol})\,}
=\big[\over{\mathcal{R}}_X^{}\hsm (P_L^e,P_R^{\bar e})\big]^{\!1/4},
~~~~
\mathcal{R}_{h_i^V}^{}
=\frac{\,h_i^V\!(\rm{unpol})\,}{h_i^V\!(\rm{pol})}
= \over{\mathcal{R}}_X^{}\hsm (P_L^e,P_R^{\bar e})\hs,
\end{align}
where the ratio $\over{\mathcal{R}}_X^{}(P_L^e,P_R^{\bar e})$ 
is defined as
\begin{align}
\label{eq:R_X}
\over{\mathcal{R}}_X^{}(P_L^e,P_R^{\bar e}) 
\hs \equiv\hs 
\frac{\,\over{\mathbb{O}}_X^{}(P_L^e,P_R^{\bar{e}})\,}
{~\over{\mathbb{O}}_X^{}(0.5,0.5)\,}
\sqrt{\frac{\,\sigma_0^{}(0.5,0.5)\,}
{\sigma_0^{}(P_L^e,P_R^{\bar{e}})}\,}\hs .
\end{align}
From the above, we derive the following estimate of the 
ratio $\over{\mathcal{R}}_X^{}$ for each observable $\mathbb{O}_X^{}$:
\beqs
\label{eq:R_ABC}
\beqa 
\over{\mathcal{R}}_A^{}  \!&\!\!=\!\!&
\frac{\,2\hs \big|c_L^{}x_L^{} P_L^eP_R^{\bar{e}}\hsm +\hsm 
c_R^{}x_R^{}(1\!-\!P_L^e)(1\!-\!P_R^{\bar{e}})\big|\,}
{|c_L^{}x_L^{} \hsm +\hsm c_R^{} x_R^{}|}
\sqrt{\frac{c_L^2\!+c_R^2}{\,P_L^e P_R^{\bar e}c_L^2\!+\!
	(1\!-\!P_L^e)(1\!-\!P_R^{\bar e})c_R^2\,}} \hs,
\hspace*{12mm}
\\
\over{\mathcal{R}}_{B,C}^{} \!&\!\!=\!\!&  
\frac{\,2\hs \big|c_L^{}x_L^{} P_L^eP_R^{\bar{e}}\hsm -\hsm 
c_R^{}x_R^{}(1\!-\!P_L^e)(1\!-\!P_R^{\bar{e}})\big|\,}
{|c_L^{}x_L^{} \hsm -\hsm c_R^{} x_R^{}|}
\sqrt{\frac{c_L^2\!+c_R^2}{\,P_L^e P_R^{\bar e}c_L^2\hsm +\hsm 
		(1\!-\!P_L^e)(1\!-\!P_R^{\bar e})c_R^2\,}},
\end{eqnarray}
\eeqs
where 
$(c^{}_L,\hs c^{}_R)\!=\!(T_3^{}\hsm -\hsm Qs_W^2, -Qs_W^2)$
denote the $Z$ coupling factors 
with the (left,\,right)-handed electrons,   
and the coupling coefficients $(x^{}_L,\hs x^{}_R)$ are given by
\beqs 
\vspace*{-1.5mm}
\label{eq:cLR-OG-OBW-OC+}
\begin{align}
(x^{}_L,\,x^{}_R) &\,=\, -Q s_W^2(1,\,1)\hs ,
\hspace*{-15mm}
& (\hs\text{for}~ &\mO_{G-}^{}\hs)\hs ,
\\
(x^{}_L,\,x^{}_R) &\,=\, (T_3^{}\!-\!Qs_W^2,\, -Qs_W^2)\hs ,
\hspace*{-15mm}
& (\hs\text{for}~ &\mO_{\!\widetilde BW}^{}\hs)\hs ,
\\
(x^{}_L,\,x^{}_R) &\,=\, -T_3(1, 0),
\hspace*{-15mm}
& (\hs\text{for}~ & \mO_{G+}^{},\mO_{C+}^{}\hs)\hs .
\end{align}
\eeqs 
%
Because the coupling coefficient $x_R^{}\!=\hsm 0\hs$ for 
$\mO_{G+}(h_4)$ and $\mO_{C+}$, we can reduce 
the significance ratio \eqref{eq:R_ABC} to the following form 
and compute its value for 
$(P_L^e,P_R^{\bar e})\!=\!(0.9,0.65)$: 
\beqs 
\beqa 
\over{\mathcal{R}}_X^{}(P_L^e,P_R^{\bar e})\!\!&=&\!\!
2P_L^e P_R^{\bar e}\sqrt{\frac{c_L^2\!+c_R^2}{\,P_L^e P_R^{\bar e}c_L^2\!+\!(1\!-\!P_L^e)(1\!-\!P_R^{\bar e})c_R^2\,}},
\\[1.5mm]
\over{\mathcal{R}}_X^{}(0.9,0.65) \!\!\!&\simeq&\!\! 2.0\hs , 
\eeqa 
\eeqs 
where $X\!=\!A,B,C$.\
Thus, we deduce the following ratios for the 
operators $(\OGP,\OCP)$ and the form factor $h_4^{}$:
\beqs 
\vspace*{-4mm}
\begin{eqnarray}
\over{\mathcal{R}}_{\cut_{G+}}^{} 
\!\!\!\!&=&\!\! \over{\mathcal{R}}_{\cut_{C+}}^{} \! =\,
\big[\over{\mathcal{R}}_X^{}(0.9,0.65)\big]^{1/4} 
\simeq 1.2\hs ,
\\
\over{\mathcal{R}}_{h_4^{}}^{} \!\!\!\!&=&\!\!\! 
\over{\mathcal{R}}_X^{}(0.9,0.65) \simeq 2.0\hs . 
\end{eqnarray}
\eeqs 
This means that the $e^-/e^+$ beam polarizations can enhance
the sensitivity reach for the cutoff scale $\cut_{G+}^{}$ by
about 20\%, and enhance the sensitivity reach for the form factor
$h_4^{}$ much more significantly, namely by about 100\%, which explains
the features shown in Tables\,\ref{tab:7}-\ref{tab:8} 
and Figs.\hs\ref{fig:8}-\ref{fig:9}.\ 
%


We further note that the coupling coefficients $x_{L,R}^{}\!=\!c_{L,R}^{}$ 
for the nTGC operator $\OBW$ and form factor 
$h_3^Z$.\ We find that to enhance the polarization effects for
probing $\OBW$ and $h_3^Z$, the most sensitive observable is 
$\mathbb{O}_A^{}$.\  Thus,  we simplify 
the significance ratio \eqref{eq:R_ABC} to the following form 
and compute its value for 
$(P_L^e,P_R^{\bar e})\!=\!(0.9,0.65)$: 
\begin{align}
\over{\mathcal{R}}_A(P_L^e,P_R^{\bar e}) &=\,
2\hs\sqrt{\frac{\,P_L^e P_R^{\bar e}c_L^2\!+\!
(1\!-\!P_L^e)(1\!-\!P_R^{\bar e})c_R^2~}{c_L^2\!+c_R^2}},
\\[1mm]
\over{\mathcal{R}}_A(0.9,0.65) &\simeq 1.2 \,.
\end{align}
With these, we deduce the following ratios for the 
operator $\OBW$ and the form factor $h_3^{Z}$:
\beqs 
\vspace*{-1mm}
\begin{eqnarray}
\over{\mathcal{R}}_{\cut_{\widetilde{B}W}}^{} 
\!\!\!\!&=&\!\! 
\big[\over{\mathcal{R}}_A^{}(0.9,0.65)\big]^{1/4} \simeq 1.05 \hs ,
\\
\over{\mathcal{R}}_{h_3^{\hsm Z}}^{} \!\!\!\!&=&\!\!\! 
\over{\mathcal{R}}_X^{}(0.9,0.65) \simeq 1.2\,. 
\end{eqnarray}
\eeqs 
This shows that the beam polarizations can increase mildly 
the sensitivity reach for the cutoff scale 
$\cut_{\widetilde{B}W}^{}$ by
about 5\%, and increase the sensitivity reach for the form factor
$h_3^{Z}$ by a larger amount of 20\%, which agree with 
the features shown in Tables\,\ref{tab:7}-\ref{tab:8} 
and Figs.\hs\ref{fig:8}-\ref{fig:9}.\
Finally, we note that 
the enhancement ratio \eqref{eq:R_X} does not apply to
the cases of $\mO_{G-}$ and $h_3^\ga$ 
because there $\mathbb{O}_B^{}$ is the most sensitive observable 
for the unpolarized case and $\mathbb{O}_A^{}$ is the most 
sensitive observable for the polarized case.\
Thus, we define the corresponding ratio 
$\mathcal{R}_{AB}^{}(P_L^e,P_R^{\bar e})$
of significances between the polarized and unpolarized cases
and compute its value $\mathcal{R}_{AB}^{}(0.9,0.65)$: 
%
\beqs
\vspace*{-4mm}
\begin{align}
\label{eq:R_AB}
\over{\mathcal{R}}_{AB}^{}(P_L^e,P_R^{\bar e}) &\simeq 
\frac{\,\over{\mathbb{O}}_A^{}(P_L^e,P_R^{\bar{e}})\,}
{~\over{\mathbb{O}}_B^{}(0.5,0.5)\,}
\sqrt{\frac{\,\sigma_0^{}(0.5,0.5)\,}
{\sigma_0^{}(P_L^e,P_R^{\bar{e}})}\,}\hs ,
\\[1mm]
\over{\mathcal{R}}_{AB}^{}(0.9,0.65) &\simeq 2.6 \,.
\end{align}
\eeqs
From these we derive the significance ratios for the 
operator $\OGM$ and the form factor $h_3^{\ga}$:
\beqs 
\vspace*{-1mm}
\begin{eqnarray}
\mathcal{R}_{\cut_{G-}}^{} \!\!\!\!&=&\!\! 
\big[\over{\mathcal{R}}_{AB}^{}(0.9,0.65)\big]^{1/4} \simeq 1.27 \hs ,
\\
R_{h_3^{\hsm\ga}}^{} \!\!\!\!&=&\!\!\! 
\over{\mathcal{R}}_X^{}(0.9,0.65) \simeq 2.6\,. 
\end{eqnarray}
\eeqs 
We see that the beam polarization effects can raise 
the sensitivity reach for the cutoff scale 
$\cut_{G-}^{}$ by
about 27\%, and raise the sensitivity reach for the form factor
$h_3^{\ga}$ by about 160\%, which agree with 
the results presented in Tables\,\ref{tab:7}-\ref{tab:8} 
and Figs.\hs\ref{fig:8}-\ref{fig:9}.\

\section{\hspace*{-2.5mm}Conclusions}
\label{sec:DandC}
\label{sec:6}
\vspace*{1mm}

Neutral triple-gauge couplings (nTGCs) provide an important window
for probing new physics beyond the SM.\ 
In this work, we have studied systematically 
the prospective experimental 
sensitivities to nTGCs at the 13\,TeV LHC
and a future 100\,TeV $pp$ collider, 
using the SMEFT approach to classify and characterize the
nTGCs that can arise from gauge-invariant dimension-8 operators. 

\vspace*{1mm}

In Section\,\ref{sec:2.1} we first considered a set of
CP-conserving dimension-8 nTGC operators and the related contact
operators in Eq.\eqref{eq:nTGC-d8}.\
Then, in Section\,\ref{sec:2.2}
we derived their contributions to the scattering amplitudes
of the partonic process $\bar{q}{\hs}q \hsm\ito\hsm Z \gamma\hs$
in Eqs.\eqref{eq:T8} and \eqref{eq:T8-OG-OBW-OC+}.\ 
With these, we computed the corresponding total cross sections 
including the SM contribution, the interference term of
$O(1/\Lambda^4)$, and the squared term of 
$O(1/\Lambda^8)$, as in 
Eqs.\eqref{eq:CS-qq-Zgamma} and \eqref{eq:CS-qqZA-d8other},
where $\cut$ is the new physics cutoff scale 
defined in Eq.\eqref{cj}.\
We further presented in 
Eq.\eqref{eq:f-phi*-OGP}, Eq.\eqref{eq:f-phi*-Ojx}, 
and Fig.\,\ref{fig:2}
their contributions to the differential angular distributions, 
in comparison with that of the SM.\
In Section\,\ref{sec:2.3} we analyzed the perturbative 
unitarity bounds on the nTGCs, as shown in Table\,\ref{tab:0}
and Fig.\hs\ref{fig:0}, which are much weaker than the 
collider limits presented in Sections\,\ref{sec:4}-\ref{sec:5}.\
Hence, the perturbation expanison is well justified for the current
collider analyses.

\vspace*{1mm}

In Section\,\ref{sec:3} we presented a new form factor formulation 
of the neutral triple gauge vertices (nTGVs)  $Z\gamma V^*$ 
(with $V\hsm\!=\hsm\! Z,\ga$), 
by mapping them to the dimension-8 nTGC operators
of the SMEFT that incorporate the spontaneously-broken
electroweak gauge symmetry 
SU(2)$_{\rm L}^{}\otimes$U(1)$_{\rm Y}^{}$ of the SM.
This differs from the conventional form factor parametrization  
of nTGCs that takes into account only the unbroken 
U(1)$_{\rm{em}}$ gauge symmetry\,\cite{nTGC1}\cite{Degrande:2013kka}.\ 
Using the SMEFT approach, we have found that {\it a new momentum-dependent 
nTGC term with form factor $h_5^V$ has to be added and the mapping
with the dimension-8 SMEFT interactions enforces
new nontrivial relations \eqref{eq:h4-h5}-\eqref{eq:h4Z-h4A}
between the form factors $(h_4^V\hsmx ,\hs h_5^V)$ 
and between the form factors $(h_4^Z\hsm ,\,h_4^\gamma)$.}\  
The new form factor $h_5^V$ was not included in all the previous
form factor analyses of nTGVs.\ We have demonstrated that 
including the new form factor $h_5^V$ is crucial for a fully consistent
form factor formulation of nTGVs and ensures the exact cancellation
of the spuriously large unphysical terms of $O(E^5)$
in the scattering amplitudes of 
$\hs q{\hs}\bar{q} \ito Z \gamma\hs$,
as shown in Eqs.\eqref{eq:FT8-ZL} and \eqref{eq:E5cancel-h4h5}.\
In consequence, among the six general nTGC form factors 
$(h_3^V\hsmx ,\,h_4^V\hsmx ,\,h_5^V)$ in Eq.\eqref{eq:ZAV*-FormF},  
we have proven that {\it only three of them, 
$(h_3^Z\hsmx ,\,h_3^\gamma,\,h_4^{})$
with $\hs h_4^{}\!\equiv\! h_4^Z\hs$,  
are independent, and the correct nTGC form factor formula is
given by Eq.\eqref{eq:FF2-nTGC}.}\ 
We have further presented the explicit correspondence 
between the nTGC form factors and the cutoff scales of the 
dimension-8 nTGC operators in 
Eqs.\eqref{eq:h-dim8}-\eqref{eq:[Lambda-4]-r34}.  

\vspace*{1mm}

In Section\,\ref{sec:4}, we have systematically studied
the sensitivity reaches for probing the new physics scales
of the nTGC operators and for probing the nTGC form factors
in the reactions $\,pp\hs (q{\hs}\bar{q})\!\to\!Z\ga\!\to\! 
\ell\bar{\ell}\ga,\hs\nu\bar{\nu}\ga$\,
at the LHC and the future $pp$\,(100{\hs}TeV) collider.\ 
We have presented analyses of sensitivity reaches
using the interference contributions of $O(\cut^{-4})$ 
in Section\,\ref{sec:4.1} and 
including the squared contributions up to $O(\cut^{-8})$
in Section\,\ref{sec:4.2}.\
We have evaluated the prospective 
$2{\hs}\sigma$ and $5{\hs}\sigma$ sensitivities 
of the LHC and the future 100\,TeV $pp$ collider 
to the different nTGCs, 
and have combined the sensitivity reaches 
of the leptonic decay channel 
$Z \ito \ell^+ \ell^-$ (Secs.\,\ref{sec:4.1}-\ref{sec:4.2})  
and the invisible decay channel 
$Z\ito \nu{\hs}\bar{\nu}\hs$
(Sec.\,\ref{sec:4.3}).\ 
We have presented our findings in Tables\,\ref{tab:1} to \ref{tab:3}
for the dimension-8 operator $\hs\OGP$ 
and the equivalent operator $\hs\OCM$, 
and in Table\,\ref{tab:4} for the
other dimension-8 operators $(\OBW,\,\OGM,\,\OCP)$.\ 
These sensitivity reaches are further summarized 
in our Fig.\,\ref{fig:8}(a).\ 
From Table\,\ref{tab:3}, we see that 
the $2{\hs}\sigma$\,($5{\hs}\sigma$) 
sensitivity to the scale of the operator 
$\OGP$ could reach 4.4\,TeV (3.9\,TeV) at the 13\,TeV LHC
with 3\,ab$^{-1}$ integrated luminosity, and reach 
30\,TeV\,(26\,TeV)
at the 100\,TeV $pp$ collider with 30\,ab$^{-1}$, 
whereas the estimated sensitivity reaches on the scales
of the dimension-8 operators $(\OBW,\,\OGM,\,\OCP)$ 
shown in Table\,\ref{tab:4} are somewhat smaller.\ 
Then, in Section\,\ref{sec:4.4}
we have presented the LHC sensitivity reaches on the 
three independent form factors
$(h_4^{},\,h_3^Z\hsmx,\,h_3^\gamma)\hs$ in Table\,\ref{tab:5},
with a summary of these sensitivities 
given in Fig.\,\ref{fig:9}(a).\ 
We see that the sensitivities for probing the form factor
$h_4^{}$ are generally higher than those of the other two
form factors $(h_3^Z,\,h_3^\gamma)$ by about a factor of 
$5\!\times\!10^{-2}$ at the LHC and by about a factor of 
$10^{-2}$ at the 100\,TeV $pp$ collider.\
We emphasise that if the dimension-8 SMEFT relations between the different form factors are not taken into account,
one would find unrealistically strong sensitivities 
due to the uncancelled large unphysical energy-dependent terms
associated with the form factor $h_4^{}\hs$,
as seen by comparing Eq.\eqref{eq:CS12-Xh4V-size}
with Eq.\eqref{eq:CS12-h4h3-size}.\ 
Then, we explicitly demonstrated in Table\,\ref{tab:6} that
the sensitivities to $h_4^Z$ and $h_4^\gamma$ in the conventional
form factor approach (marked in blue color) are (erroneously)
higher than the correct sensitivities 
(marked in red color and extracted from Table\,\ref{tab:5}) 
by about a factor of
$5\!\times\!10^{-2}$ at the LHC and by about a factor of 
$10^{-2}$ at the $pp${\hs}(100{\hs}TeV) collider.\ 
Hence, it is important to use the consistent 
form factor approach for the nTGC analysis 
as we advocated in Section\,\ref{sec:3}. 
After these comparisons, in Section\,\ref{sec:4.5} we  analyzed 
the 2-parameter correlations for both the nTGC
form factors and for the nTGC dimension-8 operators.\ 
We presented in Fig.\ref{fig:4} the correlations of 
each pair of the form factors $(h_4^{},\hs h_3^V)$ 
and $(h_3^Z,\hs h_3^\gamma)$ at hadron colliders, 
where the $(h_4^{},\hs h_3^V)$ contours 
in the plots\,(a) and (b) have rather weak correlations
due to the extra energy-suppression factor 
of Eq.\eqref{eq:rho-h3h4}, and 
the plots\,(c) and (d) demonstrate large correlations 
between the form factors $(h_3^Z,\hs h_3^\gamma)$.\
Then, we presented 
the correlations of each pair of the nTGC operators
$(\OGP,\,\OBW)$ and $(\OGP,\,\OGM)$ 
in Figs.\,\ref{fig:5}(a)-(b) which are suppressed by
large energy factor $\hs 1/\!\sqrt{\sbar\,}\,$ 
as shown in  Eq.\eqref{eq:rho-OG+OV}.\ 
The correlations of each pair of the nTGC operators
$(\OBW,\,\OGM)$, $(\OCP,\,\OBW)$, and $(\OCP,\,\OGM)$
are presented in Figs.\,\ref{fig:6}(a) to (d).\ 
These correlations are not suppressed by
any energy factor and are thus significant
at both the LHC and the 100\,TeV $pp$ collider.\
We demonstrated in Figs.\,\ref{fig:6}(c)-(d) that
the correlations of the operators
$(\OCP,\,\OBW)$ and $(\OCP,\,\OGM)$
are particularly strong.
Finally, in Section\,\ref{sec:4.7} we have made direct comparison 
with the published LHC measurements on nTGCs 
in the reaction 
$\hs p{\hs}p{\hs}(q{\hs}\bar{q})\hsm\ito\hsm Z\gamma\hs$
(with $Z\!\ito\nu{\hs}\bar{\nu}{\hs})$ 
by the CMS\,\cite{CMS2016nTGC-FF} and ATLAS\,\cite{Atlas2018nTGC-FF}
collaborations.\ Using the same kinematic cuts and integrated
luminosity together with an estimated detection efficiency
as in the ATLAS analysis\,\cite{Atlas2018nTGC-FF}, 
we have applied the conventional form factor formula 
\eqref{eq:FF0-nTGC} to reproduce the nTGC bounds 
in Eq.\eqref{eq:Atlas-Xh34-theory} 
and the strong correlations of $(h_3^V\hsmx,\hs h_4^V)$ 
in Figs.\,\ref{fig:10}(c)-(d), which agree well with the
ATLAS results\,\cite{Atlas2018nTGC-FF}.\ 
However, the $(h_3^V\hsmx,\hs h_4^V)$ contours 
of Figs.\,\ref{fig:10}(c)-(d)
differ substantially from those contours 
of Figs.\,\ref{fig:10}(a)-(b),
which exhibit rather weak correlations as predicted using
our new SMEFT form factor formula \eqref{eq:FF2-nTGC}.\
Hence, it is important to use the SMEFT form factor formulation
described in Section\,\ref{sec:3} to analyze the 
LHC bounds on nTGCs.

\vspace*{1mm}

We presented in Section\,\ref{sec:5}
systematic comparisons of the sensitivity reaches 
for the nTGCs between the hadron colliders 
(the LHC and the 100\,TeV $\!pp$ collider) and 
$e^+ e^-$ colliders with different energies.\ 
Table\,\ref{tab:7} summarizes the comparisons for probing the
nTGCs of dimension-8 operators $(\OGP,\,\OBW,\,\OGM,\,\OCP)$, 
whereas Table\,\ref{tab:8} summarizes the comparisons for probing 
the nTGC form factors $(h_4^{},\,h_3^Z,\,h_3^\gamma)$.\ 
We have summarized the above comparisons of sensitivity reaches
between the hadron colliders and lepton colliders in 
Figs.\,\ref{fig:5} and \ref{fig:8}.\ 
Then, in Table\,\ref{tab:9}, we have further demonstrated that
using naively the conventional form factor formula
without including the nontrivial constraints of the
dimension-8 SMEFT approach would cause erroneous
sensitivities to $(h_4^Z,\,h_4^\gamma)$ (marked in blue color)
that are stronger than the correct sensitivities (marked in red color
and extracted from Table\,\ref{tab:8}) at the $e^+e^-$ colliders
by a factor of $\hs O(10)\hs$ for the collision energy  
$\sqrt{s\,} \!\leqq\!1$\,TeV 
and by a factor of 
$\hs O(10^2)\hs$ for $\sqrt{s\,} \hsm =\hsm (3 - 5)\hs$TeV.\ 
Hence, it is important to use the consistent form factor approach
of Section\,\ref{sec:3}
for nTGC analyses at $e^+e^-$ colliders.\
In general, from the comparisons of
Tables\,\ref{tab:7}-\ref{tab:8} and 
Figs.\,\ref{fig:8}-\ref{fig:9},
we find that the LHC sensitivity reaches 
on the nTGCs are similar to those at 
the $e^+ e^-$ colliders with collision energy 
$\hsm\sqrt{s\,} \!\hsm\leqq\hsm\!1$\,TeV~\cite{Ellis:2020ljj}.\  
On the other hand, a higher-energy $e^+ e^-$ collider with 
$\sqrt{s\,} \hsm =\! (3\! -\! 5)\hs$TeV
would have greater sensitivities than the LHC to probing 
the new physics scales of the nTGC operators and the
corresponding nTGC form factors.\ 
However, we have shown that 
the sensitivity reaches of the 100\,TeV $pp\hs$ 
collider would be even higher.

\vspace*{1mm}

Overall, we have found that nTGCs provide a powerful means  
for probing any possible new physics beyond the SM
that could generate the dimension-8 nTGC operators in the SMEFT.\ 
We have found that both $\hs pp\hs$ and $e^+ e^-$ colliders
have significant roles to play. 
We advocate as a first step that the ATLAS and CMS experiments 
at the LHC apply the dimension-8 SMEFT approach proposed here 
to analyze the nTGCs, 
in preference to the conventional form factor approach 
that does not take into account the full electroweak gauge symmetry
SU(2)$_{\rm L}^{}\otimes\hs$U(1)$_{\rm Y}^{}$ of the SM.

\vspace*{5mm}
\noindent
{\bf\large Acknowledgements}
\\[1mm]
The work of JE was supported in part by United Kingdom STFC Grants ST/P000258/1 and ST/T000759/1,
in part by the Estonian Research Council via a Mobilitas Pluss grant, and in part
by the TDLI distinguished visiting fellow programme.\ 
The work of HJH and RQX was
supported in part by the National NSF of China (under grants Nos.\,11835005 and 12175136) and by the National Key R\,\&\,D 
Program of China (No.\,2017YFA0402204).\ 
RQX has been supported by an International Postdoctoral 
Exchange Fellowship.


\vspace*{3mm}

\end{document}